\documentclass[letterpaper,twocolumn,10pt]{article}
\usepackage{usenix,epsfig,endnotes}
\usepackage{amsmath}
\usepackage[utf8]{inputenc}
\usepackage[T1]{fontenc}
\usepackage{amsmath}
\usepackage[english]{babel}
\usepackage[table]{xcolor}
\usepackage{graphicx}
\usepackage{listings}
\usepackage{caption}
\usepackage{subcaption}
\usepackage{color}
\usepackage[linesnumbered,ruled,vlined]{algorithm2e}
\usepackage{url}
\usepackage{listings}
\usepackage{color}
\usepackage{textcomp}

\definecolor{dkgreen}{rgb}{0,0.6,0}
\definecolor{gray}{rgb}{0.5,0.5,0.5}
\definecolor{mauve}{rgb}{0.58,0,0.82}

\begin{document}

%don't want date printed
\date{}

%make title bold and 14 pt font (Latex default is non-bold, 16 pt)
% Ibdxnet: A low latency and high throughput InfiniBand network transport for Java
\title{\Large \bf Ibdxnet: Leveraging InfiniBand in Highly Concurrent Java Applications}

%for single author (just remove % characters)
\author{
{\rm Stefan~Nothaas}\\
stefan.nothaas@hhu.de
\and
{\rm Kevin~Beineke}\\
kevin.beineke@hhu.de
\and
{\rm Michael~Schöttner}\\
michael.schoettner@hhu.de
\and
Institut für Informatik, Heinrich-Heine-Universität Düsseldorf\\
Universitätsstr. 1, 40225 Düsseldorf, Germany
} % end author

\maketitle

% Use the following at camera-ready time to suppress page numbers.
% Comment it out when you first submit the paper for review.
%\thispagestyle{empty}

%%%%%%%%%%%%%%%%%%%%%%%%%%%%%%%%%%%%%%%%%%%%%%%%%%%%%%%%%%%%%%%%%%%%%%%%%%%%%%%%%%%%%

\subsection*{Abstract}
Todays big data applications generate hundreds or even thousands of terabytes of data. Commonly, Java based applications are used for further analysis. A single commodity machine, for example in a data center or typical cloud environment, cannot store and process the vast amounts of data making distribution mandatory. Thus, the machines have to use interconnects to exchange data or coordinate data analysis. However, commodity interconnects used in such environments, e.g. Gigabit Ethernet, cannot provide high throughput and low latency compared to alternatives like InfiniBand to speed up data analysis of the target applications. In this report, we describe the design and implementation of Ibdxnet, a low-latency and high-throughput transport providing the benefits of InfiniBand networks to Java applications. Ibdxnet is part of the Java-based DXNet library, a highly concurrent and simple to use messaging stack with transparent serialization of messaging objects and focus on very small messages (< 64 bytes). Ibdxnet implements the transport interface of DXNet in Java and a custom C++ library in native space using JNI. Several optimizations in both spaces minimize context switching overhead between Java and C++ and are not burdening message latency or throughput. Communication is implemented using the messaging verbs of the ibverbs library complemented by an automatic connection management in the native library. We compared DXNet with the Ibdxnet transport to the MPI implementations FastMPJ and MVAPICH2. For small messages up to 64 bytes using multiple threads, DXNet with the Ibdxnet transport achieves a bi-directional message rate of 10 million messages per second and surpasses FastMPJ by a factor of 4 and MVAPICH by a factor of 2. Furthermore, DXNet scales well on a high load all-to-all communication with up to 8 nodes achieving a total aggregated message rate of 43.4 million messages per second for small messages and a throughput saturation of 33.6 GB/s with only 2 kb message size.

%%%%%%%%%%%%%%%%%%%%%%%%%%%%%%%%%%%%%%%%%%%%%%%%%%%%%%%%%%%%%%%%%%%%%%%%%%%%%%%%%%%%%
%%%%%%%%%%%%%%%%%%%%%%%%%%%%%%%%%%%%%%%%%%%%%%%%%%%%%%%%%%%%%%%%%%%%%%%%%%%%%%%%%%%%%
%%%%%%%%%%%%%%%%%%%%%%%%%%%%%%%%%%%%%%%%%%%%%%%%%%%%%%%%%%%%%%%%%%%%%%%%%%%%%%%%%%%%%
\section{Introduction}

Interactive applications, especially on the web \cite{facebook2, Liu:2016:ECI:2964797.2964815}, simulations \cite{doi:10.1093/bioinformatics/btt055} or online data analysis \cite{Desikan:2005:IPR:1062745.1062885, 6547630, DOI:10.1007/978-3-319-55699-4_20} have to process terabytes of data often consisting of small objects. For example, social networks are storing graphs with trillions of edges resulting in a per object size of less than 64 bytes for the majority of objects \cite{Ching:2015:OTE:2824032.2824077}. Other graph examples are brain simulations with billions of neurons and thousands of connections each \cite{IntroducingGraph500} or search engines for billions of indexed web pages \cite{Gulli:2005:IWM:1062745.1062789}. To provide high interactivity to the user, low latency is a must in many of these application domains. Furthermore, it is also important in the domain of mobile networks moving state management into the cloud \cite{Kablan:2015:SNF:2785989.2785993}.

Big data applications are processing vast amounts of data which require either an expensive supercomputer or distributed platforms, like clusters or cloud environments \cite{HASHEM201598}. High performance interconnects, such as InfiniBand, are playing a key role to keep processing and response times low, especially for highly interactive and always online applications. Today, many cloud providers, e.g. Microsoft, Amazon or Google, offer instances equipped with InfiniBand.

InfiniBand offers messaging verbs and RDMA, both providing one way single digit microsecond latencies. It depends on the application requirements whether messaging verbs or RDMA is the better choice to ensure optimal performance \cite{Su:2017:RRF:3064176.3064189}.

In this report, we focus on Java-based parallel and distributed applications, especially big data applications, which commonly communicate with remote nodes using asynchronous and synchronous messages \cite{Ching:2015:OTE:2824032.2824077, Ekanayake:2016:SJH:2972969.2972972, Dean:2008:MSD:1327452.1327492, Zaharia:2016:ASU:3013530.2934664}. Unfortunately, accessing InfiniBand verbs from Java is not a built-in feature of the commonly used JVMs. There are several external libraries, wrappers or JVMs with built-in support available but all trade performance for transparency or require proprietary environments (\S \ref{related_work_java_ib}). To use InfiniBand from Java, one can rely on available (Java) MPI implementations. But, these are not providing features such as serialization for messaging objects and no automatic connection management (\S \ref{related_work_mpi}).

We developed the network subsystem DXNet (\S \ref{dxnet}) which provides transparent and simple to use sending and event based receiving of synchronous and asynchronous messages with transparent serialization of messaging objects \cite{dxnet}. It is optimized for high concurrency on all operations by implementing lock-free synchronization. DXNet is implemented in Java and open source and available at Github \cite{dxnetgithub}.

In this report, we propose Ibdxnet, a transport for the DXNet network subsystem. The transport uses reliable messaging verbs to implement InfiniBand support for DXNet and provides low latency and high throughput messaging for Java.

Ibdxnet implements scalable and automatic connection and queue pair management, the \textit{msgrc} transport engine, which uses InfiniBand messaging verbs, and a JNI interface. We present best practices applied to ensure scalability across multiple threads and nodes when working with InfiniBand verbs by elaborating on the implementation details of Ibdxnet. We carefully designing an efficient and low latency JNI layer to connect the native Ibdxnet subsystem to the Java based IB transport in DXNet. The IB transport uses the JNI layer to interface with Ibdxnet, extends DXNet's outgoing ring buffer for InfiniBand usage and implements scalable scheduling of outgoing data for many simultaneous connections. We evaluated DXNet with the IB transport and Ibdxnet, and compared then to two MPI implementations supporting InfiniBand: the well known MVAPICH2 and the Java based FastMPJ implementations.

Though, MPI is discussed in related work (\S \ref{related_work_mpi}) and two implementations are evaluated and compared to DXNet (\S \ref{eval}), DXNet, the IB transport nor Ibdxnet are implementing the MPI standard. The term \textit{messaging} is used by DXNet to simply refer to exchanging data in the form of messages (i.e. additional metadata identifies message on receive). DXNet does not implement any by the standard defined MPI primitives. Various low-level libraries to use InfiniBand in Java are not compared in this report, but in a separate one.

The report is structured in the following way: In Section \ref{dxnet}, we present a summary of DXNet and its aspects important to this report. In Section \ref{related_work}, we discuss related work which includes a brief summary of available libraries and middleware for interfacing InfiniBand in Java applications. MPI and selected implementations supporting InfiniBand are presented as available middleware solutions and compared to DXNet. Lastly, we discuss target applications in the field of Big-Data which benefit from InfiniBand usage. Section \ref{ib_basics} covers InfiniBand basics which are of concern for this report. Section \ref{java_and_native} discusses JNI usage and presents best practices for low latency interfacing with native code from Java using JNI. Section \ref{overview_infiniband_transport} gives a brief overview of DXNet's multi layered stack when using InfiniBand. Implementation details of the native part Ibdxnet are given in Section \ref{ibdxnet_native} and the IB transport in Java are presented in Section \ref{transport_impl_java}. Section \ref{eval} presents and compares the experimental results of MVAPICH2, FastMPJ and DXNet. Conclusions are presented in Section \ref{conclusions}.

%%%%%%%%%%%%%%%%%%%%%%%%%%%%%%%%%%%%%%%%%%%%%%%%%%%%%%%%%%%%%%%%%%%%%%%%%%%%%%%%%%%%%
%%%%%%%%%%%%%%%%%%%%%%%%%%%%%%%%%%%%%%%%%%%%%%%%%%%%%%%%%%%%%%%%%%%%%%%%%%%%%%%%%%%%%
%%%%%%%%%%%%%%%%%%%%%%%%%%%%%%%%%%%%%%%%%%%%%%%%%%%%%%%%%%%%%%%%%%%%%%%%%%%%%%%%%%%%%
\section{DXNet}
\label{dxnet}
DXNet is a network library for Java targeting, but not limited to, highly concurrent big data applications. DXNet implements an \textbf{asynchronous event driven messaging} approach with a simple and easy to use application interface. \textbf{Messaging} describes \textbf{transparent sending and receiving of complex (even nested) data structures} with implicit serialization and de-serialzation. Furthermore, DXNet provides a built in primitive for transparent \textbf{request-response communication}.

DXNet is optimized for highly multi-threaded sending and receiving of small messages by using \textbf{lock-free data structures, fast concurrent serialization, zero copy and zero allocation}. The core of DXNet provides \textbf{automatic connection and buffer management}, serialization of message objects and an interface for implementing different transports. Currently, an Ethernet transport using Java NIO sockets and an InifiniBand transport using \textit{ibverbs} (\S \ref{ibdxnet_native}) is implemented.

\begin{figure}[!t]
	\centering
	\includegraphics[width=3.0in]{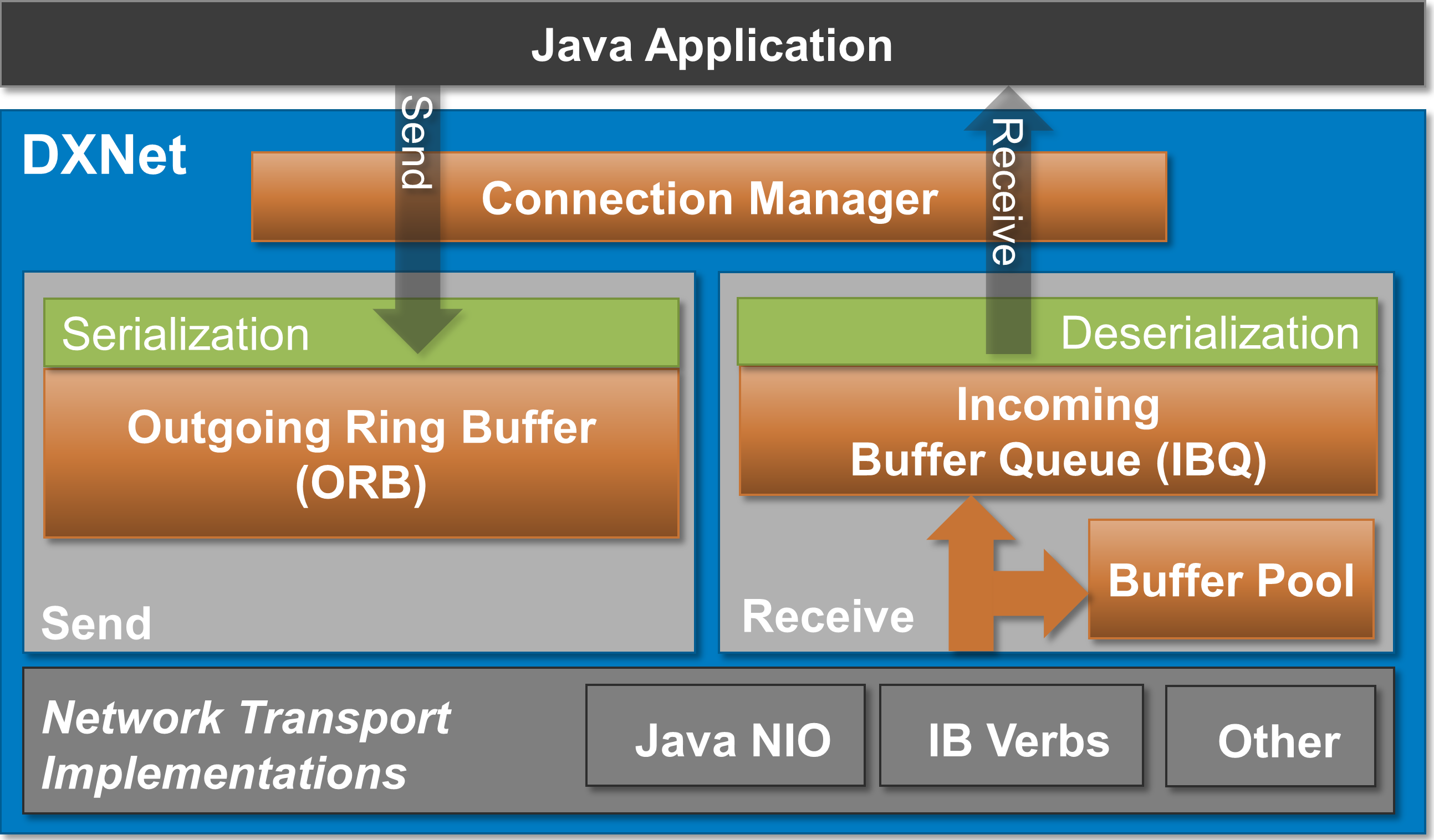}
	\caption{Simplified DXNet Architecture}
	\label{dxnet_simple_fig}
\end{figure}

The following subsections describe the most important aspects of DXNet and its core which are depicted in Figure \ref{dxnet_simple_fig} and relevant for further sections of this report. A more detailed insight is given in a dedicated paper \cite{dxnet}. The source code is available at Github \cite{dxnetgithub}.

%%%%%%%%%%%%%%%%%%
\subsection{Automatic Connection Management}
\label{dxnet_con_man}
To relieve the programmer from explicit connection creation, handling and cleanup, DXNet implements automatic and transparent connection creation, handling and cleanup. Nodes are addressed using an \textbf{abstract and unique 16-bit nodeID}. Address mappings must be registered to allow associating the nodeIDs of each remote node with a corresponding implementation dependent endpoint (e.g. socket, queue pair). To provide scalability with up to hundreds of simultaneous connections, our event driven system does not create one thread per connection. A \textbf{new connection is created automatically} once the first message is either sent to a destination or received from one. Connections are closed once a configurable connection limit is reached using a recently used strategy. Faulty connections (e.g. remote node not reachable anymore) are handled and cleaned up by the manager. Error handling on connection errors or timeouts is propagated to the application using exceptions.

%%%%%%%%%%%%%%%%%%
\subsection{Sending of Messages}
\label{dxnet_send}
\textbf{Messages} are serialized Java objects and sent \textbf{asynchronously} without waiting for a completion. A message can be targeted towards one or multiple receivers. Using the message type \textbf{Request}, it is sent to one receiver, only. When sending a request, the sender waits until \textbf{receiving a corresponding response} message (transparently handled by DXNet) or skips waiting and collects the response later.

We expect applications calling DXNet concurrently with \textbf{multiple threads} to send messages. Every message is automatically and concurrently serialized into the \textbf{Outgoing Ring Buffer (ORB)}, a natively allocated and lock-free ring buffer. \textbf{Messages are automatically aggregated} which increases send throughput. The ORB, one per connection, is allocated in native memory to allow \textbf{direct and zero-copy access} by the low-level transport. A transport runs a decoupled dedicated thread which removes the serialized and ready to send data from the ORB and forwards it to the hardware.

%%%%%%%%%%%%%%%%%%
\subsection{Receiving of Messages}
\label{dxnet_receive}
The network transport handles incoming data by writing it to \textbf{pooled native buffers} to avoid burdening the Java garbage collection. Depending on how a transport writes and reads data, the buffers might contain fully serialized messages or just fragments. Every received buffer is pushed to the ring buffer based \textbf{Incoming Buffer Queue (IBQ)}. Both, the buffer pool as well as the IBQ are shared among all connections. \textbf{Dedicated handler threads} pull buffers from the IBQ and process them asynchronously by de-serializing them and creating Java message objects. The messages are passed to \textbf{pre-registered callback methods} of the application.

%%%%%%%%%%%%%%%%%%
\subsection{Flow Control}
DXNet implements its own \textbf{flow control (FC)} mechanism to avoid flooding a remote node with many (very small) messages. This would result in an increased overall latency and lower throughput if the receiving node cannot keep up with processing incoming messages. On sending a message, the per connection dedicated FC checks if a configurable threshold is exceeded. This threshold describes the \textbf{number of bytes sent by the current node but not fully processed by the receiving node}. Once the configurable threshold is exceeded, the receiving node slices the number of bytes received into equally sized windows (window size configurable) and sends the number of windows confirmed back to the source node. Once the sender receives this confirmation, the number of bytes sent but not processed is \textbf{reduced by the number of received windows multiplied with the configured window size}. If an application send thread was previously blocked due to exceeding this threshold, it can now continue with processing.

%%%%%%%%%%%%%%%%%%
\subsection{Transport Interface}
\label{dxnet_transport_interface}
DXNet provides a transport interface allowing implementations of different transport types. On initialization of DXNet, one of the implemented transports can be selected. Afterwards when using DXNet, the transport is transparent for the application. The following tasks must be handled by every transport implementation:

\begin{itemize}
 \item Connection: Create, close and cleanup
 \item Get ready to send data from ORB and send it (ORB triggers callback once data is available)
 \item Handle received data by pushing it to the IBQ
 \item Manage flow control when sending/receiving data
\end{itemize}

Every other task that is not exposed directly by one of the following methods must be handled internally by the transport. The core of DXNet relies on the following methods of abstract Java classes/interfaces which must be implemented by every transport:

\begin{itemize}
 \item Connection: open, close, dataPosted
 \item ConnectionManager: createConnection, closeConnection
 \item FlowControl: sendFlowControlData, getAndResetFlowControlData
\end{itemize}

We elaborate on further details about the transport interface in Section \ref{transport_impl_java} where we describe the transport implementation for Ibdxnet.

%%%%%%%%%%%%%%%%%%%%%%%%%%%%%%%%%%%%%%%%%%%%%%%%%%%%%%%%%%%%%%%%%%%%%%%%%%%%%%%%%%%%%
%%%%%%%%%%%%%%%%%%%%%%%%%%%%%%%%%%%%%%%%%%%%%%%%%%%%%%%%%%%%%%%%%%%%%%%%%%%%%%%%%%%%%
%%%%%%%%%%%%%%%%%%%%%%%%%%%%%%%%%%%%%%%%%%%%%%%%%%%%%%%%%%%%%%%%%%%%%%%%%%%%%%%%%%%%%
\section{Related Work}
\label{related_work}
Related work discusses different topics which are of interest to DXNet with the IB transport and Ibdxnet. First, we present a summary of our evaluation results of existing solutions to use InfiniBand in Java applications (\S \ref{related_work_java_ib}). These results were important before developing Ibdxnet. Next, we compare DXNet and the MPI standard (\S \ref{related_work_mpi}) followed by MPI implementations supporting InfiniBand (\S \ref{related_work_mpi_impls}) and UPX (\S \ref{rel_work_other}). To our knowledge, this concludes the list of available middleware offering higher level networking primitives comparable to DXNet's. In the last Subsection \ref{rel_work_big_data}, we discuss big-data systems and applications supporting InfiniBand for target applications of interest to DXNet.

%%%%%%%%%%%%%%%%%%
\subsection{Java and InfiniBand}
\label{related_work_java_ib}
Before developing Ibdxnet and the InfiniBand transport for DXNet, we evaluated available (low-level) solutions for leveraging InfiniBand hardware in Java applications. This includes using NIO sockets with \textbf{IP over InfiniBand (IPoIB)} \cite{ipoib}, \textbf{jVerbs} \cite{Stuedi:2013:JUL:2523616.2523631}, \textbf{JSOR} \cite{jsor}, \textbf{libvma} \cite{libvma} and \textbf{native c-verbs with ibverbs}. Extensive experiments analyzing throughput and latency of both messaging verbs and RDMA were conducted to determine a suitable candidate for using InfiniBand with Java applications and are published in a separate report.

Summerized, the results show that transparent solutions like IPoIB, libvma or JSOR, which allow existing socket-based applications to send and receive data transparently over InfiniBand hardware, are not able to deliver an overall adequate throughput and latency. For the verbs-based libraries, jVerbs gets close to the native ibverbs performance but, like JSOR, requires a proprietary JVM to run. Overall, none of the analyzed solutions, other than ibverbs, are delivering an adequate performance. Furthermore, we want DXNet to stay independent of the JVM when using InfiniBand hardware. Thus, we decided to use the native ibverbs library with the Java Native Interface to avoid the known performance issues of the evaluated solutions.

\subsection{MPI}
\label{related_work_mpi}
The message passing interface \cite{Forum:1994:MMI:898758} defines a standard for high level networking primitives to send and receive data between local and remote processes, typically used for HPC applications.

An application can send and receive primitive data types, arrays, derived or vectors of primitive data types, and indexed data types using MPI. The synchronous primitives \textit{MPI\_Send} and \textit{MPI\_Recv} perform these operations in blocking mode. The asynchronous operations \textit{MPI\_Isend} and \textit{MPI\_Irecv} allow non blocking communication. A status handle is returned with each started asynchronous operation. This can be used to check the completion of the operation or to actively wait for one or multiple completions using \textit{MPI\_Wait} or \textit{MPI\_Waitall}. Furthermore, there are various collective primitives which implement more advanced operations such as scatter, gather or reduce.

Sending and receiving of data with MPI requires the application to issue a receive for every send with a target buffer that can hold at least the amount of data sent by the remote. DXNet relieves the application from this responsibility. Application threads can send messages with variable size andDXNet manages the buffers used for sending and receiving. The application does not have to issue any receive operations and wait for data to arrive actively. Incoming messages are dispatched to pre-registered callback handlers by dedicated handler threads of DXNet.

DXNet supports transparent serialization and de-serialization of complex (even nested) data types (Java objects) for messages. MPI primitives for sending and receiving data require the application to use one of the available data types supported and doesn't offer serialization for more complex datatypes such as objects. However, the MPI implementation can benefit from the lack of serialization by avoiding any copying of data, entirely. Due to the nature of serialization, DXNet has to create a (serialized) "copy" of the message when serializing it into the ORB. Analogously, data is copied when a message is created from incoming data during de-serialization.

Messages in DXNet are sent asynchronously while requests offer active waiting or probing for the corresponding response. These communication patterns can also be applied by applications using MPI. The communication primitives currently provided by DXNet are limited to messages and request-response. Nevertheless, using these two primitives, other MPI primitives, such as scatter, gather or reduce, can be implemented by the application if required.

DXNet does not implement multiple protocols for different buffer sizes like MPI with eager and rendezvous. A transport for DXNet might implement such a protocol but our current implementations for Ethernet and InfiniBand do not. The aggregated data available in the ORB is either sent as a whole or sliced and sent as multiple buffers. The transport on the receiving side passes the stream of buffers to DXNet and puts them into the IBQ. Afterwards, the buffers are re-connected to a stream of data by the MCC before extracting and processing the messages.

An instance using DXNet runs within one process of a Big Data application with one or multiple application threads. Typically, one DXNet instance runs per cluster node. This allows the application to dynamically scale the number of threads up or down within the same DXNet instance as needed. Furthermore, fast communication between multiple threads within the same process is possible, too.

Commonly, an MPI application runs a single thread per process. Multiple processes are spawned according to the number of cores per node with IPC fully based on MPI. MPI does offer different thread modes which includes issuing MPI calls using different threads in a process. Typically, this mode is used in combination with OpenMP \cite{openmp}. However, it is not supported by all MPI implementations which also offer InfiniBand support (\S \ref{related_work_mpi_impls}). Furthermore, DXNet supports dynamic up and down scaling of instances. MPI implementations support up-scaling (for non singletons) but down scaling is considered an issue for many implementations. Processes cannot be removed entirely and might cause other processes to get stuck or crash.

Connection management and identifying remote nodes are similar with DXNet and MPI. However, DXNet does not come with deployment tools such as \textit{mpirun} which assigns the ids/ranks to identify the instances. This intentional design decision allows existing applications to integrate DXNet without restrictions to the bootstrapping process of the application. Furthermore, DXNet supports dynamically adding and removing instances. With MPI, an application must be created by using the MPI environment. MPI applications must be run using a special coordinator such as \textit{mpirun}. If executed without a communicator, an MPI world is limited to the current process it is created in which doesn't allow communication with any other instances. Separate MPI worlds can be connected but the implementation must support this feature. To our knowledge, there is no implementation (with InfiniBand support) that currently supports this.

%%%%%%%%%%%%%%%%%%
\subsection{MPI Implementations Supporting InfiniBand}
\label{related_work_mpi_impls}
This section only considers MPI implementations supporting InfiniBand directly. Naturally, IPoIB can be used to run any MPI implementation supporting Ethernet networks over InfiniBand. But, as previously discussed (\S \ref{related_work_java_ib}), the network performance is very limited when using IPoIB.

\textbf{MVAPICH2} is a MPI library \cite{4343853} supporting various network interconnects, such as Ethernet, iWARP, Omni-Path, RoCE and InfiniBand. MVAPICH2 includes features like RDMA fast path or RDMA operations for small message transfers and is widely used on many clusters over the world. \textbf{Open MPI} \cite{openmpi} is an open source implementation of the MPI standard (currently full 3.1 conformance) supporting a variety of interconnects, such as Ethernet using TCP sockets, RoCE, iWARP and InfiniBand. 

\textbf{mpiJava} \cite{mpijava} implements the MPI standard by a collection of wrapper classes that call native MPI implementations, such as MVAPICH2 or OpenMPI, through JNI. The wrapper based approach provides efficient communication relying on native libraries. However, it is not threadsafe and, thus, is not able to take advantage of multi-core systems using multithreading.

\textbf{FastMPJ} \cite{Exposito:2014aa} uses Java Fast Sockets \cite{Taboada:2008:JFS:1456731.1457122} and ibvdev to provide a MPI implementation for parallel systems using Java. Initially, \textbf{ibvdev} \cite{Exposito2012} was implemented as a low-level communication device for \textbf{MPJ Express} \cite{MPJExpress}, a Java MPI implementation of the mpiJava 1.2 API specification. ibvdev implements InfiniBand support using the low-level verbs API and can be integrated into any parallel and distributed Java application. FastMPJ optimizes MPJ Express collective primitives and provides efficient non-blocking communication. Currently, FastMPJ supports issuing MPI calls using a single thread, only.

%%%%%%%%%%%%%%%%%%
\subsection{Other Middleware}
\label{rel_work_other}
\textbf{UCX} \cite{7312665} is a network stack designed for next generation systems for applications with an highly multi-threaded environment. It provides three independent layers: UCS is a service layer with different cross platform utilities, such as atomic operations, thread safety, memory management and data structures. The transport layer UCT abstracts different hardware architectures and their low-level APIs, and provides an API to implement communication primitives. UCP implements high level protocols such as MPI or PGAS programming models by using UCT.

UCX aims to be a common computing platform for multi-threaded applications. However, DXNet does not and, thus, does not include its own atomic operations, thread safety or memory management for data structures. Instead, it relies on the multi-threading utilities provided by the Java environment. DXNet does abstract different hardware like UCX but only network interconnects and not GPUs or other co-processors. Furthermore, DXNet is a simple networking library for Java applications and does not implement MPI or PGAS models. Instead, it provides simple asynchronous messaging and synchronous request-response communication, only.

%%%%%%%%%%%%%%%%%%
\subsection{Target Applications using InfiniBand}
\label{rel_work_big_data}
Providing high throughput and low latency, InfiniBand is a technology which is widely used in various big-data applications.

\textbf{Apache Hadoop} \cite{Islam:2012:HPR:2388996.2389044} is a well known Java big-data processing framework for large scale data processing using the MapReduce programming model. It uses the Hadoop Distributed File System for storing and accessing application data which supports InfiniBand interconnects using RDMA. Also implemented in Java, \textbf{Apache Spark} is a framework for big-data processing offering the domain-specific-language Spark SQL, a stream processing and machine learning extension and the graph processing framework GraphX. It supports InfiniBand hardware using an additional RDMA plugin \cite{sparkrdma}.

Numerous key-value storages for big-data applications have been proposed that use InfiniBand and RDMA to provide low latency data access for highly interactive applications.

\textbf{RAMCloud} \cite{Ousterhout:2015:RSS:2818727.2806887} is a distributed key-value storage optimized for low latency data access using InfiniBand with messaging verbs. Multiple transports are implemented for network communication, e.g. using reliable and unreliable connections with InfiniBand and Ethernet with unreliable connections. \textbf{FaRM} \cite{179767} implements a key-value and graph storage using a shared memory architecture with RDMA. It performs well with a throughput of 167 million key-value lookups and 31 us latency using 20 machines. \textbf{Pilaf} \cite{Mitchell:2013:UOR:2535461.2535475} also implements a key-value storage using RDMA for get operations and messaging verbs for put operations. \textbf{MICA} \cite{179747} implements a key-value storage with a focus on NUMA architectures. It maps each CPU core to a partition of data and communicates using a request-response approach using unreliable connections. \textbf{HERD} \cite{Kalia:2014:URE:2619239.2626299} borrows the design of MICA and implements networking using RDMA writes for the request to the server and messaging verbs for the response back to the client.

%%%%%%%%%%%%%%%%%%%%%%%%%%%%%%%%%%%%%%%%%%%%%%%%%%%%%%%%%%%%%%%%%%%%%%%%%%%%%%%%%%%%%
%%%%%%%%%%%%%%%%%%%%%%%%%%%%%%%%%%%%%%%%%%%%%%%%%%%%%%%%%%%%%%%%%%%%%%%%%%%%%%%%%%%%%
%%%%%%%%%%%%%%%%%%%%%%%%%%%%%%%%%%%%%%%%%%%%%%%%%%%%%%%%%%%%%%%%%%%%%%%%%%%%%%%%%%%%%

\section{InfiniBand and ibverbs Basics}
\label{ib_basics}
This section covers the most important aspects of the InfiniBand hardware and the native ibverbs library which are relevant for this report. Abbreviations introduced here (most of them commonly used in the InfiniBand context) are used throughout the report from this point on.

The \textbf{host channel adapter (HCA)} connected to the PCI bus of the host system is the network device for communicating with other nodes. The offloading engine of the HCA processes outgoing and incoming data asynchronously and is connected to other nodes using copper or optical cables via one or multiple switches. The \textbf{ibverbs} API provides the interface to communicate with the HCA either by exchanging data using Remote Direct Memory Access (RDMA) or messaging verbs. 

A \textbf{queue pair (QP)} identifies a physical connection to a remote node when using \textbf{reliable connected (RC)} communication. Using non connected \textbf{unreliable datagram (UD)} communication, a single QP is sufficient to send data to multiple remotes. A QP consists of one \textbf{send queue (SQ)} and one \textbf{receive queue (RQ)}. On RC communication, a QP's SQ and RQ are always cross connected with a target's QP, e.g. node 0 SQ connects to node 1 RQ and node 0 RQ to node 1 SQ.

If an application wants to send data, it posts a \textbf{work request (WR)} containing a pointer to the buffer to send and the length to the SQ. A corresponding WR must be posted on the RQ of the connected QP on the target node to receive the data. This WR also contains a pointer to a buffer and its size to receive any incoming data to.

Once the data is sent, a \textbf{work completion (WC)} is generated and added to a \textbf{completion queue (CQ)} associated with the SQ. A WC is also generated for the corresponding WCQ of the remote's RQ receiving the data, once the data arrived. The WC of the send task tells the application that the data was successfully sent to the remote (or provides error information otherwise). On the remote receiving the data, the WC indicates that the buffer attached to the previously posted WR is now filled with the remote's data.

When serving multiple connections, every single SQ and RQ does not need a dedicated CQ. A single CQ can be used as a \textbf{shared completion queue (SCQ)} with multiple SQs or RQs. Furthermore, when receiving data from multiple sources, instead of managing many RQs to provide buffers for incoming data, a \textbf{shared receive queue (SRQ)} can be used on multiple QPs instead of single RQs.

When attaching a buffer to a WR, it is attached as a \textbf{scatter gather element (SGE)} of a \textbf{scatter gather list (SGL)}. For sending, the SGL allows the offloading engine to gather the data from many scattered buffers and send it as one WR. For receiving, the received data is scattered to one or multiple buffers by the offloading engine.

%%%%%%%%%%%%%%%%%%%%%%%%%%%%%%%%%%%%%%%%%%%%%%%%%%%%%%%%%%%%%%%%%%%%%%%%%%%%%%%%%%%%%
%%%%%%%%%%%%%%%%%%%%%%%%%%%%%%%%%%%%%%%%%%%%%%%%%%%%%%%%%%%%%%%%%%%%%%%%%%%%%%%%%%%%%
%%%%%%%%%%%%%%%%%%%%%%%%%%%%%%%%%%%%%%%%%%%%%%%%%%%%%%%%%%%%%%%%%%%%%%%%%%%%%%%%%%%%%
\section{Low Latency Data Exchange Between Java and C}
\label{java_and_native}
In this section, we describe our experiences with and best practices for the Java Native Interface (JNI) to avoid performance penalties for latency sensitive applications. These are applied to various implementation aspects of the IB transport which are further explained in their dedicated sections.

Using JNI is mandatory if the Java space has to interface with native code, e.g. for IO operations or when using native libraries. As we decided to use the low-level ibverbs library to benefit from full control, high flexibility and low latency (\S \ref{related_work_java_ib}), we had to ensure that interfacing with native code from Java does not introduce too much overhead compared to the already existing and evaluated solutions.

The Java Native Interface (JNI) allows Java programmers to call native code from C/C++ libraries. It is a well known method to interface with native libraries that are not available in Java or access IO using system calls or other native libraries. When calling code of a native library, the library has to expose and implement a predefined interface which allows the JVM to connect the native functions to native declared Java methods in a Java class. With every call from Java to the native space and vice versa, a context switch is required to be executed by the JVM environment. This involves tasks related to thread and cache management adding latency to every native call. This increases the duration of such a call and is crucial, especially regarding the low latency of IB.

\textbf{Exchanging data with a native library without adding considerable overhead is challenging}. For single primitive values, passing parameters to functions is convenient and does not add any considerable overhead. However, access to Java classes or arrays from native space requires synchronization with the JVM (and its garbage collector) which is very expensive and must be avoided. Alternatively, one can use ByteBuffers allocated as DirectByteBuffers which allocates memory in native memory. Java can access the memory through the ByteBuffer and the native library can get the native address of the array and the size with the functions \texttt{GetDirectBufferAddress} and \texttt{GetDirectBufferCapacity}. However, these two calls increase the latency by tenth to even hundreds of microseconds (with high variation).

This problem can be solved by \textbf{allocating a native buffer in the native space, passing its address and size} to the Java space and \textbf{access it using the Unsafe API} or wrap it as a newly allocated (Direct) ByteBuffer. The latter requires reflection to access the constructor of the DirectByteBuffer and set the address and size fields. We decided to use the Unsafe API because we map native structs and don't require any of the additional features the ByteBuffer provides. The native address is cached which allows fast exchange of data from Java to native and vice versa. To improve convenience when accessing fields of a data structure, a helper class with getter and setter wrapper methods is created to access the fields of the native struct.

\begin{figure}[!t]
	\centering
	\includegraphics[width=3.3in]{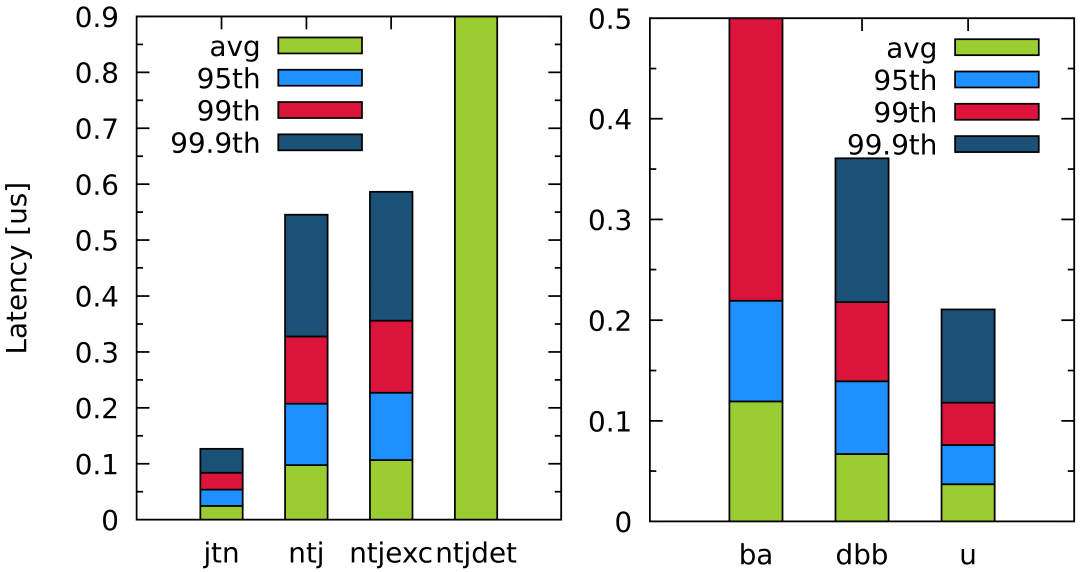}
	\caption{Microbenchmarks to evaluate JNI call overhead and data exchange overhead using different types of memory access}
	\label{jni_overhead}
\end{figure}

We evaluated different means of passing data from Java to native and vice versa as well as the function/method call overhead. Figure \ref{jni_overhead} shows the results of the microbenchmarks used to evaluate JNI call overhead as well as overhead of different memory access methods. The results displayed are the averages of three runs of each benchmark executing the operation 100,000,000 times. A warm-up of 1,000 operations preceeds each benchmark run. For JNI context switching, we measured the latency introduced of Java to native (jtn), native to Java (ntj), native to Java with exception checking (ntjexc) and native to Java with thread detaching (ntjdet). For exchanging data between Java and native, we measured the latency introduced by accessing a 64 byte buffer in both spaces for a primitive Java byte array (ba), Java DirectByteBuffer (dbb) and Unsafe (u). The benchmarks were executed on a machine with Intel Core i7-5820K CPU and Java 1.8 runtime.

The results show that the average single costs for context switching are neglectable with an average switching time of only up to 0.1 \textmu s. We exchange data using primitive function arguments, only. Data structures are mapped and accessed as C-structs in the native space. In Java, we access the native C-structs using a helper class which utilizes the Unsafe library \cite{javaunsafe} as this is the fastest method in both spaces.

These results influenced the important design decision to \textbf{run native threads, attached once as daemon threads to the JVM}, which call to Java instead of Java threads calling native methods (\S \ref{send_thread}, \S \ref{receive_thread}). Furthermore, we avoid using any of the JNI provided helper functions where possible \cite{Liang:1999:JNI:520155}. For example: attaching a thread to the JVM involves expensive operations like creating a new Java thread object and various state changes to the JVM environment. Avoiding them on every context switch is crucial to latency and performance on every call.

Lastly, we minimized the number of calls to the Java space by combining multiple tasks into a single cross-space call instead of yielding multiple calls. For inter space communication, we highly rely on communication via buffers mapped to structs in native space and wrapper classes in Java (see above). This is highly application dependable and not always possible. But if possible and applied, this can improve the overall performance.

We applied this technique of combining multiple tasks into a single cross-space call to sending and receiving of data to minimize latency and context switching overhead. The native send and receive threads implement the most latency critical logic in the native space which is not simply wrapping ibverbs functions to be exposed to Java (\S \ref{send_thread} and \ref{receive_thread}).. The counterpart to the native logic is implemented in Java (\S \ref{transport_impl_java}). In the end, we are able to reduce sending and receiving of data to a single context switching call.

%%%%%%%%%%%%%%%%%%%%%%%%%%%%%%%%%%%%%%%%%%%%%%%%%%%%%%%%%%%%%%%%%%%%%%%%%%%%%%%%%%%%%
%%%%%%%%%%%%%%%%%%%%%%%%%%%%%%%%%%%%%%%%%%%%%%%%%%%%%%%%%%%%%%%%%%%%%%%%%%%%%%%%%%%%%
%%%%%%%%%%%%%%%%%%%%%%%%%%%%%%%%%%%%%%%%%%%%%%%%%%%%%%%%%%%%%%%%%%%%%%%%%%%%%%%%%%%%%
\section{Overview Ibdxnet and Java InfiniBand Transport}
\label{overview_infiniband_transport}

\begin{figure}[!t]
	\centering
	\includegraphics[width=2.0in]{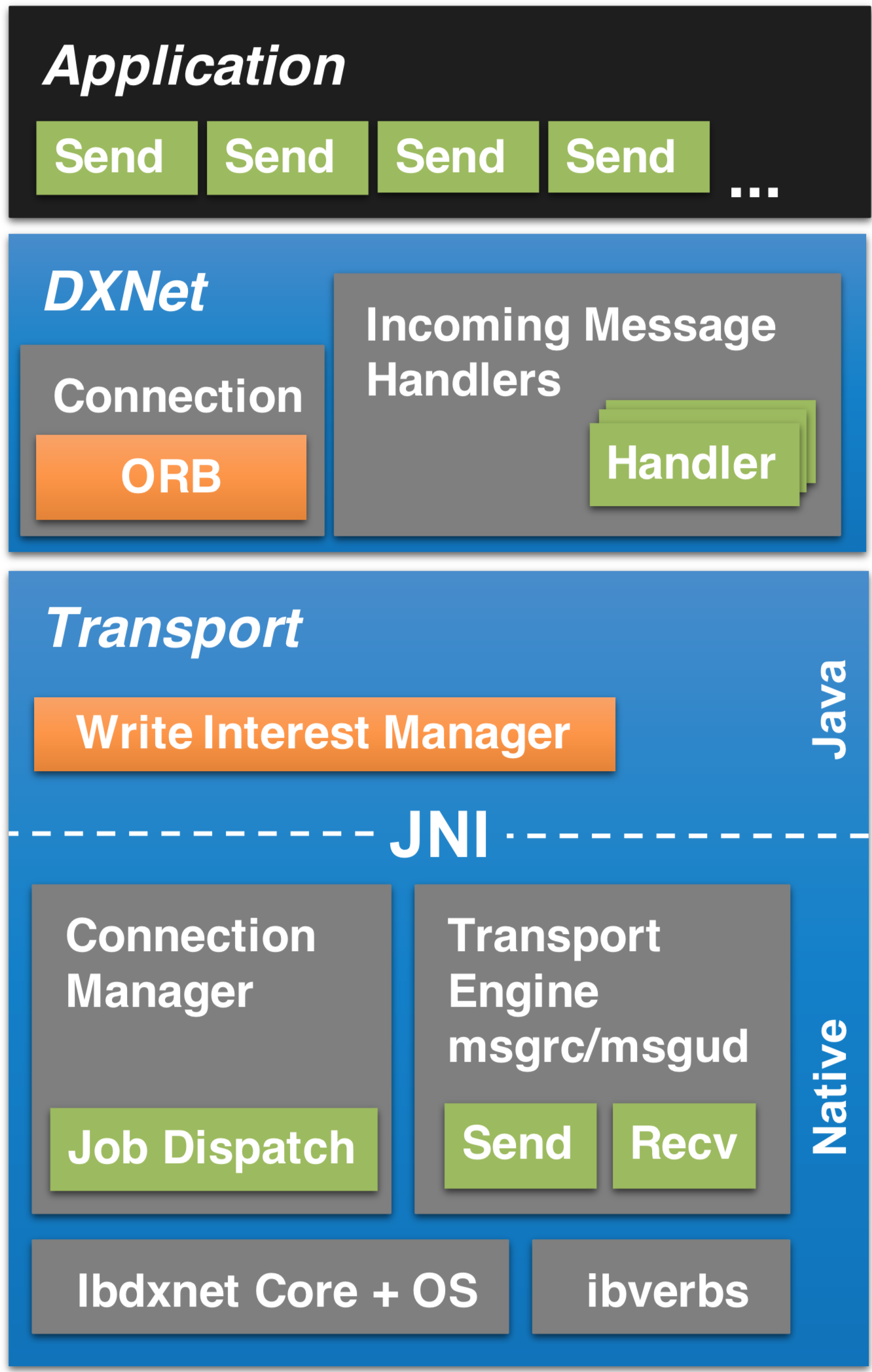}
	\caption{Layered architecture of DXNet with the IB transport and Ibdxnet (using the msgrc engine). Threads are colored green.}
	\label{dxnet_ib_simple}
\end{figure}

This section gives a brief top-down introduction of the full transport implementation. Figure \ref{dxnet_ib_simple} depicts the different components and layers involved when using InfiniBand with DXNet. The \textbf{Java InfiniBand transport (IB transport)} (\S \ref{transport_impl_java}) implements DXNet's transport interface (\S \ref{dxnet_transport_interface}) and uses JNI to connect to the native counterpart. \textbf{Ibdxnet} uses the native ibverbs library to access the hardware and provides a separate subsystem for connection management, sending and receiving data. Furthermore, it implements a set of functions for the Java Native Interface to connect to the Java implementation.

%%%%%%%%%%%%%%%%%%%%%%%%%%%%%%%%%%%%%%%%%%%%%%%%%%%%%%%%%%%%%%%%%%%%%%%%%%%%%%%%%%%%%
%%%%%%%%%%%%%%%%%%%%%%%%%%%%%%%%%%%%%%%%%%%%%%%%%%%%%%%%%%%%%%%%%%%%%%%%%%%%%%%%%%%%%
%%%%%%%%%%%%%%%%%%%%%%%%%%%%%%%%%%%%%%%%%%%%%%%%%%%%%%%%%%%%%%%%%%%%%%%%%%%%%%%%%%%%%
\section{Ibdxnet: Native InfiniBand Subsystem with Transport Engine}
\label{ibdxnet_native}

\begin{figure}[!t]
	\centering
	\includegraphics[width=3.0in]{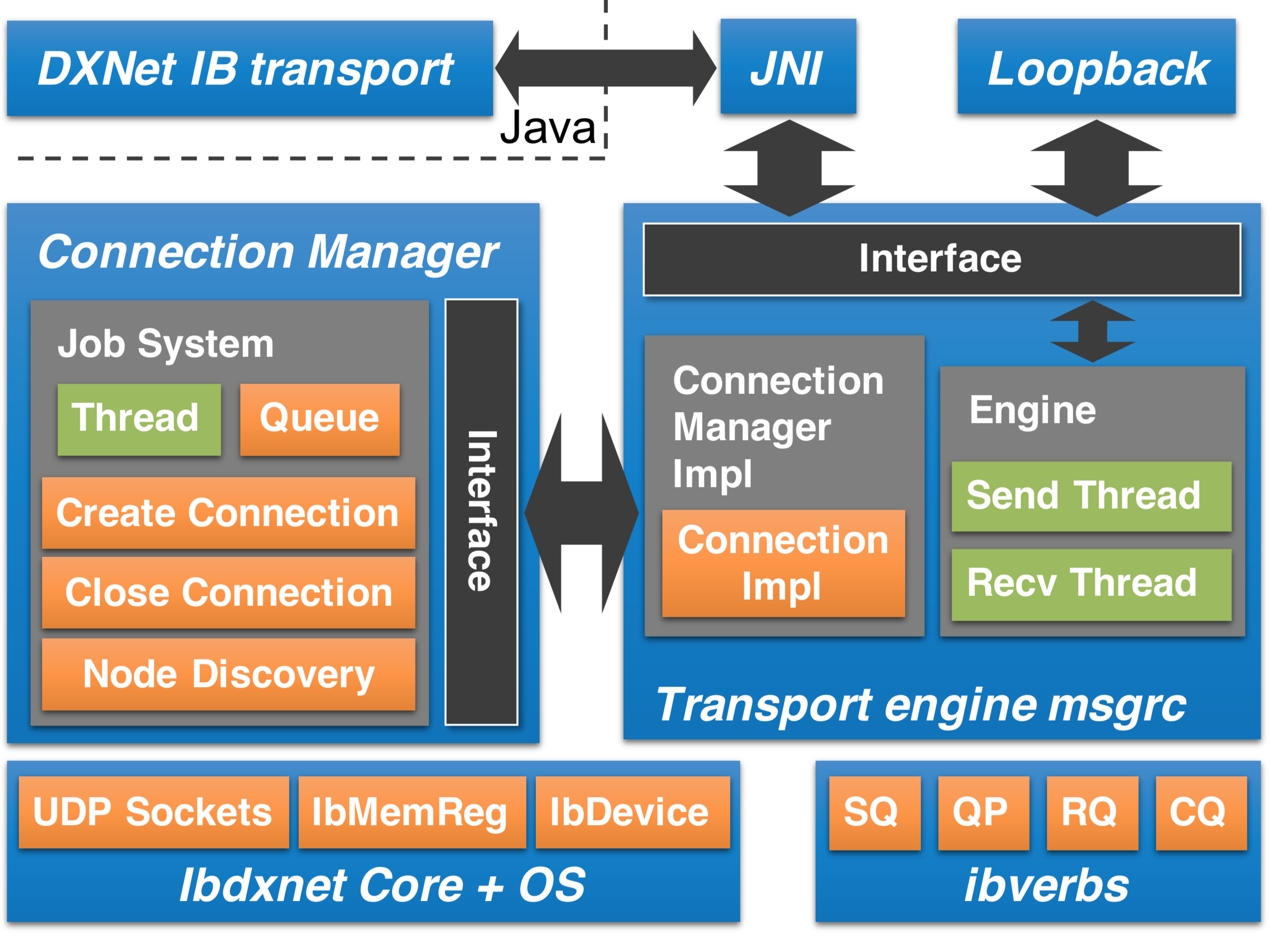}
	\caption{Simplified architecture of Ibdxnet with the msgrc transport engine}
	\label{ibdxnet_simple}
	\vspace{-20pt}
\end{figure}

This section elaborates on the implementation details of our native InfiniBand subsystem \textbf{Ibdxnet} which is used by the IB transport implementation in DXNet to utilize InfiniBand hardware. Ibdxnet provides the following key features: a basic foundation with re-usable components for implementations using different means of communication (e.g. messaging verbs, RDMA) or protocols, automatic connection management and transport engines using different communication primitives. Figure \ref{ibdxnet_simple} shows an outline of the different components involved.

Ibdxnet provides an \textbf{automatic connection and QP manager} (\S \ref{scalable_connection_management}) which can be used by every transport engine. An interface for the connection manager and a connection object allows implementations for different transport engines. The engine \textbf{msgrc} (see Figure \ref{ibdxnet_simple}) uses the provided connection management and is based on RC messaging verbs. The engine \textbf{msgud} using UD messaging verbs is already implemented and will be discussed and extensively evaluated in a separate publication.

A \textbf{transport engine} implements its own protocol to send/receive data and exposes a low-level interface. It creates an abstraction layer to hide direct interaction with the ibverbs library. Through the low-level interface, a transport implementation (\S \ref{transport_impl_java}) provides data-to-send and forwards received data for further processing. For example: the low-level interface of the msgrc engine does not provide concurrency control or serialization mechanisms for messages. It accepts a stream of data in one or multiple buffers for sending and provides buffers creating a stream of data on receive (\S \ref{msgrc}). This engine is connected to the Java transport counterpart via JNI and uses the existing infrastructure of DXNet (\S \ref{transport_impl_java}).

Furthermore, we implemented a \textbf{loopback} like stand alone transport for debugging and measuring performance of the native engine, only. The loopback transport creates a continuous stream of data for sending to one or multiple nodes and throws away any data received. This ensures that sending and receiving introduce no additional overhead and allows measuring the performance of different low-level aspects of our implementation. This was used to determine the maximum possible throughput with Ibdxnet (\S \ref{eval_dxnet_nodes_tp}).

In the following sections, we explain the implementation details of Ibdxnet's connection manager (\S \ref{scalable_connection_management}) and the messaging engine msgrc (\S \ref{msgrc}). Additionally, we describe best practices for using the ibverbs API and optimizations for optimal hardware utilization. Furthermore, we elaborate on how Ibdxnet connects to the IB transport in Java using JNI and how we implemented low overhead data exchange between Java and native space.

%%%%%%%%%%%%%%%%%%
\subsection{Dynamic, Scalable and Concurrent Connection Management}
\label{scalable_connection_management}
Efficient connection management for many nodes is a challenging task. For example, hundreds of application threads want to send data to a node but the connection is not yet established. Who creates the connection and synchronizes access of other threads? How to avoid synchronization overhead or blocking of threads that want to get an already established connection? How to manage the lifetime of a connection? 

These challenges are addressed by a \textbf{dedicated connection manager in Ibdxnet}. The connection manager handles all tasks required to establish and manage connections and hides them from the higher level application. For our higher level Java transport (\S \ref{transport_con_man}), complexity and latency is reduced for connection setup by avoiding context switching. 

First, we explain how nodes are identified, the contents of a connection and how online/offline nodes are discovered and handled. Next, we describe how existing connections are accessed and non-existing connections are created on the fly during application runtime. We explain the details how a connection creation job is handled by the internal job manager, how connection data is exchanged with the remote in order to create a QP. At last, we briefly describe our previous attempt which failed to address the above challenges properly.

\begin{figure}[!t]
	\centering
	\includegraphics[width=3.0in]{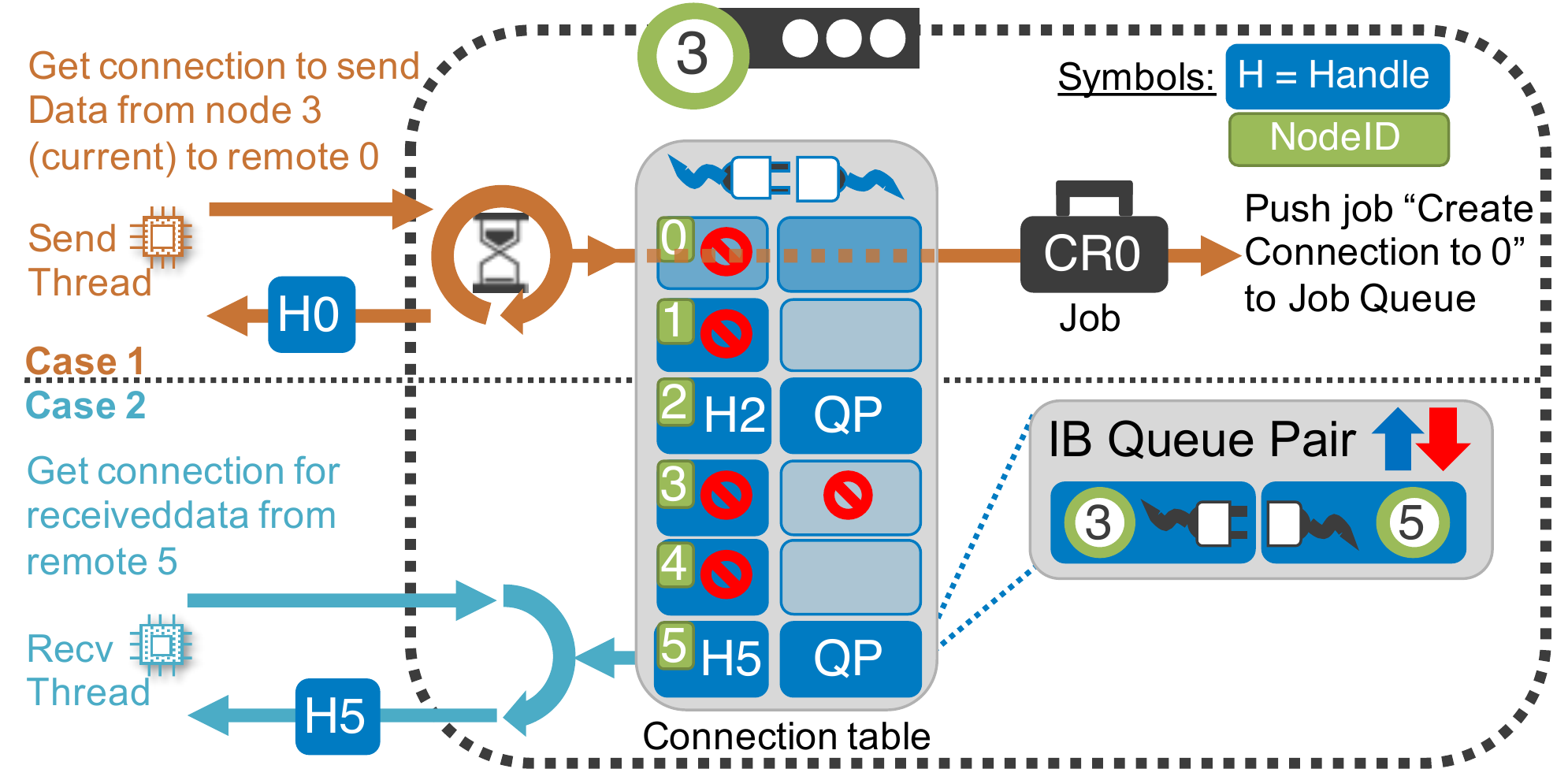}
	\caption{Connection manager: Creating non-existing connections (send thread: node 1 to node 0) and re-using existing connections (recv thread: node 1 to node 5).}
	\label{ibdxnet_conman1}
\end{figure}

\begin{figure}[!t]
	\centering
	\includegraphics[width=3.0in]{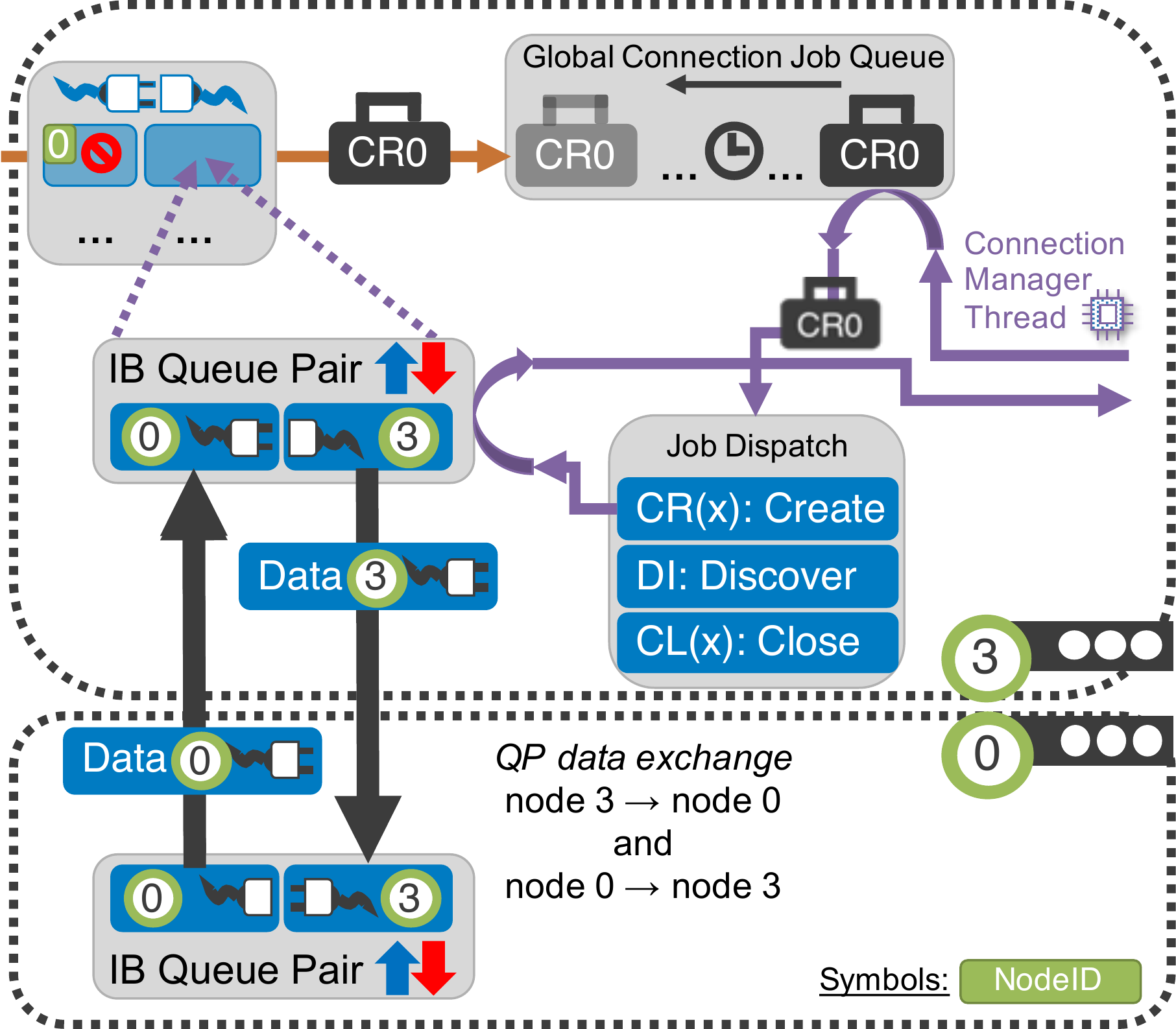}
	\caption{Connection manager: Automatic connection creation with QP data exchange (node 3 to node 0). The job \textit{CR0} is added to the back of the queue to initiate this process. The dedicated thread processes the queue by removing jobs from the front and processing them according to their type.}
	\label{ibdxnet_conman2}
\end{figure}

A node is identified by \textbf{a unique 16-bit integer nodeID (NID)}. The NID is assigned to a node on start of the connection manager and cannot be changed during runtime. A connection consists of the source NID (the current node) and the destination NID (the target remote node). Depending on the transport implementation, an existing connection holds one or multiple ibverbs QPs, buffers and other data necessary to send and receive data using that connection. The connection manager provides a \textbf{connection interface for the transport engines} which allows them to implement their own type of connection. The following example describes a connection with a single QP, only.

Before a connection to a remote node can be established, the remote node must be discovered and known as available. The job type \textbf{node discovery} (further details about the job system follow in the next paragraphs) detects online/offline nodes using UDP sockets over Ethernet. On startup, a list of node hostnames is provided to the connection manager. The list can be extended by adding/removing entries during runtime for dynamic scaling. The discovery job tries to contact all non-discovered nodes of that list in regular intervals. When a node was discovered, it is removed from the list and marked as discovered. A connection can only be established with an already discovered node. If a connection to the node was already created and is lost (e.g. node crash), the NID is added back to the list in order to re-discovered the node on the next iteration of the job. Node discovery is mandatory for InfiniBand in order to exchange QP information on connection creation.

Figure \ref{ibdxnet_conman1} shows how existing connections are accessed and new connections are created when two threads, e.g. a send and a receive thread, are accessing the connection manager. The send thread wants to send new data to node 0 and the receive thread has received some data (e.g. from a SRQ). It has to forward it for further processing which requires information stored in each connection (e.g. a queue for the incoming data). If the connection is already established (the receive thread gets the connection to node 5), a connection handle (\textit{H5}) is returned to the calling thread. If no connection has been established so far (the send thread wants to get the connection to node 0), \textbf{a job to create the specific connection} (\textit{CR0} = create to node 0) is added to the internal job queue. The calling thread has to wait until the job is dispatched and the connection is created before being able to send the data.

Figure \ref{ibdxnet_conman2} shows how connection creation is handled by the internal job thread. The job \textit{CR0} (yielded by the send thread from the previous example in figure \ref{ibdxnet_conman1}) is pushed to the back of the job queue. The job queue might contain jobs which affect different connections, i.e. there is no per connection dedicated queue. \textbf{The dedicated connection manager thread} is processing the queue by removing a job from the front and dispatching by type. There are three types of jobs: create a connection to a node with a NID, discover other connection managers, close an existing connection to node.

To create a new connection with a remote node, the current node has to create an ibverbs QP with a SQ and RQ. Both queues are cross-connected to a remote QP (send with recv, recv with send) which requires data exchange using another communication channel (Sockets over Ethernet). For the job \textit{CR0}, the thread creates a new QP on the current node (3) and exchanges its QP data with the remote it wants to connect to (0) using UDP sockets. The remote (0) also creates a QP and uses the received connection information (of 3). It replies with its own QP data (0 to 3) to complete QP creation. The newly established connection is added to the connection table and is now accessible (by the send and receive thread).

At last, we briefly describe our lessons learned from our first attempt for an automatic connection manager. It was relying on active connection creation. The first thread calling the connection manager to acquire a connection creates it on the fly, if it does not exist. The calling thread executes connection exchange, waits for the remote data and finishes connection creation. This requires coordination of all threads accessing the connection manager either to create a new connection or getting an existing one. It introduced a very complex architecture with high synchronization overhead and latency especially when many threads are concurrently accessing the connection manager. Furthermore, it was error prone and difficult to debug. We encountered severe performance issues when creating connections with one hundred nodes in a very short time range (e.g. all-to-all communication). This resulted in connection creation times of up to half a minute. Even with a small setup of 4 to 8 nodes, creating a connection could take up to a few seconds if multiple threads tried to create the same or different connections simultaneously.

%%%%%%%%%%%%%%%%%%QP
\subsection{msgrc: Transport Engine for Messaging using RC QPs}
\label{msgrc}
This section describes the \textbf{msgrc} transport engine. It uses reliable QPs to implement messaging using a dedicated send and receive thread. The engine's interface allows a transport to provide a stream of data (to send) in form of variable sized buffers and provides a stream of data (received) to a registered callback handler. 

This interface is rather low-level and the backend does not implement any means of serialization/deserialization for sending/receiving of complex data structures. In combination with DXNet (\S \ref{dxnet}), the logic for these tasks resides in the Java space with DXNet and is shared with other transports such as the NIO Ethernet transport \cite{dxnetnio}. However, there are no restrictions to implement these higher level components for the msgrc engine natively, if required. Further details on how the msgrc engine is connected with the Java transport counterpart are given in Section \ref{transport_impl_java}.

The following subsections explain the general architecture and interface of the transport, sending and receiving of data using dedicated threads and how various features of InfiniBand were used for optimal hardware utilization.

%%%%%%%%%
\subsubsection{Architecture}
\label{engine_architecture}
This section explains the basic architecture as well as the low-level interface of the engine. Figure \ref{ibdxnet_simple} includes the msgrc transport and can be referred to for an abstract representation of the most important components. The engine relies on our dedicated connection manager (\S \ref{scalable_connection_management}) for connection handling. We decided to use one dedicated thread for sending (\S \ref{send_thread}) and one for receiving (\S \ref{receive_thread}) to benefit from the following advantages: a clear separation of responsibilities resulting in a less complex architecture, no scheduling of send/receive jobs when using a single thread for both and higher concurrency because we can run both threads on different CPU cores concurrently. The architecture allows us to create decoupled pipeline stages using lock-free queues and ring buffers. Thereby, we avoid complex and slow synchronization between the two threads and with hundreds of threads concurrently accessing shared resources.

\subsubsection{Engine interface}
\label{engine_interface}

\begin{lstlisting}[caption={Structures and callback of the msgrc engine's send interface},label=send_interface, xleftmargin=4.0ex]
struct NextWorkPackage {
    uint32_t posBackRel;
    uint32_t posFrontRel;
    uint8_t flowControlData;
    uint16_t nodeId;
};

struct PrevWorkPackageResults {
    uint16_t nodeId;
    uint32_t numBytesPosted;
    uint32_t numBytesNotPosted;
    uint8_t fcDataPosted;
    uint8_t fcDataNotPosted;
};

struct CompletedWorkList {
    uint16_t numNodes;
    uint32_t bytesWritten[NODE_ID_MAX_NUM_NODES];
    uint8_t fcDataWritten[NODE_ID_MAX_NUM_NODES];
    uint16_t nodeIds[];
};

NextWorkPackage* GetNextDataToSend(PrevWorkPackageResults* prevResults, CompletedWorkList* completionList);
\end{lstlisting}

The low-level interface allows fine-grained control for the target transport over the engine. The interface for sending data is depicted in Listing \ref{send_interface} and receiving is depicted in Listing \ref{recv_interface}. Both interfaces create an abstraction hiding connection and QP management as well as how the hardware is driven with the ibverbs library. For sending data, the interface provides the callback \textit{GetNextDataToSend}. This function is called by the send thread to pull new data to send from the transport (e.g. from the ORB, see \ref{transport_send}). When called, an instance of each of the two structures \textit{PrevWorkPackageResults} and \textit{CompletedWorkList} are passed to the implementation of the callback as parameters: the first contains information about the previous call to the function and how much data was actually sent. If the SQ is full, no further data can be sent. Instead of introducing an additional callback, we combine getting the next data with returning information about the previous send call to reduce call overhead (important for JNI access). The second parameter contains data about completed work requests, i.e. data sent for the transport. This must be used in the transport to mark data processed (e.g. moving the pointers of the ORB).

\begin{lstlisting}[caption={Structure and callback of the msgrc engine's receive interface},label=recv_interface, xleftmargin=4.0ex]
struct ReceivedPackage {
    uint32_t count;
    struct Entry {
        uint16_t sourceNodeId;
        uint8_t fcData;
        IbMemReg* data;
        void* dataRaw;
        uint32_t dataLength;
    } m_entries[];
};

struct IncomingRingBuffer {
    uint32_t m_usedEntries;
    uint32_t m_front;
    uint32_t m_back;
    uint32_t m_size;

    struct Entry {
        con::NodeId m_sourceNodeId;
        uint8_t m_fcData;
        uint32_t m_dataLength;
        core::IbMemReg* m_data;
        void* m_dataRaw;
    } m_entries[];
};

uint32_t Received(IncomingRingBuffer* ringBuffer);

void ReturnBuffer(IbMemReg* buffer);
\end{lstlisting}

If data is received, the receive thread calls the callback function \textit{Received} with an instance of the \textit{IncomingRingBuffer} structure as its parameter. This parameter holds a list of received buffers with their source NID. The transport can iterate this list and forward the buffers for further processing such as de-serialization. If the transport has to return the number of elements processed and, thus, is able to control the amount of buffers it can process. Once the received buffers are processed by the transport, they must be returned back to the \textit{RecvBufferPool} by calling \textit{ReturnRecvBuffer} to allow re-using them for further receives.

%%%%%%%%%%%%%%%%%%
\subsubsection{Sending of Data}
\label{send_thread}

\begin{figure}[!t]
	\centering
	\includegraphics[width=3.3in]{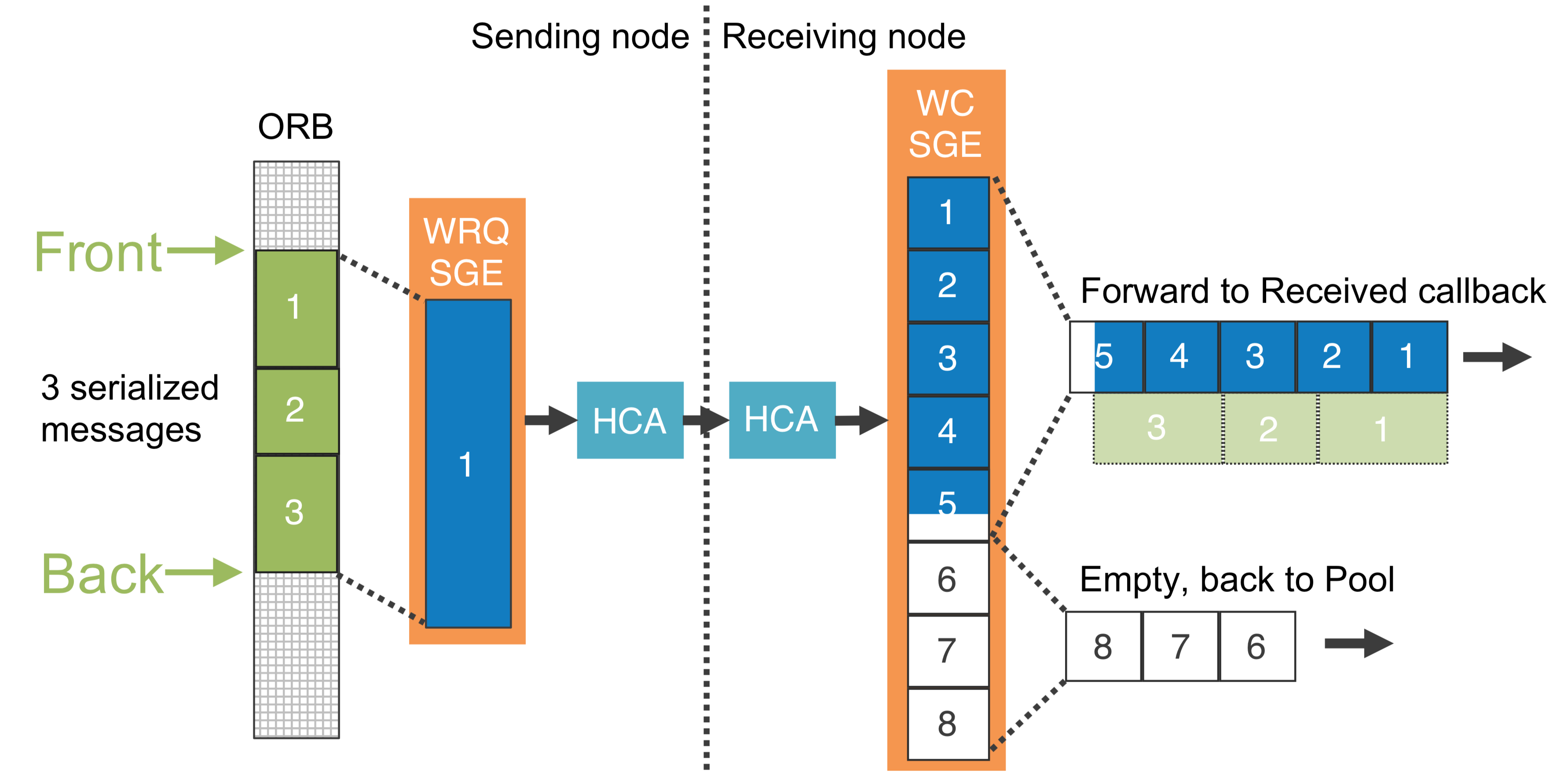}
	\caption{Example for sending and receiving data using scatter gather elements: Get data (aggregated messages) from ORB, send 1 SGE. Receive data scattered to multiple receive buffers.}
	\label{msgrc_sge}
\end{figure}

This section explains the data and control flow of the \textbf{dedicated send thread} which \textbf{asynchronously} drives the engine for sending data. Listing \ref{send_thread_code} depicts a simplified version of the contents of its main loop with the relevant aspects for this section. Details of the functions involved in the main flow are explained further below.

\begin{lstlisting}[caption={Send thread main flow (simplified)},label=send_thread_code, xleftmargin=4.0ex]
workPackage = GetNextDataToSend(prevWorkResults, completionList);
Reset(prevWorkResults);
Reset(completionList);

if (workPackage != NULL) {
	connection = GetConnection(workPackage.nodeId);
	prevWorkResults = SendData(connection, workPackage);
	ReturnConnection(connection);
}

completionList = PollCompletions();
\end{lstlisting}

The loop starts with getting a \textit{workPackage}, the next data to send (line 1), using the engine's low-level interface (\S \ref{engine_interface}). The instance \textit{prevWorkResults} contains information about posted and non-posted data from the previous loop iteration. The instance \textit{completionList} holds data about completed sends. Both instances are reseted/nulled (line 2-3) for re-use in the current iteration. 

If the \textit{workPackage} is valid (line 5), i.e. data to send is available, the \textit{nodeId} from that package is used to get the \textit{connection} to the send target from the connection manager (line 6). The \textit{connection} and \textit{workPackage} are passed to the \textit{SendData} function (line 7). It processes the \textit{workPackage} and returns how much data was processed, i.e. posted to the SQ of the connection, and how much data could not be processed. The latter happens if the SQ is full and must be kept track of to not lose any data. Afterwards, the thread returns the \textit{connection} to the connection manager (line 8).

At the end of a loop iteration, the thread polls the SCQ to remove any available WCs. \textbf{We share the completion queue among all SQs/connections to avoid iterating over many connections for a task}. The loop iteration ends and the thread starts from the beginning by calling \textit{GetNextDataToSend} and provides the work results of our previous iteration. Data about polled WCs from the SCQ are stored in the \textit{completionList} and forwarded via the interface (to the transport).

If no data is available (line 5), lines 6-8 are skipped and the thread executes a completion poll, only. This is important to ensure that any outstanding WCs are processed and passed to the transport (via the \textit{completionList} and calling \textit{GetNextDataToSend}). Otherwise, if no data is sent for a while, the transport will not receive any information about previously processed data. This leads to false assumptions about the available buffer space for sending data, e.g. assuming that data fits into the buffer but actually does not because the processed buffer space is not free'd, yet.

In the following paragraphs, we further explain how the functions \textit{SendData} and \textit{PollCompletions} make optimal use of the ibverbs library and how this cooperates with the interleaved control flow of the main thread loop explained above.

The \textbf{SendData} function is responsible for preparing and posting of FC data and normal data (payload). FC data, which determines the number of flow control windows to confirm, is a small number (< 128) and, thus, does not require a lot of space. We post it as part of the \textbf{immediate data}, which can hold up to 4 bytes of data, with the WR instead of using a separate side channel, e.g. another QP. This avoids overhead of posting and polling of another QP which \textbf{benefits overall performance, especially with many simultaneous connections}. With FC data using 1 byte of the immediate data field, we use further 2 bytes to include the NID of the source node. This allows us to identify the source of the incoming WC on the remote. Otherwise, identifying the source would be very inconvenient. The only information provided with the incoming WC is the sender's unique physical QP id. In our case, this id must be mapped to the corresponding NID of the sender. However, this introduces an indirection every time a package arrives which hurts performance.

For sending normal data (payload), the provided \textit{workPackage} holds two pointers, front and back, which enclose a memory area of data to send. This memory area belongs to a buffer (e.g. the ORB) which was registered with the protection domain on start to allow access by the HCA. Figure \ref{msgrc_sge} depicts an example with three (aggregated) ready to send messages in the ORB. We create a WR for the data to send and provide a single \textbf{SGE which takes the pointers of the enclosed memory area}. The HCA will directly read from that area without further copying of the data (zero copy). For buffer wrap arounds, two SGEs are created and attached to one WR: one SGE for the data from the front pointer to the end of the buffer, another SGE for the data from the start of the buffer to the back pointer. If the size of the area to send (sum of all SGEs) exceeds the maximum configurable receive size, the data to send must be sliced into multiple WRs. \textbf{Multiple WRs are chained to a link list to minimize call overhead} when posting them to the SQ using \textit{ibv\_post\_send}. This greatly increases performance compared to posting multiple standalone WRs with single calls.

The number of SGEs of a WR can be 0, if no normal data is available to send but FC data is available. To send FC data only, we write it to the immediate data field of a WR along with our source NID and post it without any SGEs attached which results in a 0 length data WR. 

The \textbf{PollCompletions} function calls \textit{ibv\_poll\_cq}, \textbf{once, to poll for any completions available} on the SCQ. A SCQ is used instead of per connection CQs to avoid iterating the CQs of all connections which impacts performance. The send thread keeps track of the number of posted WRs and, thus, knows how many WCs are outstanding and expected to arrive on the SCQ. If none are being expected, polling is skipped. \textit{ibv\_poll\_cq} is called once per PollCompletion call, only, and every call tries to poll WCs in batches to keep the call overhead minimal.

Experiments have shown that most calls to \textit{ibv\_poll\_cq}, even on high loads, will return empty, i.e. no WRs have completed. Thus, polling the SCQ until at least one completion is received is the wrong approach and greatly impacts overall performance. If the SQ of another connection is not full and there is data available to send, this method wastes CPU resources on busy polling instead of processing further data to send. The performance impact (resulting in low throughput) increases with the number of simultaneous connections being served. Furthermore, this increases the chance of SQs running empty because time is wasted on waiting for completions instead of keeping all SQs filled. \textbf{Full SQs ensure that the HCA is kept busy which is the key to optimal performance}.

%%%%%%%%%%%%%%%%%%
\subsubsection{Receiving of Data}
\label{receive_thread}

\begin{lstlisting}[caption={Receive thread main flow (simplified)},label=recv_thread_code, xleftmargin=4.0ex]
workCompletions = PollCompletions();

if (recvQueuePending < ibqSize) {
    Refill();
}

if (workCompletions > 0) {
	ProcessCompletions(workCompletions);
}

if (!IncomingRingBufferIsEmpty()) {
    DispatchReceived();
}
\end{lstlisting}

Analogous to Section \ref{send_thread}, this section explains the data and control flow of the \textbf{dedicated receive thread} which \textbf{asynchronously} drives the engine for receiving data. Listing \ref{recv_thread_code} depicts a simplified version of its main loop with the relevant aspects for this section. Details of the functions involved in the main flow are explained further below.

Data is received using a SRQ and SCQ instead of multiple receive and completions queues. This avoids iterating over all open connections and checking for data availability which introduces overhead with increasing number of simultaneous connections. Equally sized buffers for receiving data (configurable size and amount) are pooled and returned for re-use by the transport, once processed (\S \ref{engine_interface}).

The loop starts by calling \textit{PollCompletions} (line 1) to poll the SCQ for WCs. Before processing the WCs returned, the SRQ is refilled by calling \textit{Refill} (line 4), if the SRQ is not filled, yet. Next, if any WCs were polled previously, they are processed by calling \textbf{ProcessCompletions} (line 8). This step pushes them to the \textbf{Incoming Ring Buffer (IRB)}, a temporary ring buffer, before dispatching them. Finally, if the IRB is not empty (line 11), the thread tries to forward the contents of the IRB by calling \textit{DispatchReceived} via the interface to the transport (\S \ref{engine_interface}).

The following paragraphs are further elaborating on how \textit{PollCompletions}, \textit{Refill}, \textit{ProcessCompletions} and \textit{DispatchReceived} make optimal use of the ibverbs library and how this cooperates with the interleaved control flow of the main thread loop explained above.

The \textbf{PollCompletions} function is very similar to the one explained in Section \ref{send_thread} already. WCs are polled in batches of max. currently available IRB space and buffered before being processed.

The \textbf{Refill} function adds new receive WRs to the SRQ, if the SRQ is not completely filled and receive buffers from the receive buffer pool are available. Every WR consists of a configurable number of SGEs which make up the maximum receive size. This is also the limiting size the send thread can post with a single WR (sum of sizes of SGE list). Using this method, the receive thread does not have to take care of any software slicing of received data because the HCA scatters one big chunk of send data transparently to multiple (smaller) receive buffers on the receiver side. At last, \textit{Refill} chains the WRs to a linked list which is posted on a single call to \textit{ibv\_post\_srq\_recv} for minimal overhead.

If WCs are buffered from the previous call to \textit{PollCompletions}, the \textbf{ProcessReceived} function iterates this list of WCs. For each WC of the list, it gets the source NID and FC data from the immediate data field. If the recv length of this WC is non zero, the attached SGEs contain the received data scattered to the receive buffers of the SGE list.

As the receive thread does not know or have any means of determining the size of the next incoming data, the challenge is optimal receive buffer usage with minimal internal fragmentation. Here, fragmentation describes the amount of receive buffers provided with a WR as SGEs in relation to the amount of received data written to that block of buffers. The less data written to the buffers, the higher the fragmentation. In the example shown in figure \ref{msgrc_sge}, the three aggregated and serialized messages are received in five buffers but the last buffer is not completely used.

This fragmentation cannot be avoided but handled to avoid negative results like empty buffer pools or low per buffer utilization. Receive buffers/SGEs of a WR that do not contain any received data, because the amount of received data is less than the total size of the list of buffers of the SGE list, are pushed back to the buffer pool. All receive buffers of the SGE list that contain valid received data are pushed to the IRB (in the order they were received). 

Depending on the target application, the fragmentation degree can be lowered if one configures the receive buffer and pool sizes accordingly. Applications typically sending small messages are performing well with small receive buffer sizes. However, throughput might decrease slightly for applications sending mainly big messages on small receive buffer sizes requiring more WRs per send data send (data sliced into multiple WRs).

If the IRB contains any elements, the \textbf{DispatchReceived} function tries to forward them to the transport via the \textit{Received} callback (\S \ref{engine_interface}). The callback returns the number of elements it consumed from the IRB and, thus, is allowed to consume none or up to what's available. The consumed buffers are returned asynchronously to the receive buffer pool by transport, once it finished processing them.

%%%%%%%%%%%%%%%%%%	
\subsubsection{Load Adaptive Thread Parking}
\label{thread_parking}

The send and receive threads must be kept busy running their loops to send and receive data as fast as possible to ensure low latency. However, pure busy polling without any sleeping or yielding introduces high CPU load and occupying two cores of the CPU permanently. This is unnecessary during periods when the network is not used frequently. We do not want the send and receive threads to waste CPU resources and, therewith, decrease the overall node performance. Experiments have shown that simply adding sleep or yield operations highly impacts network latency and throughput and introduces high fluctuations \cite{dxnet}.

To solve this, we used a simple but efficient wait pattern we call \textit{load adaptive thread parking}. After a defined amount of time (e.g. 100 ms) of polling and no data available, the thread enters a yield phase and calls yield on every loop iteration if no data is available. After another timeframe passed (e.g. 1 sec), the thread enters a parking phase calling sleep/park with a minimum value of 1 ns on every loop iteration reducing CPU load significantly. The lowest value possible (1 ns) ensure that the scheduler of the operating system sends the thread sleeping for the shortest period of time possible. Once data is available, the current phase is interrupted and the timer is reset. This ensures busy looping for the next iterations keeping latency for successive messages and on high loads low. For further details including evaluation results refer to our DXNet publication \cite{dxnet}.

%%%%%%%%%%%%%%%%%%%%%%%%%%%%%%%%%%%%%%%%%%%%%%%%%%%%%%%%%%%%%%%%%%%%%%%%%%%%%%%%%%%%%
%%%%%%%%%%%%%%%%%%%%%%%%%%%%%%%%%%%%%%%%%%%%%%%%%%%%%%%%%%%%%%%%%%%%%%%%%%%%%%%%%%%%%
%%%%%%%%%%%%%%%%%%%%%%%%%%%%%%%%%%%%%%%%%%%%%%%%%%%%%%%%%%%%%%%%%%%%%%%%%%%%%%%%%%%%%
\section{IB Transport Implementation in DXNet (Java)}
\label{transport_impl_java}

\begin{figure}[!t]
	\centering
	\includegraphics[width=2.8in]{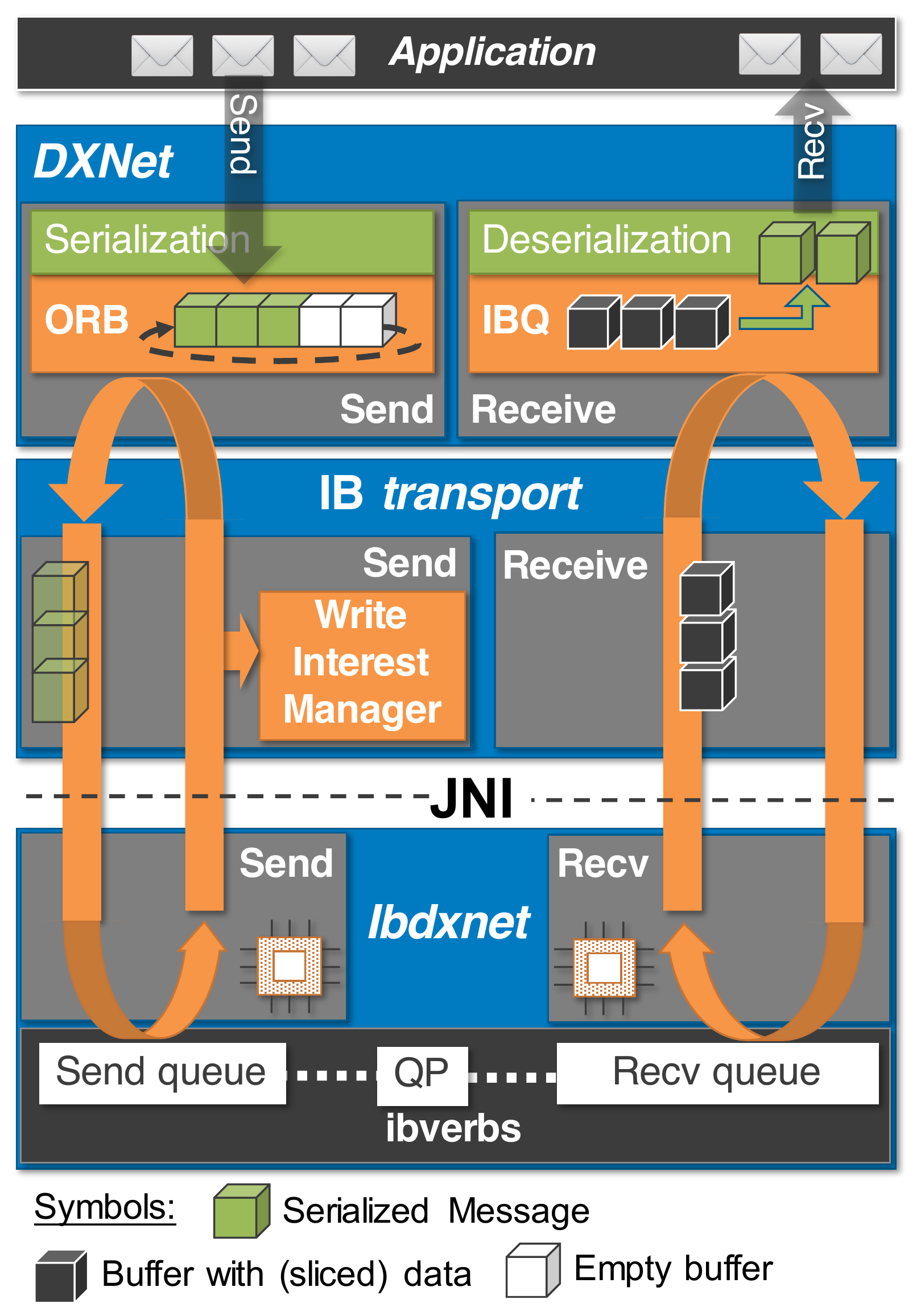}
	\caption{Components of Ibdxnet, IB transport and DXNet involved in data and control flow (simplified).}
	\label{transport_java}
\end{figure}

This section describes the transport implementation for DXNet in Java which utilizes the low-level transport engines, e.g. msgrc (\S \ref{msgrc}), provided by Ibdxnet (\S \ref{ibdxnet_native}). We describe the native interface which implements the low-level interface exposed by the engine (\S \ref{engine_interface}) and how it is used in the DXNet IB transport for higher level connection management (\S \ref{transport_con_man}), sending serialized data from the ORB (\S \ref{transport_send}) and handling incoming receive buffers from remote nodes (\S \ref{transport_recv}).

Figure \ref{transport_java} depicts the involved components with the main aspects of their data and control flow which are referred to in the following subsection.

If an application wants to send one or multiple messages, it calls DXNet which serializes them into the ORB and signals the WriteInterestManager (WIM) about available data (\S \ref{dxnet_send}). The native send thread checks the WIM for data to send periodically and, if available, gets it from the ORB. Depending on the size, the data to send might be sliced into multiple elements which are posted to the SQ as one or multiple work requests (\S \ref{send_thread}).

Received data on the recv queue is written to one or multiple buffers (depending on the amount of data) from a native buffer pool (\S \ref{receive_thread}). Without further processing, the buffers are forwarded to the Java space and pushed to the IncomingBufferQueue (IBQ). DXNet's de-serialization is processing the buffers in order and creates messages (Java objects) which are dispatched to pre-registered callbacks using dedicated message handler threads (\S \ref{dxnet_receive}).

%%%%%%%%%%%%%%%%%%	
\subsection{Connection Handling}
\label{transport_con_man}
To implement new transports in DXNet, it provides an interface to create specific connection types for the transport to implement. The DXNet core, which is shared across all transport implementations, manages the connections for the target application by automatically creating new connections on demand or closing connections if a configurable threshold is exceeded (\S \ref{dxnet_con_man}). 

For the IB transport implementation, the derived connection does not have to store further data or implement functionality. This is already stored and handled by the connection manager of Ibdxnet. It reduces overall architectural complexity by avoiding split functionality between Java and native space. Furthermore, it avoids context switching between Java and native code. 

Only the NID of either the target node to send to or the source node of the received data is exchanged between the Java and native space and vice versa. Thus, \textbf{Connection setup} in the transport implementation in Java is limited to creating the Java connection object for DXNet's connection manager. \textbf{Connection close and cleanup} is similar with an additional callback to the native library to signal a connection was closed to Ibdxnet's connection management.

%%%%%%%%%%%%%%%%%%	
\subsection{Dispatch of Ready-to-send Data}
\label{transport_send}

\begin{figure}[!t]
	\centering
	\includegraphics[width=3.0in]{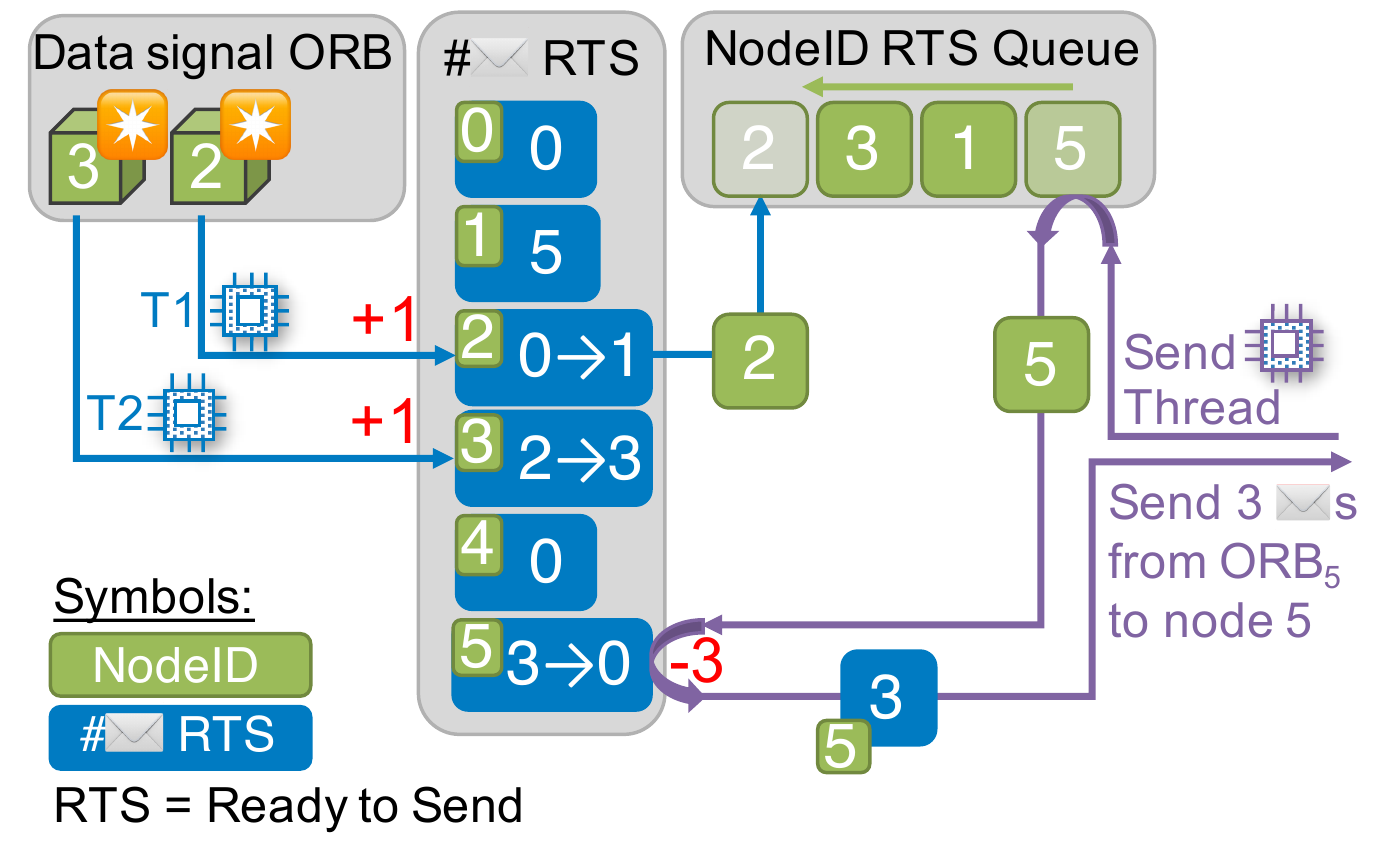}
	\caption{Internals of the Write Interest Manager (WIM).}
	\label{ibdxnet_wim}
\end{figure}

The engine msgrc is running dedicated threads for sending data. The send thread pulls new data from the transport via the \textit{GetNextDataToSend} function of the low-level interface (\S \ref{engine_interface}, \S \ref{send_thread}). In order to allow this and other callbacks (for connection management and receiving data) to be available to the IB transport, a lightweight JNI binding with the aspects explained in Section \ref{java_and_native} was created. The transports implement the \textit{GetNextDataToSend} function exposed by the JNI binding. To get new data to send, the send thread calls the JNI binding which is implemented in the IB transport in Java.

Next, we elaborate on the implementation of \textit{GetNextDataToSend} in the IB transport, how the send thread gets data to send and how the different states for the data (posted, not posted, send completed) are handled in combination with the existing ORB data structure.

Application threads using DXNet and sending messages are concurrently serializing them into the ORB (\S \ref{dxnet_send}). Once serialization completes, the thread signals the transport that there is ready to send (RTS) data in the ORB. For the IB transport, this signal \textbf{adds a write interest to the dedicated Write Interest Manager (WIM)}. The WIM manages interest tokens using a lock-free list (based on a ring buffer) and a per connection atomic counter for both, RTS normal data from the ORB and FC data. Each type has a separate atomic counter, but, if not explicitly stated, we refer to them as one for ease of comprehension.

The list contains the nodeIDs of the connections that have RTS data in the order they were added. The atomic counter is used to keep track of the number of interests signalled, i.e. the number of times the callback was triggered for the selected NID. 

Figure \ref{ibdxnet_wim} depicts this situation with two threads (T1 and T2) which finished serializing data to the ORBs of two independent connections (3 and 2). The table with atomic counters keeps track of the number of signaled interests for RTS data/messages per connection. By calling \textit{GetNextDataToSend}, the send thread from Ibdxnet checks a lock-free list which contains nodeIDs of the connections with at least one write interest available. The nodeIDs are added in order to the list but only if it is not already in the list. This is detected by checking if the atomic counter returned 0 after a fetch and add operation. This mechanism ensures that data from many connection is processed in a round robin fashion. Furthermore, avoiding duplicates in the queue sets an upper bound for memory requirement which is \textit{sizeof(nodeID) * maxNumConnections}. Otherwise, the queue can grow depending on the load and number of active connections. If the queue of the WIM is empty, the send thread aborts and returns to the native space.

The send thread uses the NID it removed from the queue to get and reset the number of interests of the corresponding atomic counter. If there are any interests available for FC data, the send thread processes them by getting the FC from the connection and getting, but not yet removing, the stored FC data. For interests concerning normal data, the send thread gets the ORB from the connection and reads the current front and back pointers. The pointers of the ORB are not modified, only read (details below). With this data, along with the NID of the connection, the send thread returns to the native space for processing (\S \ref{send_thread}).

Every time the send thread returns to the Java space to get more data to send, it carries the parameters \textit{prevWorkResults}, which contains data about the previous send operation, and \textit{completionList}, which contains data about completed WRs, i.e. data send confirmations (\S \ref{send_thread}). For performance reasons, this data resides in native memory as structs and is mapped and accessed using DirectByteBuffers (\S \ref{java_and_native}).

The asynchronous workflow used to send and receive data by posting WRs and polling WCs must be adopted by updating the ORB and FC accordingly. Depending on the fill level of the SQ, the send thread might not be able to post all normal data or FC it retrieved in the previous iteration. The \textit{prevWorkResults} parameter contains this information about how much normal and FC data was processed and could not be processed. This information must be preserved for the next send operation to avoid sending data multiple times. For the ORB however, we cannot move the front pointer because this frees up the memory which is not confirmed to be sent, yet.

\begin{figure}[!t]
	\centering
	\includegraphics[width=3.0in]{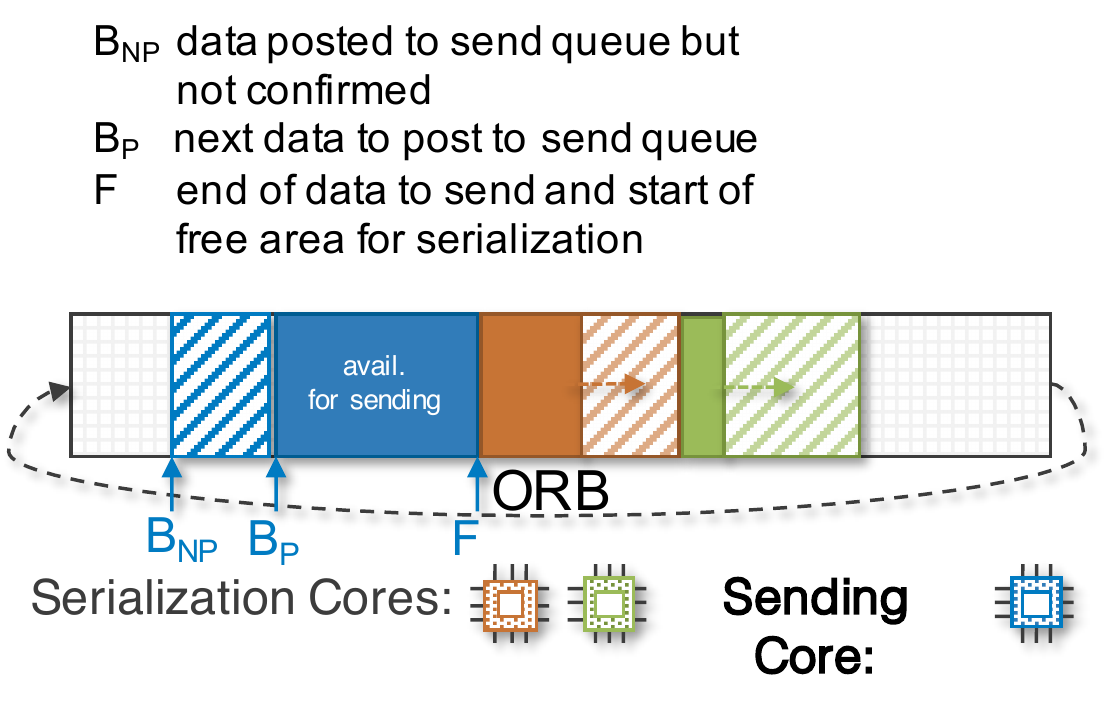}
	\caption{Extended outgoing ring buffer used by IB transport.}
	\label{ib_ringbuffer}
\end{figure}

Thus, we introduce a second front pointer, front posted, which is only known to and modified by the send thread and allows it to keep track of already posted data. Figure \ref{ib_ringbuffer} depicts the most important aspects of the enhanced ORB which is used for the IB transport. In total, this creates three virtual areas of memory designated to the following states:
\begin{itemize}
 \item Data posted but not confirmed: front to front posted
 \item Data RTS and not posted: front posted to back
 \item Free memory for send threads to serialize to: back to front
\end{itemize}

Using the parameter \textit{prevWorkResults}, the front posted pointer is moved by the amount of data posted. Any non processed data remains unprocessed (front posted not moved to cover entire area of RTS data). For data provided with the parameter \textit{completionList}, the front pointer is updated according to the number of bytes now confirmed to be sent. A similar but less complex approach is applied to updating FC.

%%%%%%%%%%%%%%%%%%
\subsection{Process Incoming Buffers}
\label{transport_recv}

The dedicated receive thread of msgrc is pushing received data to the low-level interface. Analogous to how RTS data is pulled from the IB transport via the JNI binding, the receive thread uses a received function provided by the binding to push the received buffers to the IB transport into Java space. All received buffers are stored as a batch in the \textit{recvPackage} data structure (\S \ref{engine_interface}) to minimize context switching overhead. For performance reasons, this data resides in native memory as structs and is mapped and accessed using DirectByteBuffers (\S \ref{java_and_native}).

The receive thread iterates the package in Java space, dispatches received FC data to each connection and pushes the received buffers (including the connection of the source node) to the IBQ (\S \ref{dxnet_receive}). The buffers are handled and processed asynchronously by the MessageCreationCoordinator and one or multiple MessageHandlers of the DXNet core (all of them are Java threads). Once the buffers are processed (de-serializing its contents), the Java threads return them asynchronously to the transport engines receive buffer pool (\S \ref{receive_thread}).

%%%%%%%%%%%%%%%%%%%%%%%%%%%%%%%%%%%%%%%%%%%%%%%%%%%%%%%%%%%%%%%%%%%%%%%%
%%%%%%%%%%%%%%%%%%%%%%%%%%%%%%%%%%%%%%%%%%%%%%%%%%%%%%%%%%%%%%%%%%%%%%%%
%%%%%%%%%%%%%%%%%%%%%%%%%%%%%%%%%%%%%%%%%%%%%%%%%%%%%%%%%%%%%%%%%%%%%%%%
%%%%%%%%%%%%%%%%%%%%%%%%%%%%%%%%%%%%%%%%%%%%%%%%%%%%%%%%%%%%%%%%%%%%%%%%
%%%%%%%%%%%%%%%%%%%%%%%%%%%%%%%%%%%%%%%%%%%%%%%%%%%%%%%%%%%%%%%%%%%%%%%%
%%%%%%%%%%%%%%%%%%%%%%%%%%%%%%%%%%%%%%%%%%%%%%%%%%%%%%%%%%%%%%%%%%%%%%%%

%%%%%%%%%%%%%%%%%%%%%%%%%%%%%%%%%%%%%%%%%%%%%%%%%%%%%%%%%%%%%%%%%%%%%%%%    
\section{Evaluation}
\label{eval}

For better readability, we refer to DXNet with the IB transport Ibdxnet and msgrc engine as DXNet from here onwards.

We implemented commonly used microbenchmarks to compare DXNet to two MPI implementations supporting InfiniBand: MVAPICH2 and FastMPJ. We decided to compare against two MPI implementations for the following reasons: To the best of our knowledge, there is no other system available that offers all features of DXNet and big data applications implementing their dedicated network stack do not offer it as a separate application/library like DXNet does. MPI can be used to partially cover some features of DXNet but not all (\S \ref{related_work}). We are aware that MPI is targeting a different application domain, mainly HPC, whereas DXNet is targeting big data. However, MPI was already used in big data applications as well and several aspects related to the network stack and the technologies are overlapping in both application domains.

Bandwidth with two nodes is compared using typical uni- and bi-directional benchmarks. We also compared scalability using an all-to-all benchmark (worst-case scenario) with up to 8 nodes. Latency is compared by measuring the RTT with a request-response communication pattern. These benchmarks are executed single threaded to compare all three systems.

Furthermore, we compared how DXNet and MVAPICH2 perform in a multi-threaded environment which is typical for Big Data but not HPC applications. However, we can only compare it using three benchmarks. Latency multi-threaded is not possible since it would require MVAPICH2 to implement additional infrastructure to store and map requests with responses and dynamic dispatching callbacks to handlers of incoming data to multiple receive threads (similar to DXNet). MVAPICH2 does not provide such a processing pipeline. FastMPJ cannot be compared at all here because it only supports single threaded environments. Table \ref{eval_benchmarks} summerizes the systems and benchmarks executed.

All benchmarks were executed on up to 8 nodes of our private cluster, each with a single socket Intel Xeon CPU E5-1650 v3, 6 cores running at 3.50 GHz per core clock speed and 64 GB RAM. The nodes are running Ubuntu 16.04 with kernel version 4.4.0-57. All nodes are equipped with a Mellanox MT27500 HCA, connected with 56 Gbps links to a single Mellanox SX6015 18 port switch. For Java applications, we used the Oracle JVM version 1.8.0\_151.

\begin{table}[]
\centering
\begin{tabular}{|l|c|c|c|}
\hline
                 & \multicolumn{1}{l|}{FastMPJ} & \multicolumn{1}{l|}{MVAPICH2} & \multicolumn{1}{l|}{DXNet} \\ \hline
Uni-dir. TP ST   & x                            & x                             & x                          \\ \hline
Bi-dir. TP ST    & x                            & x                             & x                          \\ \hline
Latency ST       & x                            & x                             & x                          \\ \hline
All-to-all TP ST & x                            & x                             & x                          \\ \hline
Uni-dir. TP MT   &                              &                               & x                          \\ \hline
Bi-dir. TP MT    &                              & x                             & x                          \\ \hline
Latency MT       &                              &                               & x                          \\ \hline
All-to-all MT    &                              &                               & x                          \\ \hline
\end{tabular}
\caption{Summary of benchmarks and systems. TP = throughput, ST = single threaded, MT = multi-threaded}
\label{eval_benchmarks}
\end{table}

%%%%%%%%%%%%%%%%%%%%%%%%%%%%%%%%%%%%%%%%%%%%%%%%%%%%%%%%%%%%%%%%%%%%%%%%
\subsection{Benchmarks}
\label{benchmarks}

The \textit{osu} benchmarks included with MVAPICH2 implement typical micro benchmarks to measure uni- and bi-directional bandwidth and uni-directional latency which reflect basic usage of any network stack for point-to-point communication. \textit{osu\_latency} is used as a foundation and extended with recording of all RTTs to determine the 95th, 99th and 99.9th percentile after execution. The latency measured is the full RTT when the source is sending a request to the destination up to when the corresponding response is received by the source. For evaluating throughput, the benchmarks \textit{osu\_bw} and \textit{osu\_bibw} were combined to a single benchmark and extended to enable all-to-all bi-directional execution with more than two nodes. We consider this a relevant benchmark to show if the system is capable of handling multiple connections under high load. This is a common situation found in big data applications as well as backend storages \cite{yahoo}. On all-to-all, every node receives from all other nodes and sends messages to all other nodes in a round robin fashion. The bi-directional and all-to-all results presented are the aggregated send throughputs of all participating nodes. We added options to support multi-threaded sending and receiving using a configurable number of send and receive threads. As the per-processor core count increases, the multi-threading aspect becomes more and more important. Furthermore, our target application domain big data relies heavily on multi-threaded environments.

For the evaluation of FastMPJ, we ported the \textit{osu} benchmarks to Java. The benchmarks for evaluating a multi-threaded MPI process were omitted because FastMPJ does not support multi-threaded processes. DXNet comes with its own benchmarks already implemented which are comparable to the \textit{osu} benchmarks.

The \textit{osu} benchmarks use a configurable parameter \textit{window\_size} (WS) which denotes the number of messages sent in a single batch. Since MPI does not support implicit message aggregation like DXNet, we executed all MPI experiments with increasing WS to determine bandwidth peaks and saturation under optimal conditions and ensure a fair comparison to DXNet's built in aggregation. No MPI collectives are required for the benchmarks and, thus, aren't evaluated.

All benchmarks are executed three times and their variance is displayed using error bars. Throughputs are specified in GB/s, latencies/RTTs in us and message rates in mmps (million messages per second). All throughput benchmarks send 100 million messages and all latency benchmarks 10 million messages. The total number of messages is incrementally halved starting with 4 kb message size to avoid unnecessary long running benchmark runs. All throughputs measured are based on the total amount of sent payload bytes. This does not include any overhead like message headers or envelopes that are required by the systems for message identification or routing.

Furthermore, we included the results of the ib perf tools \textit{ib\_write\_bw} and \textit{ib\_write\_lat} as baselines to all end-to-end type benchmarks. These simple perf tools cannot be compared directly to the complex systems evaluated. But, these baselines show the best possible network performance (without any overhead by the evaluated system) and for rough comparisons of the systems across multiple plots. We chose parameters that reflect the configuration values of DXNet as close as possible (but still allow comparisons to FastMPJ and MVAPICH2 as well): receive queue size 2000 and send queue size 20 for both bandwidth and latency measurements; 100,000,000 messages for bandwidth and 10,000,000 for latency.

%%%%%%%%%%%%%%%%%%%%%%%%%%%%%%%%%%%%%%%%%%%%%%%%%%%%%%%%%%%%%%%%%%%%%%%%
\subsection{DXNet with Ibdxnet Transport}

\begin{table*}[t]
\centering
\begin{tabular}{|l|l|}
\hline
IBQ max. capacity buffer count & 8192 \\ \hline
IBQ max. capacity aggregated data size & 128 MB \\ \hline
Message handlers & varying (see experiments) \\ \hline
IB SQ size (per connection) & 20 \\ \hline
IB SRQ size & 2000 (default value for up to 100 connections) \\ \hline
Max. connection limit & 100 \\ \hline
Recv buffer pool capacity & 4 GB \\ \hline
Flow control window & 16 MB \\ \hline
Flow control threshold & 0.1 \\ \hline
Receive buffer size (for small message sizes 1 - 16 kb) & 32 kb \\ \hline
SGEs per WR (for small message sizes 1 - 16 kb) & 4 \\ \hline
Receive buffer size (for medium/large message sizes 32 kb - 1 MB) & 1 MB \\ \hline
SGEs per WR (for medium/large message sizes 32 kb - 1 MB) & 1 \\ \hline
\end{tabular}
\caption{DXNet configuration values for experiments}
\label{dxnet_config}
\end{table*}

We configured DXNet using the parameters depicted in Table \ref{dxnet_config}. The configuration values were determined with various debugging statistics and experiments, and are currently considered optimal configuration parameters.

For comparing single threaded performance, the number of application threads and message handlers (referred to as MH) is limited to one each to allow comparing it to FastMPJ and MVAPICH2. DXNet's multi-threaded architecture does not allow combining the logic of the application send thread and a message handler into a single thread. Thus, DXNet's ``single threaded'' benchmarks are always executed with one dedicated send and one dedicated receive thread.

The following subsections present the results of the various benchmarks. First, we present the results of all single threaded benchmarks with one send thread: uni- and bi-directional throughput, uni-directional latency and all-to-all with increasing node count. Afterwards, the results of the same four benchmarks are presented with multiple send threads.

%%%%%%%%%%%%%%%%%%%%%%%%%%%
\subsubsection{Uni-directional Throughput}
\label{eval_dxnet_uni_tp}

\begin{figure}[!t]
	\centering
	\includegraphics[width=3.4in]{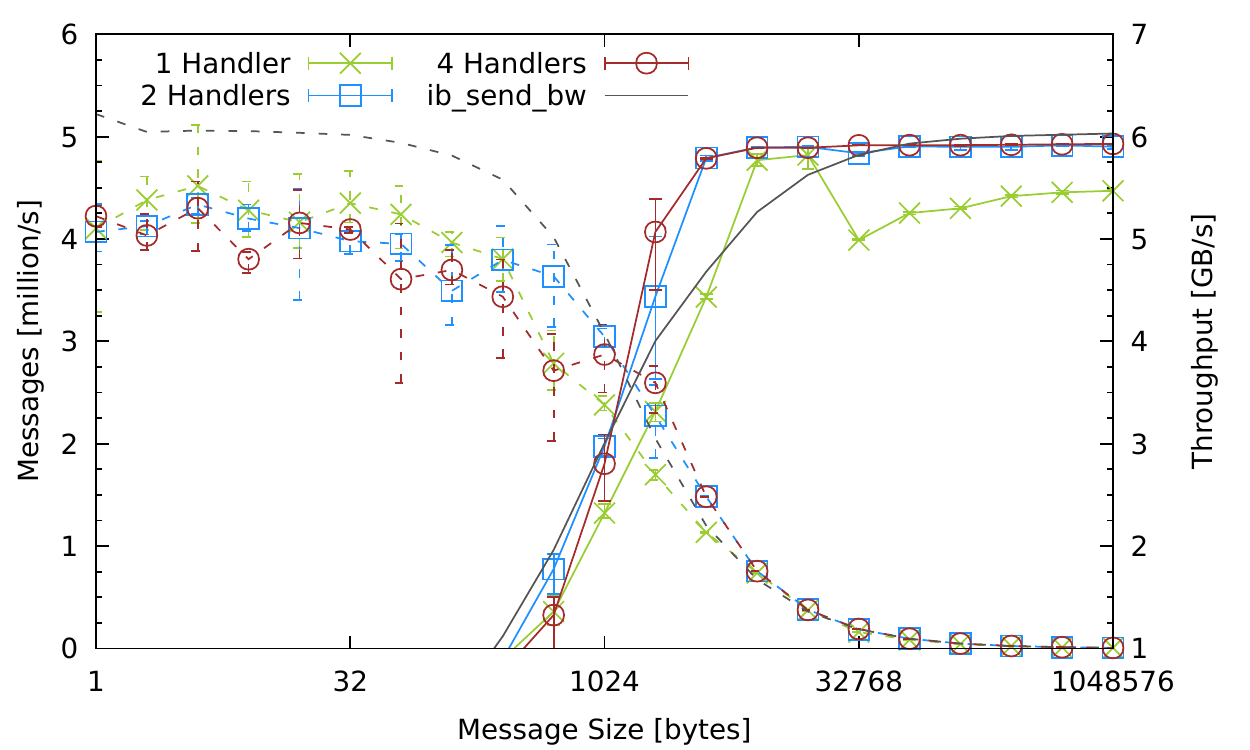}
	\caption{\textbf{DXNet}: 2 nodes, uni-directional throughput and message rate with one application send thread, increasing message size and number of message handlers}
	\label{eval_dxnet_uni_bw}
\end{figure}

The results of the uni-directional benchmark are depicted in figure \ref{eval_dxnet_uni_bw}. Considering one MH, DXNet's throughput peaks at 5.9 GB/s at a message size of 16 kb. For larger messages (32 kb to 1 MB), one MH is not sufficient to de-serialize and dispatch all incoming messages fast enough and drops to a peak bandwidth of 5.4 GB/s. However, this can be resolved by simply using two MHs. Now, DXNet's throughput peaks and saturates at 5.9 GB/s with a message size of just 4 kb and stays saturated up to 1 MB. Message sizes smaller than 4 kb also benefit significantly from the shorter receive processing times by utilizing two MHs. Further MHs can still improve performance but only slightly for a few message sizes.

For small messages up to 64 bytes, DXNet achieves peak message rates of 4.0-4.5 mmps using one MH. Multiple MHs cannot significantly increase performance for such small messages further. However, with growing message size (512 byte to 16 kb), the message rate can be increased with two message handlers.

Compared to the baseline performance of \textit{ib\_send\_bw}, DXNet's peak performance is approx. 0.5 to 1.0 mmps less. With increasing message size, this gap closes and DXNet even surpasses the baseline 1 kb to 32 kb message sizes when using multiple threads. DXNet peaks close to the baseline's peak performance of 6.0 GB/s. The results with small message sizes are fluctuating independent of the number of MHs. This can be observed on all other benchmarks with DXNet measuring message/payload throughput as well. It is a common issue which can be observed when running high load throughput benchmarks using the bare ibverbs API as well.

This benchmark shows that DXNet is capable of handling a vast amount of small messages efficiently. The application send thread and, thus, the user does not have to bother with aggregating messages explicitly because DXNet handles this transparently and efficiently. The overall performance benefits from multiple message handlers increasing receive throughput. Large messages do impact performance with one MH because the de-serialization of data consumes most of the processing time during receive. However, simply adding at least another MH solves this issue and further increases performance.

%%%%%%%%%%%%%%%%%%%%%%%%%%%
\subsubsection{Bi-directional Throughput}
\label{eval_dxnet_bi_tp}

\begin{figure}[!t]
	\centering
	\includegraphics[width=3.4in]{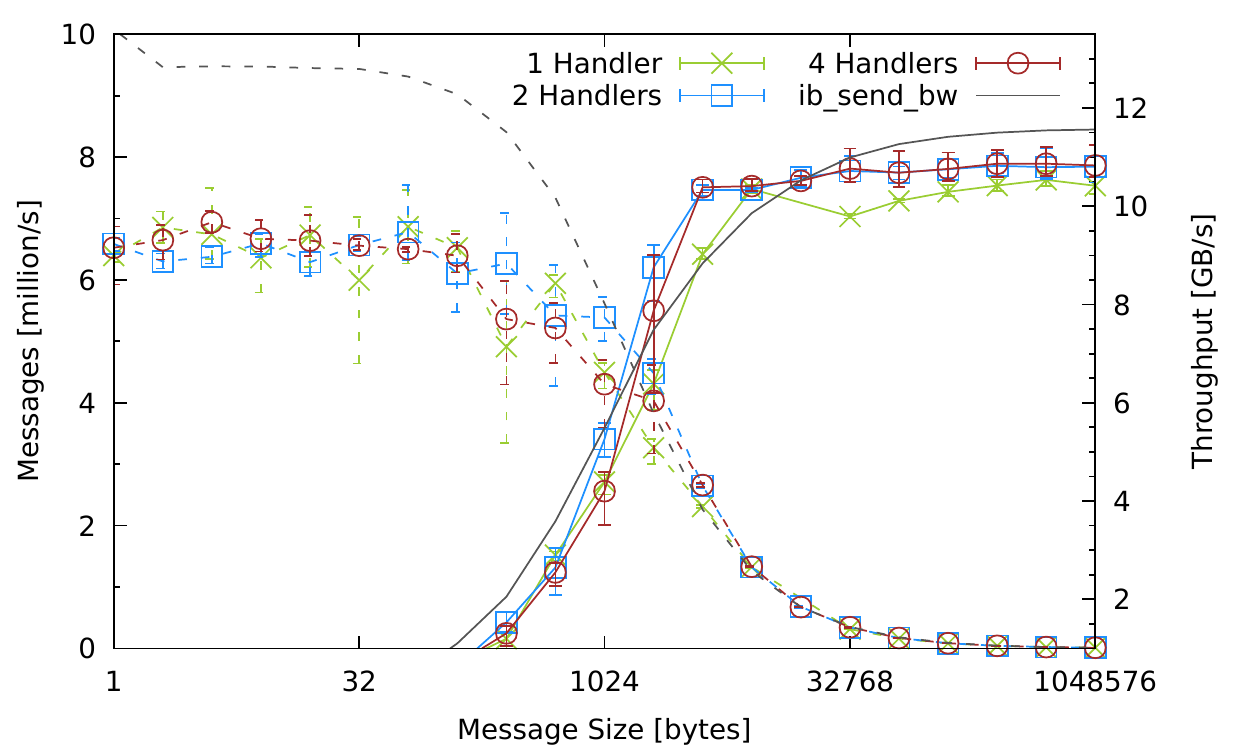}
	\caption{\textbf{DXNet}: 2 nodes, bi-directional throughput and message rate with one application send thread, increasing message size and number of message handlers}
	\label{eval_dxnet_bi_bw}
\end{figure}

Figure \ref{eval_dxnet_bi_bw} depicts the results of the bi-directional benchmark with one send thread. With one MH, the aggregated throughput peaks at approx. 10.4 GB/s for 8 kb. Using two message handlers, the fluctuations starting with 16 kb messages using one MH can be resolved (as already explained in \ref{eval_dxnet_uni_tp}). Further increasing the performance using four MHs is not possible in this benchmark and actually degrades it for 512 byte to 2 kb message sizes. DXNet's throughput peaks at approx. 10.4 GB/s and saturates with a message size of 32 kb.

The peak aggregated message rate for small messages up to 64 bytes is varying from approx. 6 to 6.9 mmps with one MH. Using more MHs cannot improve performance significantly for this benchmark. Due to the multi-threaded and highly pipelined architecture of DXNet, these variations cannot be avoided, especially when exclusively handling many small messages.

Compared to the baseline performance of \textit{ib\_send\_bw}, there is still room for improvement for DXNet's performance on small message sizes (up to 2.5 mmps difference). For medium message sizes, \textit{ib\_send\_bw} yields slightly higher throughput for up to 1 kb message size. But, DXNet surpasses \textit{ib\_send\_bw} on 1 kb to 16 kb message size. DXNet's peak performance is approx. 1.1 GB/sec less than \textit{ib\_send\_bw}'s (11.5 GB/sec).

Overall, this benchmark shows that DXNet can deliver great performance especially for small messages similar to the uni-directional benchmark (\S \ref{eval_dxnet_uni_tp}).

%%%%%%%%%%%%%%%%%%%%%%%%%%%
\subsubsection{Uni-directional Latency}
\label{eval_dxnet_uni_lat}

\begin{figure}[!t]
	\centering
	\includegraphics[width=3.4in]{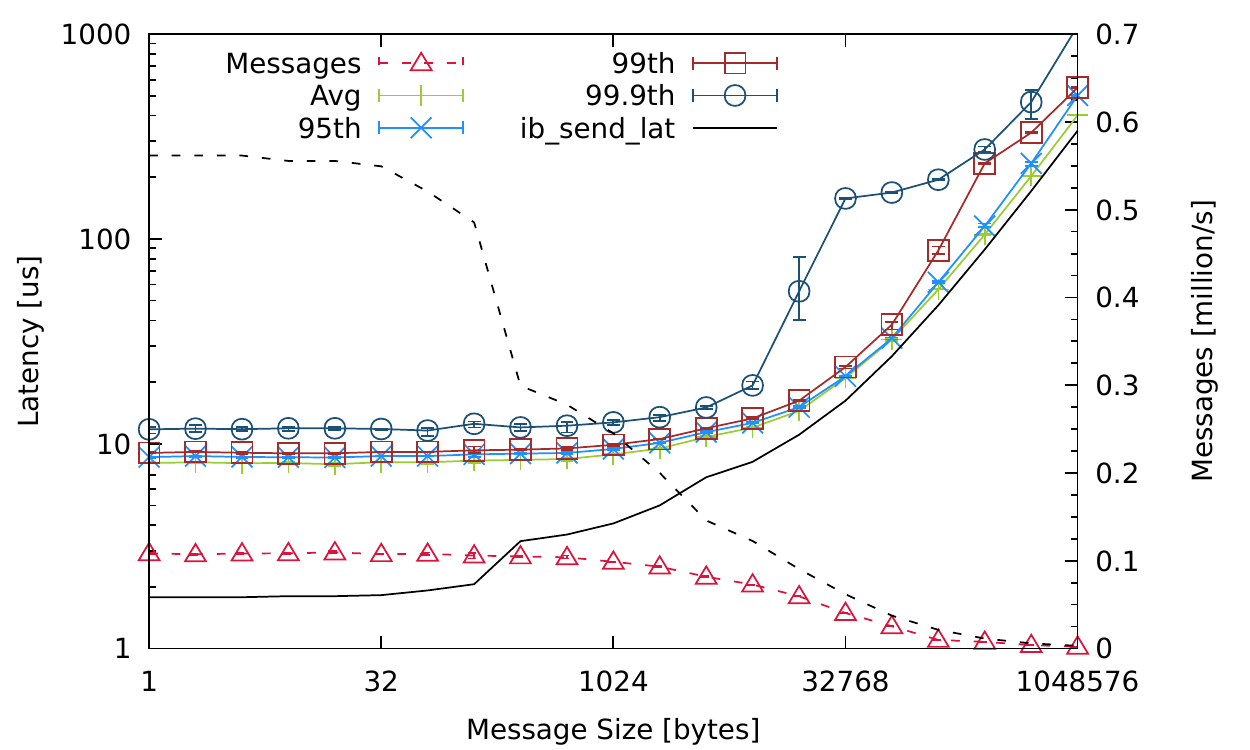}
	\caption{\textbf{DXNet}: 2 nodes, uni-directional RTT and message rate with one application send thread, increasing message size}
	\label{fig_eval_dxnet_uni_lat}
\end{figure}

Figure \ref{fig_eval_dxnet_uni_lat} depicts the average RTTs as well as the 95th, 99th and 99.9th percentile of the uni-directional latency benchmark with one send thread and one MH. For message sizes up to 512 bytes, DXNet achieves an avg. RTT of 7.8 to 8.3 \textmu s, a 95th percentile of 8.5 to 8.9 \textmu s, a 99th percentile of 8.9 to 9.2 and 99.9th percentile of 11.8 to 12.7 \textmu s. This results in a message rate of approx 0.1 mmps. As expected, starting with 1 kb message size, latency increases with increasing message size.

The RTT can be broken down into three parts: DXNet, Ibdxnet and hardware processing. Taking the lowest avg. of 7.8 \textmu s, DXNet requires approx. 3.5 \textmu s of the total RTT (the full breakdown is published in our other publication \cite{dxnet}) and the hardware approx. 2.0 \textmu s (assuming avg. one way latency of 1 \textmu s for the used hardware). Message de- and serialization as well as message object creation and dispatching are part of DXNet. For Ibdxnet, this results in approx. 2.3 \textmu s processing time which includes JNI context switching as well as several pipeline stages explained in the earlier sections.

Compared to the baseline performance of \textit{ib\_send\_lat}, DXNet's latency is significantly higher. Obviously, additional latency cannot be avoided with such a long and complex processing pipeline. Considering the breakdown mentioned above, the native part Ibdxnet, which calls ibverbs to send and receive data, is to some degree comparable to the minimal perf tool \textit{ib\_send\_bw}. With a total of 2.3 \textmu s (of the full pipeline's 7.8 \textmu s), the total RTT is just slightly higher than \textit{ib\_send\_bw}'s 1.8 \textmu s. But, Ibdxnet already includes various data structures for state handling and buffer scheduling (\S \ref{send_thread}, \S \ref{receive_thread}) which \textit{ib\_send\_bw} doesn't. Buffers for sending data are re-used instantly and the data received is discarded immediately.

%%%%%%%%%%%%%%%%%%%%%%%%%%%
\subsubsection{All-to-all Throughput with up to 8 Nodes}
\label{eval_dxnet_nodes_tp}

\begin{figure}[!t]
	\centering
	\includegraphics[width=3.4in]{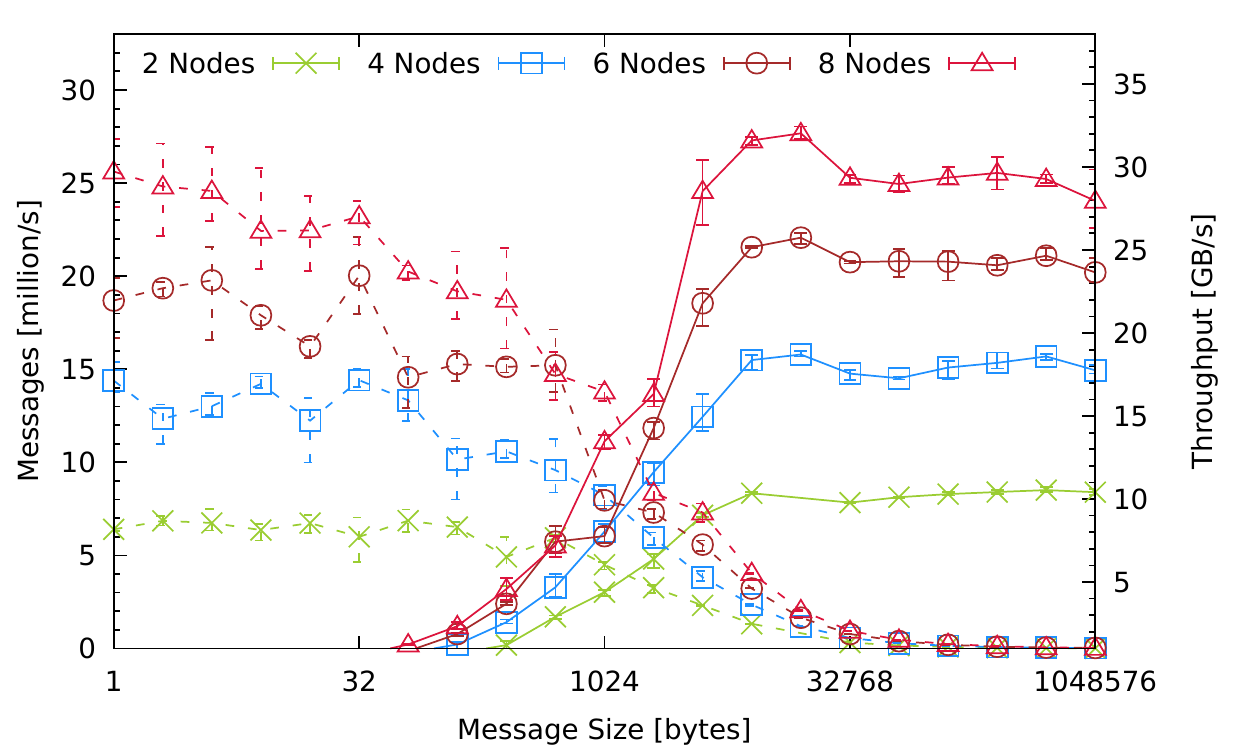}
	\caption{\textbf{DXNet}: 2 to 8 nodes, all-to-all aggregated send throughput and message rate with one application send thread, increasing message size and one message handler}
	\label{eval_dxnet_nodes}
\end{figure}

Figure \ref{eval_dxnet_nodes} shows the aggregated send throughput and message rate of all participating nodes (up to 8) executing the all-to-all benchmark with one send thread and one MH with increasing message size. For small up to 64 byte messages, peak message rates of 7.0 mmps, 14.5 mmps, 20.1 mmps and 25.6 mmps are achieved for 2, 4, 6 and 8 nodes. Throughput increases with increasing node count peaking at 8 kb message size with 10.4 GB/s for 2 nodes. The peaks for 4, 6 and 8 nodes are reached with 16 kb message size at 18.9 GB/s, 26.0 GB/s and 32.4 GB/s. Incrementally adding two nodes, throughput is increased by 8.5 GB/s (for 2 to 4 nodes), by 7.1 GB/s (for 4 to 6 nodes) and 6.4 GB/s (for 6 to 8 nodes). One would expect approx. equally large throughput increments but the gain is noticeably lowered with every two nodes added.

We tried different configuration parameters for DXNet and ibverbs like different MTU sizes, SGE counts, receive buffer sizes, WRs per SQ/SRQ or CQ size. No combination of settings allowed us to improve this situation.

We assume that the all-to-all communication pattern puts high stress on the HCA which, at some point, cannot keep up with processing outstanding requests. To rule out any software issues with DXNet first, we implemented a low-level ``loopback'' like test which uses the native part of Ibdxnet, only. The loopback test does not involve any dynamic message posting when sending data or data processing when receiving. Instead, a buffer equally to the size of the ORB is processed by Ibdxnet's send thread on every iteration and posted to every participating SQ. This ensures that all SQs are filled and are quickly refilled once at least one WR was processed. When receiving data on the SRQ, all buffers received are directly put back into the pool without processing and the SRQ is refilled. This ensures that no additional processing overhead is added for sending and receiving data. Thus, Ibdxnet's loopback test comes close to a perftool like benchmark. We executed the benchmark with 2, 4, 6 and 8 nodes which yielded aggregated throughputs of 11.7 GB/s, 21.7 GB/s, 28.3 GB/s and 34.0 GB/s. 

These results are very close to the performance of the full DXNet stack but don't rule out all software related issues, yet. The overall aggregated bandwidth could still somehow be limited by Ibdxnet. Thus, we executed another benchmark which, first, executes all-to-all communication with up to 8 nodes, then, once bandwidth is saturated, switching to a ring formation for communication without restarting the benchmark (every node sends to its successor determined by NID, only). 

Once the nodes switch the communication pattern during execution, the per node aggregated bandwidth increases very quickly and reaches a maximum aggregated bandwidth of approx. $(11.7 / 2 \times num\_nodes)$ GB/s independent of the number of nodes used. This rules out total bandwidth limitations for software and hardware. Furthermore, we can now rule out any performance issues in DXNet or even ibverbs with connection management (e.g. too many QPs allocated).

This leads to the assumption that the HCA cannot keep up with processing outstanding WRQs when SQs are under high load (always filled with WRQs). With more than 3 SQs per node, the total bandwidth drops noticably. Similar results with other systems further support this assumption (\S \ref{eval_fmpj_nodes} and \ref{eval_mva_nodes}).

%%%%%%%%%%%%%%%%%%%%%%%%%%%
\subsubsection{Uni-directional Throughput Multi-threaded}

\begin{figure}[!t]
	\centering
	\includegraphics[width=3.4in]{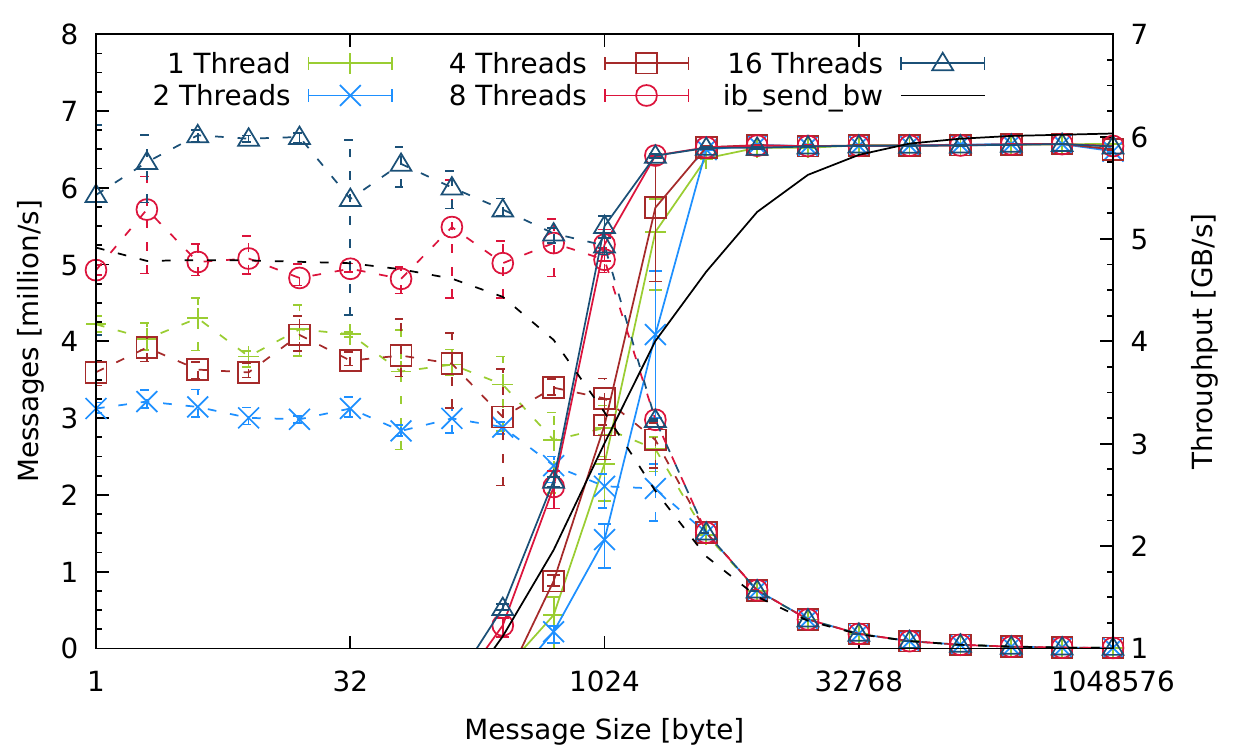}
	\caption{\textbf{DXNet}: 2 nodes, uni-directional throughput and message rate with multiple application send threads, increasing message size and 4 message handlers}
	\label{eval_dxnet_uni_bw_mt}
\end{figure}

Figure \ref{eval_dxnet_uni_bw_mt} shows the uni-directional benchmark executed with 4 MHs and 1 to 16 send threads. For 1 to 4 send threads throughput saturates at 5.9 GB/s at either 4 kb or 8 kb messages. For 256 byte to 8 kb, using one thread yields better throughput than two or sometimes four threads. However, running the benchmark with 8 and 16 send threads increases overall throughput for all messages greater 32 byte significantly with saturation starting at 2 kb message size. DXNet's pipeline benefits from the many threads posting messages to the ORB concurrently. This results in greater aggregation of multiple messages and allows higher buffer utilization for the underlaying transport.

DXNet also increases message throughput on small message sizes up to 512 byte. from approx. 4.0 mmps up to 6.7 mmps for 16 send threads. Again, performance is slightly worse with two and four compared to a single thread.

Furthermore, DXNet even surpasses the baseline performance of \textit{ib\_send\_bw} when using multiple send threads. However, the peak performance cannot be improved further which shows the current limit of DXNet for this benchmark and the hardware used.

%%%%%%%%%%%%%%%%%%%%%%%%%%%
\subsubsection{Bi-directional Throughput Multi-threaded}

\begin{figure}[!t]
	\centering
	\includegraphics[width=3.4in]{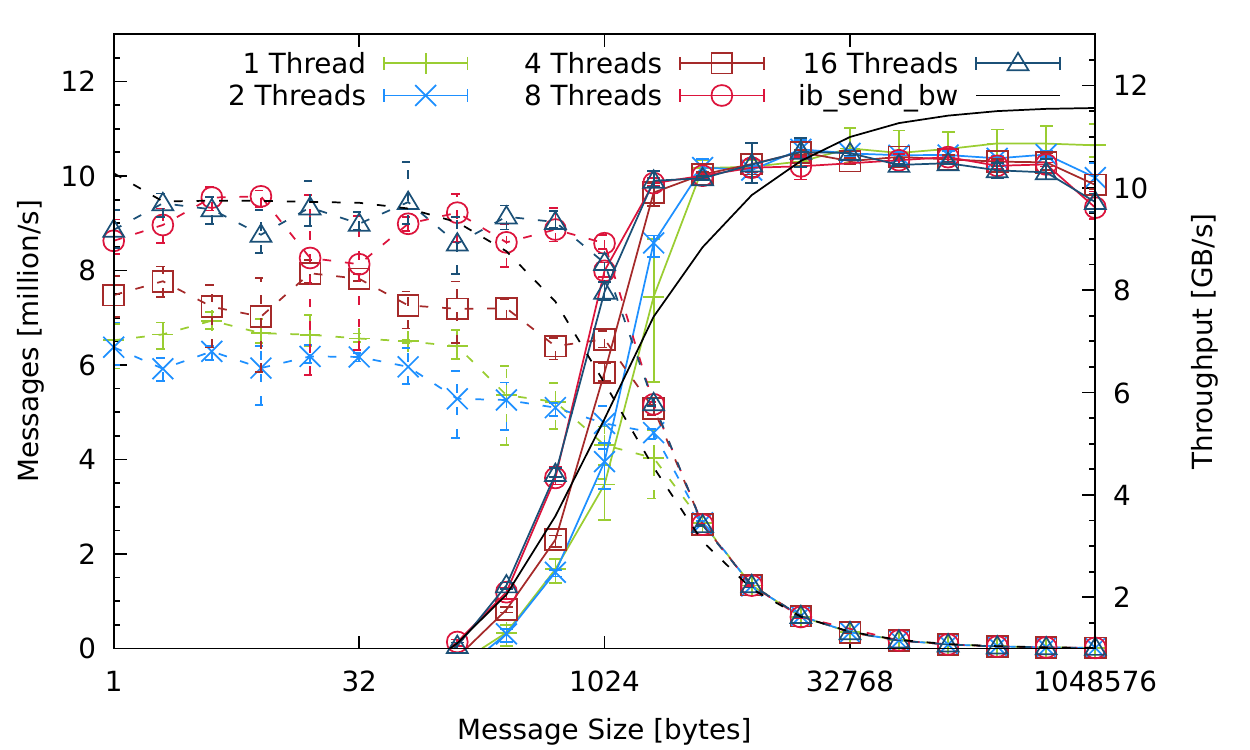}
	\caption{\textbf{DXNet}: 2 nodes, bi-directional throughput and message rate with multiple application send threads, increasing message size and 4 message handlers}
	\label{eval_dxnet_bi_bw_mt}
\end{figure}

Figure \ref{eval_dxnet_bi_bw_mt} shows the bi-directional benchmark executed with 4 MHs and 1 to 16 send threads. With more than one send thread, the aggregated throughput peaks at approx. 10.4 and 10.7 GB/s with messages sizes of 2 and 4 kb. DXNet delivers higher throughputs for all medium and small messages with increasing send thread count. The baseline performance of \textit{ib\_send\_bw} is reached on small message sizes and even surpassed with medium sized messages up to 16 kb. The peak throughput is not reached showing DXNet's current limit with the used hardware.

The overall performance with 8 and 16 send threads don't differ noticeably which indicates saturation of DXNet's processing pipeline. For small messages (less than 512 byte), the message rates also increase with increasing send thread count. Again, saturation starts with 8 send threads with a message rate of approx. 8.6 to 10.2 mmps.

DXNet is capable of handling a multi-threaded environment under high load with CPU over-provisioning and still delivers high throughput. Especially for small messages, DXNet's pipeline even benefits from the highly concurrent activity by aggregating many messages. This results in higher buffer utilization and, for the user, higher overall throughput.

%%%%%%%%%%%%%%%%%%%%%%%%%%%
\subsubsection{Uni-directional Latency Multi-threaded}

\begin{figure}[!t]
	\centering
	\includegraphics[width=3.4in]{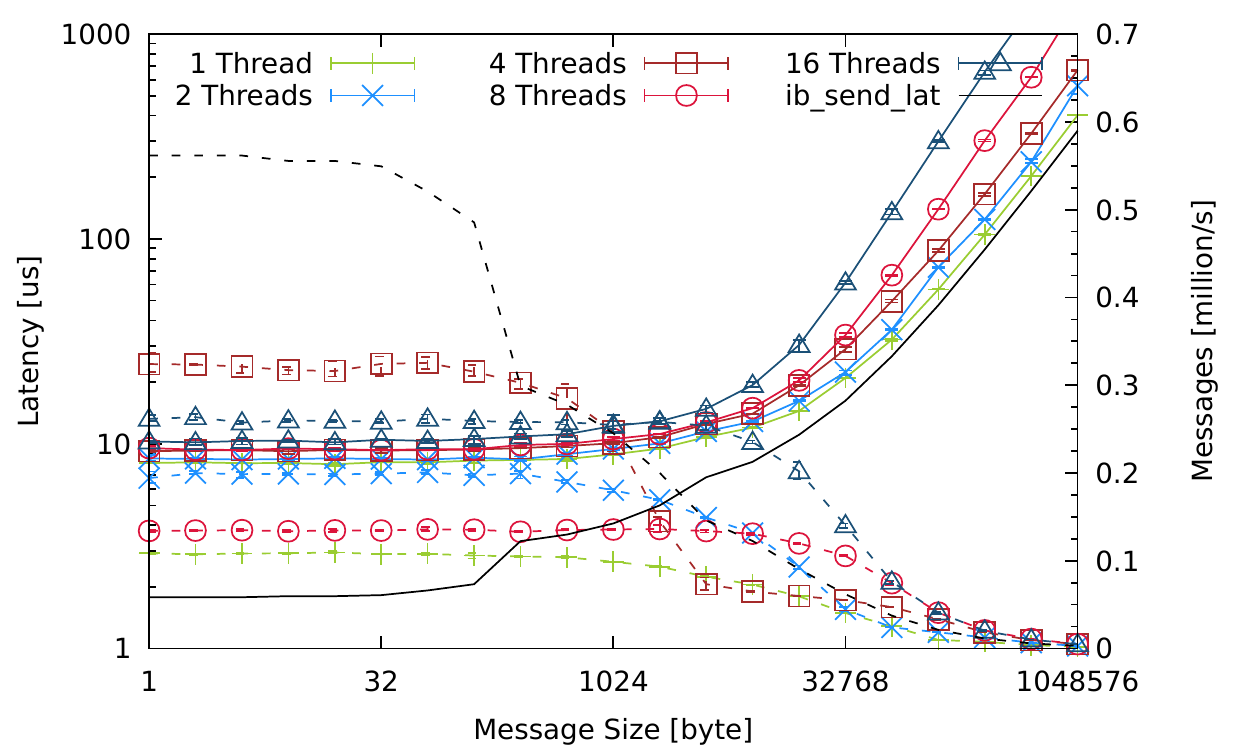}
	\caption{\textbf{DXNet}: 2 nodes, uni-directional avg. RTT and message rate with multiple application send threads, increasing message size and 4 message handlers}
	\label{eval_dxnet_uni_lat_mt}
\end{figure}

\begin{figure}[!t]
	\centering
	\includegraphics[width=3.4in]{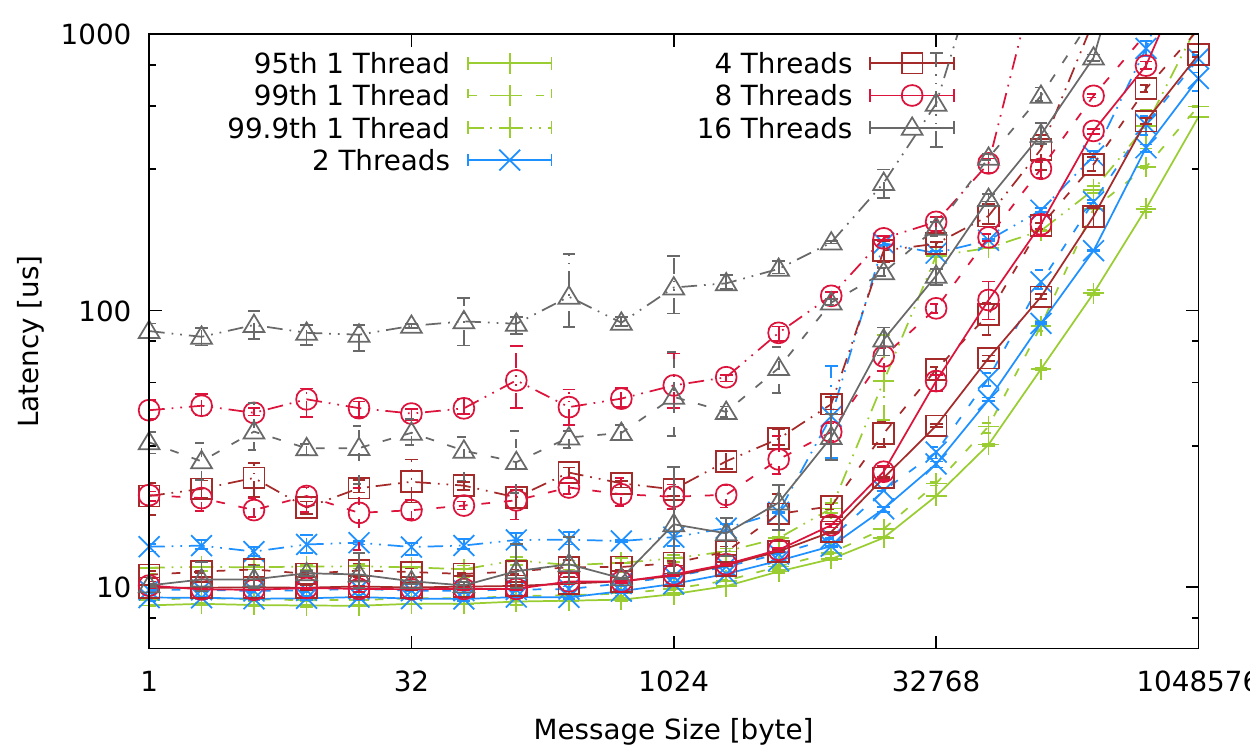}
	\caption{\textbf{DXNet}: 2 nodes, uni-directional 95th, 99th and 99.9th percentile RTT and message rate with multiple application send threads, increasing message size and 4 message handlers}
	\label{eval_dxnet_uni_lat_mt2}
\end{figure}

Figure \ref{eval_dxnet_uni_lat_mt} depicts the avg. RTT and message rate of the uni-directional latency benchmark with up to 16 send threads and 4 MHs. The 95th, 99th and 99.9th percentiles are depicated in figure \ref{eval_dxnet_uni_lat_mt2}. DXNet keeps a very stable avg. RTT of 8.1 \textmu s for message sizes of 1 to 512 bytes with one send thread. Using two send threads, this value just slightly increases. With four or more send threads the avg. RTT increases to approx 9.3 \textmu s. When the total number of threads, which includes DXNet's internal threads, MH and send threads, exceed the core count of the CPU, DXNet switches to different parking strategies for the different thread types which slightly increase latency but greatly reduce overall CPU load (\S \ref{thread_parking}).

The message rate can be increased up to 0.33 mmps with up to 4 send threads as, practically, every send thread can use a free MH out of the 4 available. With 8 and 16 send threads, the MHs on the remote must be shared and DXNet's over-provisioning is active which reduces the overall throughput. The percentiles shown in figure \ref{eval_dxnet_uni_lat_mt2} reflect this sitution very well and increase noticeably. 

With a single thread, as already discussed in \ref{eval_dxnet_uni_lat}, the difference of the avg. (7.8 to 8.3 \textmu s) and 99.9th percentile (11.8 to 12.7 \textmu s) RTT for message sizes less than 1 kb is approx. 4 to 5 \textmu s. When doubling the send thread count, the 99.9th percentiles roughly double as well. When over-provisioning the CPU, we cannot avoid the higher than usual RTT caused by the increasing amount of messages getting posted.

%%%%%%%%%%%%%%%%%%%%%%%%%%%
\subsubsection{All-to-all Throughput with up to 8 Nodes Multi-threaded}

\begin{figure}[!t]
	\centering
	\includegraphics[width=3.4in]{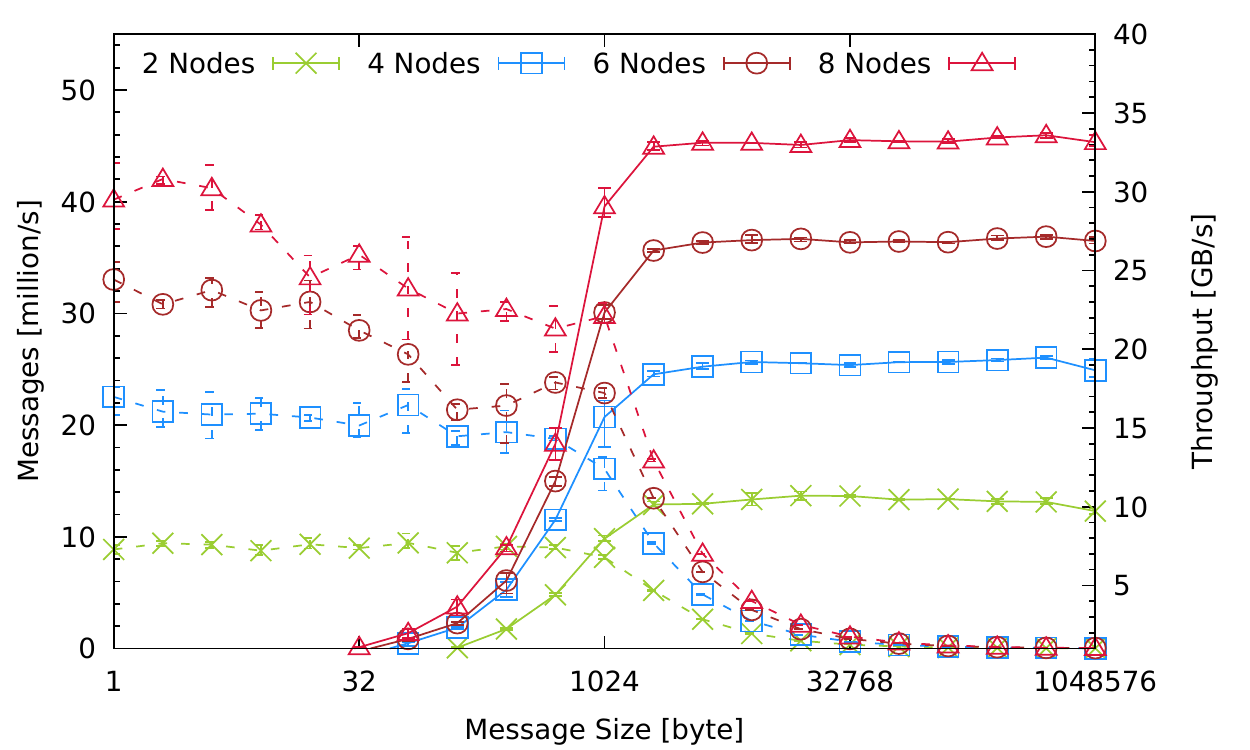}
	\caption{\textbf{DXNet}: 2 to 8 nodes, all-to-all aggregated send throughput and message rate with 16 application send threads, increasing message size and 4 message handlers}
	\label{eval_dxnet_nodes_mt}
\end{figure}

Figure \ref{eval_dxnet_nodes_mt} shows the results of the all-to-all benchmark with up to 8 nodes, 16 application send threads and 4 message handlers. Compared to the single threaded results (\S \ref{eval_dxnet_nodes_tp}), DXNet achieves slightly higher throughputs for all node counts: for two nodes, throughput saturates at 4 kb message size and peaks at 10.7 GB/s; for 4 nodes, throughput saturates at 4 kb message size and peaks at 19.5 GB/s; for 6 nodes, throughput saturates at 2 kb message size and peaks at 27.0 GB/s; for 8 nodes, throughput saturates at 2 kb message size and peaks at 33.6 GB/s. However, the message rate is improved significantly for small messages up to 64 byte with 8.4 to 10.3 mmps, 18.9 to 21.1 mmps, 27.6 to 31.4 mmps and 33.2 to 43.4 mmps for 2 to 8 nodes.

These results show that DXNet delivers high throughputs and message rates under high loads with increasing node and thread count. Small messages profit significantly through better aggregation and buffer utilization.

%%%%%%%%%%%%%%%%%%%%%%%%%%%
\subsubsection{Summary Results}

This section briefly summerizes the most important results and numbers of the previous benchmarks. All values are considered ``up to'' and show the possible peak performance in the given benchmark.

\textbf{Single-threaded}:
\begin{itemize}
 \item \textbf{Uni-directional throughput} One MH: saturation with 16 kb messages, peak throughput at 5.9 GB/s; Two MHs: saturation with 4 kb messages, peak throughput at 5.9 GB/s; Peak message rate of 4.0 to 4.5 mmps for small messages up to 64 bytes 
 \item \textbf{Bi-directional throughput} Saturation with 8 kb messages at 10.4 GB/s with one MH; Peak message rate of 6.0 to 6.9 mmps for small messages up to 64 bytes
 \item \textbf{Uni-directional latency} up to 512 byte messages: avg. 7.8 to 8.3 \textmu s, 95th percentile of 8.5 to 8.9 \textmu s, 99th percentile of 8.9 to 9.2, 99.9th percentile of 11.8 to 12.7 \textmu s; Peak message rate of 0.1 mmps.
 \item \textbf{All-to-all nodes} With 8 nodes: Total aggregated peak throughput of 32.4 GB/s, saturation with 16 kb message size; Peak message rate of 25.6 mmps for small messages up to 64 bytes.
\end{itemize}

\textbf{Multi-threaded}:
Overall, DXNet benefits from higher message aggregation through multiple outstanding messages in the ORB posted concurrently by many threads.
\begin{itemize}
 \item \textbf{Uni-directional throughput} Saturation at 5.9 GB/s at 4 kb message size
 \item \textbf{Bi-directional throughput} Overall improved throughput for many message sizes, saturation at 10.7 GB/s with 4 kb message size, message rate of 8.6 to 10.2 mmps for small messages up to 64 bytes
 \item \textbf{Uni-directional latency} Slightly higher latencies than single threaded as long as enough MHs serve available send threads. Message rate can be increased with additional send threads but at the cost of increasing avg. latency. The 99.9th percentiles roughly double when doubling the number of send threads.
 \item \textbf{All-to-all nodes}: With up to 8 nodes, 33.6 GB/s peak throughput, saturation at 2 kb message size, 33.2 to 43.4 mmps for up to 64 byte messages 
\end{itemize}

%%%%%%%%%%%%%%%%%%%%%%%%%%%%%%%%%%%%%%%%%%%%%%%%%%%%%%%%%%%%%%%%%%%%%%%%
\subsection{FastMPJ}
\label{eval_fmpj}

This section describes the results of the benchmarks executed with FastMPJ and compares them to the results of DXNet presented in the previous sections. We used FastMPJ 1.0\_7 with the device \textit{ibvdev} to run the benchmarks on InfiniBand hardware. The \textit{osu} benchmarks of MVAPICH2 were ported to Java (\S \ref{benchmarks}) and used for all following experiments. Since FastMPJ does not support multithreading in a single process, all benchmarks were executed single threaded and compared to the single threaded results of DXNet, only.

%%%%%%%%%%%%%%%%%%%%%%%%%%%%%%%%%%%%%%%%
\subsubsection{Uni-directional Throughput}
\label{eval_fmpj_uni_tp}

\begin{figure}[!t]
	\centering
	\includegraphics[width=3.4in]{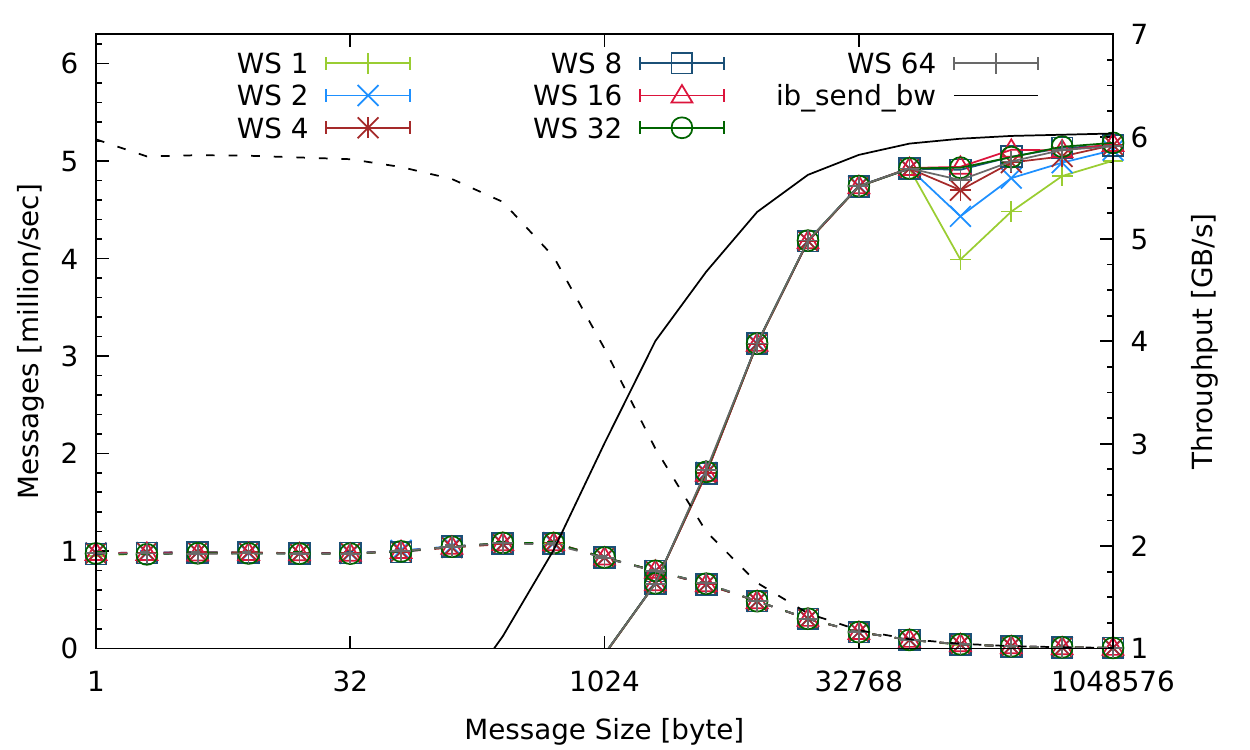}
	\caption{\textbf{FastMPJ}: 2 nodes, uni-directional throughput and message rate with increasing message and window size}
	\label{eval_fmpj_unibw}
\end{figure}

Figure \ref{eval_fmpj_unibw} shows the results of executing the uni-directional benchmark with two nodes with increasing message size. Furthermore, the benchmark was executed with increasing WS to ensure bandwidth saturation. As expected, throughput increases with increasing message size and bandwidth saturation starts at a medium message size of 64k with approx. 5.7 GB/s. The actual peak throughput is reached with large 512k message for a WS of 64 with 5.9 GB/s. 

For small message sizes up to 512 byte and independent of the WS, FastMPJ achieves a message rate of approx. 1.0 mmps. Furthermore, the results show that the WS doesn't matter for message sizes up to 64 KB. For 128 KB to 1 MB, FastMPJ profits from explicit aggregation with increasing WS. This indicates that ibvdev might include some message aggregation mechanism.

Compared to the baseline performance of \textit{ib\_send\_bw}, FastMPJ's performance is always inferior to it with a peak performance of 5.9 GB/s close to \textit{ib\_send\_bw}'s with 6.0 GB/s.

Compared to the results of DXNet (\S \ref{eval_dxnet_uni_tp}), DXNet's throughput saturates and peaks earlier at a message size of 16 kb with 5.9 GB/s. However, if using one MH, throughput drops for larger messages down to 5.4 GB/s due to increased message processing time (de-serialization). However, such a mechanism is absent from FastMPJ and DXNet can further improve performance by using two MHs. With two MHs, DXNet's throughput peaks even earlier at 5.9 GB/s with 4 kb message size. For small messages of up to 64 bytes, DXNet achieves 4.0 to 4.5 mmps compared to FastMPJ with 1.0 mmps.

%%%%%%%%%%%%%%%%%%%%%%%%%%%%%%%%%%%%%%%%
\subsubsection{Bi-directional Throughput}
\label{eval_fmpj_bi_tp}

\begin{figure}[!t]
	\centering
	\includegraphics[width=3.4in]{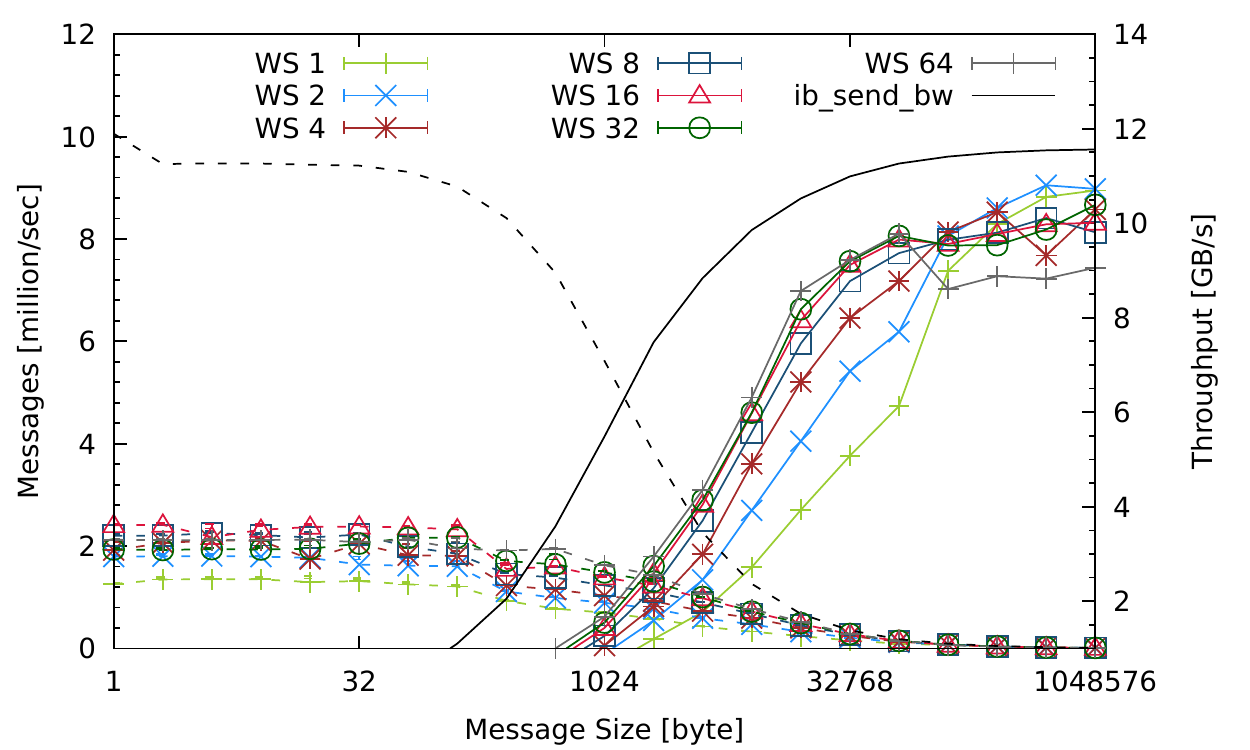}
	\caption{\textbf{FastMPJ}: 2 nodes, bi-directional (aggregated) throughput and message rate with increasing message and window size}
	\label{eval_fmpj_bibw}
\end{figure}

The results of the bi-directional benchmark are depicted in figure \ref{eval_fmpj_bibw}. Again, throughput increases with increasing message size peaking at 10.8 GB/s with WS 2 and large 512 kb messages. However, when handling messages of 128 kb and greater, throughput peaks at approx 10.2 GB/s for the WSs 4 to 32 and saturation varies depending on the WS. For WSs 4 to 32, throughput is saturated with 64 kb messages, for WSs 1 and 2 at 512 kb. Starting at 128 kb message size, WSs of 1 and 2 achieve slightly better results than the greater WSs. Especially WS 64 drops significantly with message sizes of 128 kb and greater. However, for message sizes of 64 kb to 512 kb, FastMPJ profits from explicit aggregation.

Compared to the uni-directional results (\S \ref{eval_fmpj_uni_tp}), FastMPJ does profit to some degree from explicit aggregation for small messages with 1 to 128 bytes. WS 1 to 16 allow higher message throughputs with WS 16 as an optimal value peaking at approx. 2.4 mmps for 1 to 128 byte messages. Greater WSs degrade message throughput significantly. However, this does not apply to message sizes of 256 bytes where greater explcit aggregation does always increase message throughput.

Compared to the baseline performance of \textit{ib\_send\_bw}, FastMPJ's performance is again always inferior to it with a difference in peak performance of 0.7 GB/sec (10.8 GB/s to 11.5 GB/s).

When comparing to DXNet's results (\S \ref{eval_dxnet_bi_tp}), the throughputs are nearly equal with 10.7 GB/s also at 512 kb message size. However, DXNet outperforms FastMPJ for medium sized messages by reaching a peak throughput of 10.4 GB/s for just 8 kb messages. Even with a WS of 64, FastMPJ can only achieve 6.3 GB/s aggregated throughput here. For small messages of up to 64 bytes, DXNet clearly outperforms FastMPJ with 6 to 7.2 mmops compared to 1.9 to 2.1 mmops with a WS of 16.

%%%%%%%%%%%%%%%%%%%%%%%%%%%%%%%%%%%%%%%%
\subsubsection{Uni-directional Latency}

\begin{figure}[!t]
	\centering
	\includegraphics[width=3.4in]{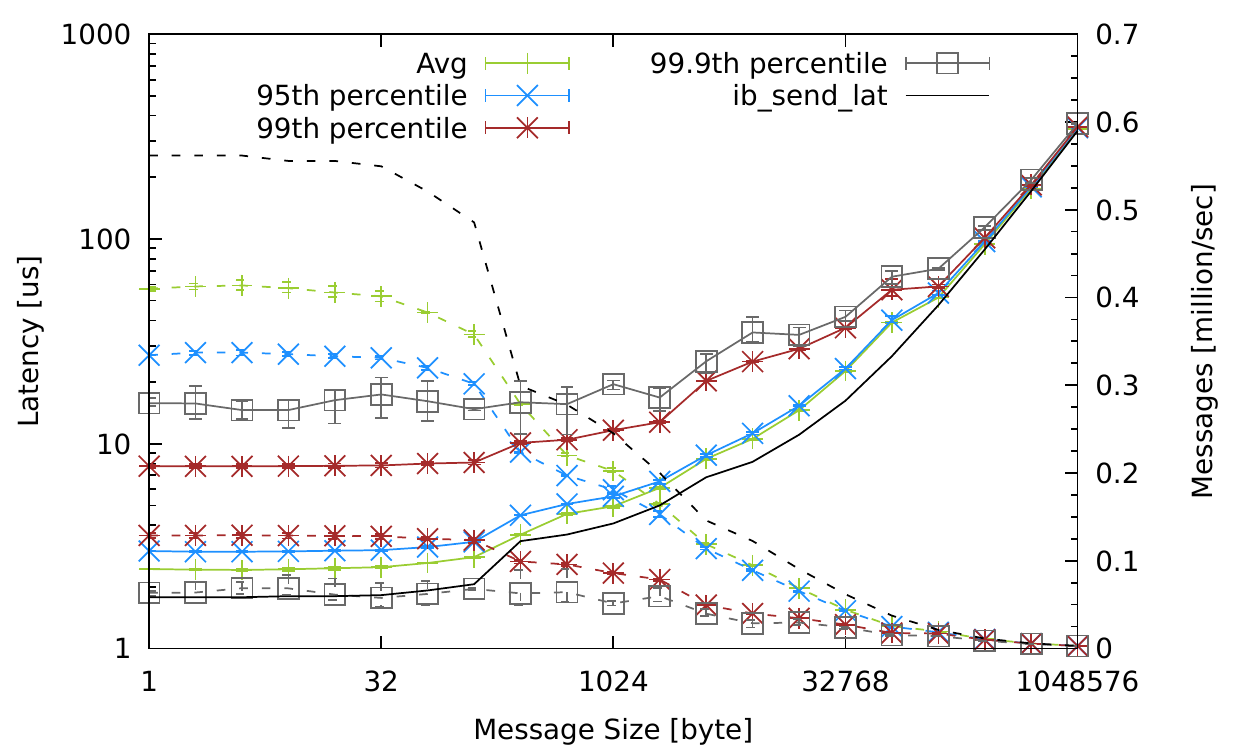}
	\caption{\textbf{FastMPJ}: 2 nodes, uni-directional latency and message rate with increasing message and window size}
	\label{eval_fmpj_lat}
\end{figure}

The results of the latency benchmark are depicted in figure \ref{eval_fmpj_lat}. FastMPJ achieves a very low average RTT of 2.4 \textmu s for up to 16 byte messages. This just slightly increases to 2.8 \textmu s for up to 128 byte messages and to 4.5 \textmu s for up to 512 byte messages. 3 \textmu s RTT is achieved by the 95th percentile for up to 64 byte which slightly increases to 5 \textmu s for up to 512 byte messages. Message sizes up to 16 bytes achieve a 7.7 \textmu s RTT for the 99th percentile with up to 10 \textmu s for up to 512 byte messages. For the 99.9th percentile, messages sizes up to 16 byte fluctuate slightly with a RTT of 14.5 to 15.5 \textmu s. This continues for 32 byte to 2 kb with a low of 16.3 \textmu s and a high of 19.5 \textmu s. The average message rate peaks at approx. 0.41 mmps for up to 16 byte messages.

Compared to the baseline performance of \textit{ib\_send\_lat}, FastMPJ's average RTT comes close to its 1.8 \textmu s and closes that gap slightly further starting with 256 byte message size. 

Comparing the avg. RTT and 95th percentile to DXNet's results (\S \ref{eval_dxnet_uni_lat}), FastMPJ outperforms DXNet by a up to four times lower RTT. This is also reflected by the message rate of 0.41 mmps for FastMPJ and 0.1 mmps for DXNet. The breakdown given Section \ref{eval_dxnet_uni_lat} explains the rather high RTTs and the amount of processing time spent by DXNet on major sections of the pipeline. However, even DXNet's avg. RTT for message sizes up to 512 byte is higher than FastMPJ's, DXNet achieves lower 99th (8.9 to 9.2 \textmu s) and 99.9th percentile (11.8 to 12.7 \textmu s) than FastMPJ.

%%%%%%%%%%%%%%%%%%%%%%%%%%%%%%%%%%%%%%%%
\subsubsection{All-to-all with Increasing Node Count}
\label{eval_fmpj_nodes}

\begin{figure}[!t]
	\centering
	\includegraphics[width=3.4in]{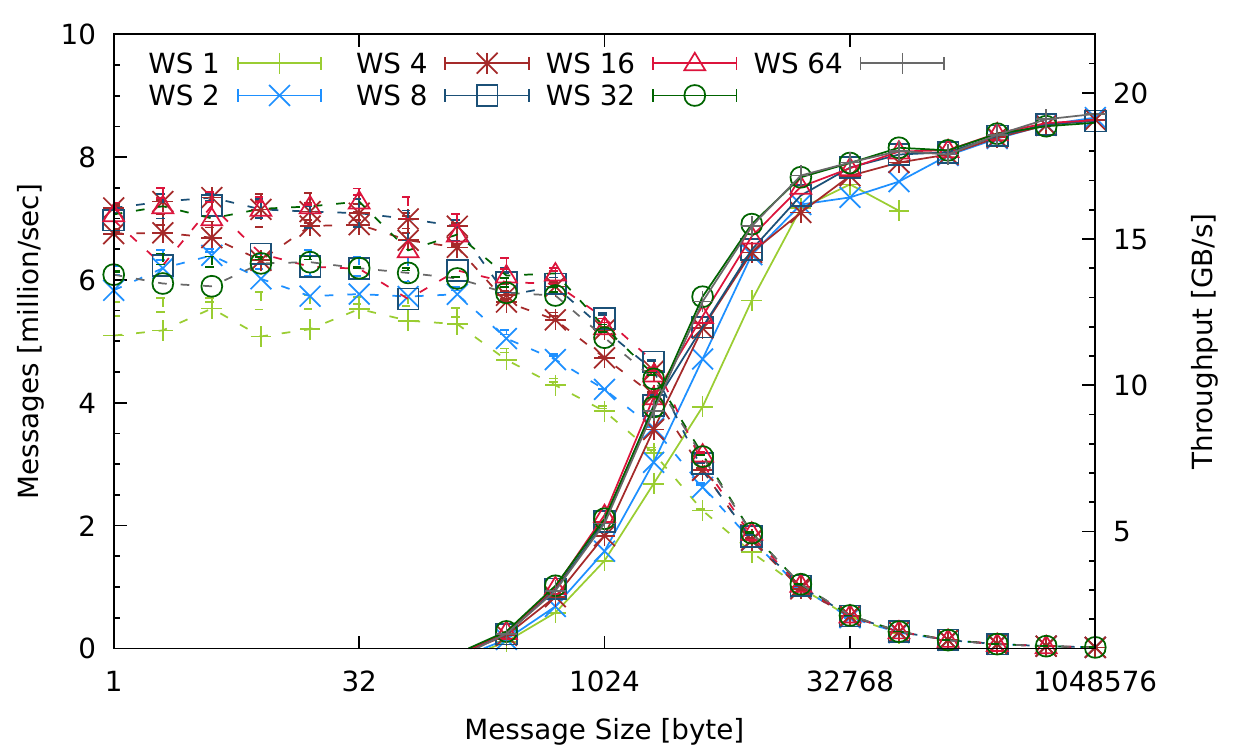}
	\caption{\textbf{FastMPJ}: 4 nodes, all-to-all (aggregated) throughput and message rate with increasing message and window size}
	\label{eval_fmpj_4nodes}
\end{figure}

\begin{figure}[!t]
	\centering
	\includegraphics[width=3.4in]{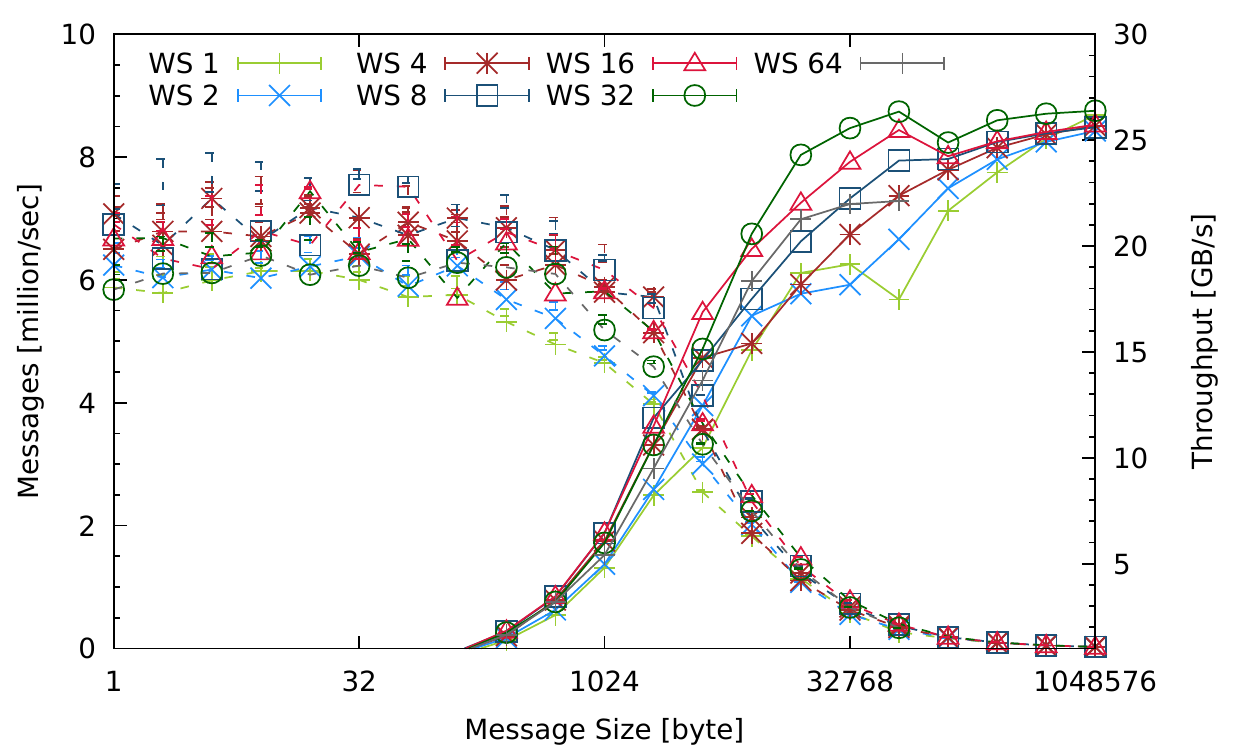}
	\caption{\textbf{FastMPJ}: 6 nodes, all-to-all (aggregated) throughput and message rate with increasing message and window size}
	\label{eval_fmpj_6nodes}
\end{figure}

\begin{figure}[!t]
	\centering
	\includegraphics[width=3.4in]{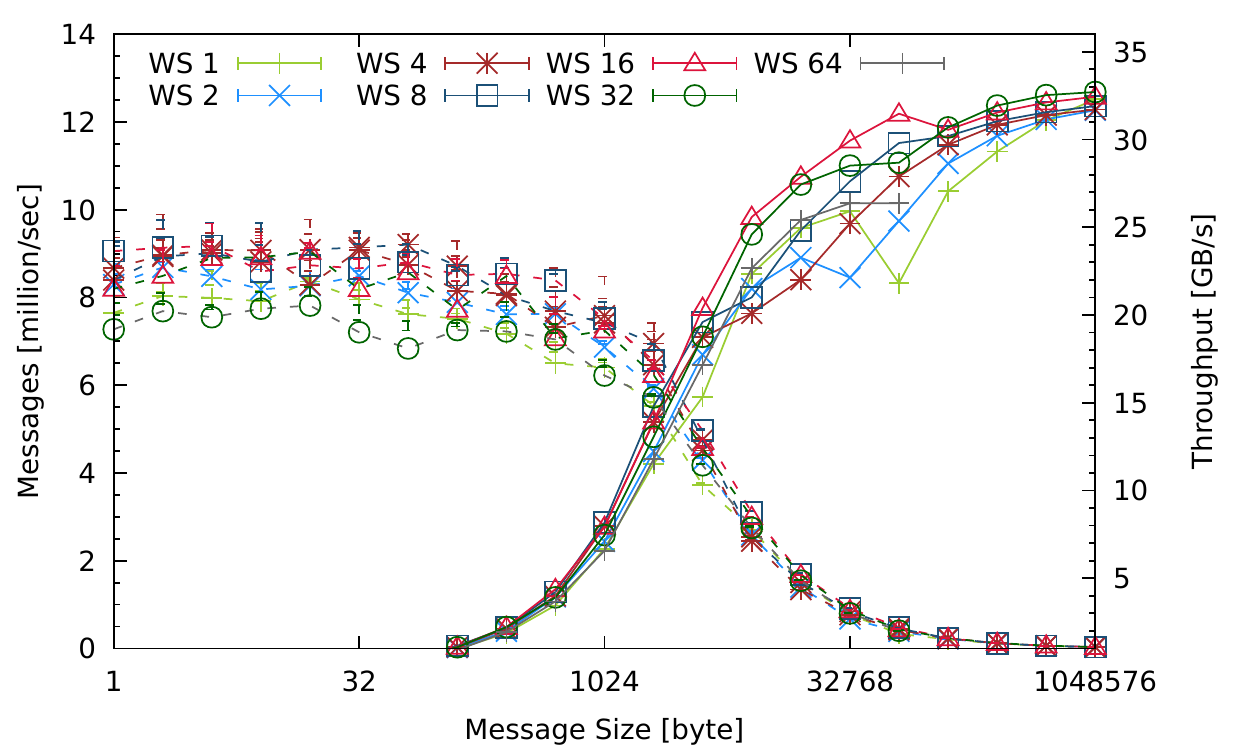}
	\caption{\textbf{FastMPJ}: 8 nodes, all-to-all (aggregated) throughput and message rate with increasing message and window size}
	\label{eval_fmpj_8nodes}
\end{figure}

Figures \ref{eval_fmpj_4nodes}, \ref{eval_fmpj_6nodes} and \ref{eval_fmpj_8nodes} show the aggregated send throughputs and message rates of the all-to-all benchmark running on 4, 6 and 8 nodes. The results for 2 nodes were already discussed in \ref{eval_fmpj_bi_tp} and are depicted in figure \ref{eval_fmpj_bibw}. The results for WS 64 and messages greater than 64 kb are absent because FastMPJ hangs (no error output) on message sizes greater than 64 kb with WS 64. We couldn't resolve this by re-running the benchmark several times and with different configuration parameters like increasing buffer sizes.

FastMPJ scales well with increasing node count on all-to-all communication with the following peak throughputs: 10.8 GB/s with WS 2 and 512 kb messages on 2 nodes, 19.2 GB/s with WS 64 and 1 MB messages on 4 nodes, 26.3 GB/s with WS 32 and 1 MB messages on 6 nodes, 32.7 GB/s with WS 32 and 1 MB messages on 8 nodes. This results in per node send throughputs of 5.1 GB/s, 4.8 GB/s, 4.38 GB/s and 4.08 GB/s. The gradually decreasing per node throughput seems to be a non software related issue as explained in Section \ref{eval_dxnet_nodes_tp}. For small messages up to 64 bytes, FastMPJ achieves the following peak message rates: 2.4 mmps for WS 16 on 2 nodes, 7.2 mmps for WS 16 on 4 nodes, 7.6 mmps for WS 16 on 6 nodes and 9.5 mmps for WS 8 on 8 nodes.

DXNet also reaches peak throughputs close to FastMPJ's (\S \ref{eval_dxnet_nodes_tp}) on all node counts. However, DXNet saturates bandwidth very early with just 8 kb and 16 kb message sizes. Furthermore, DXNet outperforms FastMPJ's message rates for small messages on all node counts by up to three times (7.0 mmps, 15.0 mmps, 21.1 mmps and 27.3 mmps).

%%%%%%%%%%%%%%%%%%%%%%%%%%%
\subsubsection{Summary Results}

This section briefly summerizes the most important results and key numbers of the previous benchmarks. All values are considered ``up to'' and show the possible peak performance in the given benchmark and are single-threaded, only. All results benefit from explicit aggregation using the WS.

\begin{itemize}
 \item \textbf{Uni-directional throughput} Saturation at 64 kb message size with 5.7 GB/s; Peak throughput at 512 kb message size with 5.9 GB/s; 1.0 mmps for message sizes up to 64 byte
 \item \textbf{Bi-directional throughput} Saturation at 64 kb message size with 10.8 GB/s; 2.4 mmps for message sizes up to 128 byte
 \item \textbf{Uni-directional latency} For up to 512 byte messages: avg. RTT of 2.4 to 4.5 \textmu s, 95th percentile of 3 to 5 \textmu s; 99th percentile of 7.7 to 10 \textmu s; 99.9th percentile of 16.3 to 19.5 \textmu s  
 \item \textbf{All-to-all nodes} With 8 nodes: Total aggregated peak throughput of 32.7 GB/s, saturation with 1 mb message size; Peak message rate of 9.5 mmps for small messages up to 64 byte
\end{itemize}

Compared to DXNet's single threaded results, it outperforms FastMPJ on small messages with a up to 4 times higher message rate on both un- und bi-directional benchmarks. However, FastMPJ achieves a lower average and 95th percentile latency on the uni-directional latency benchmark. But, even with a more complicated and dynamic pipeline, DXNet achieves lower 99th and 99.9th percentile than FastMPJ demonstrating high stability. On all-to-all communication with up to 8 nodes, DXNet reaches similar throughputs to FastMPJ's for large messages but outperforms FastMPJ's message rate by up to three times for small messages. \textbf{DXNet is always better for small messages}.

%%%%%%%%%%%%%%%%%%%%%%%%%%%%%%%%%%%%%%%%%%%%%%%%%%%%%%%%%%%%%%%%%%%%%%%%
\subsection{MVAPICH2}
\label{eval_mvapich2}

This section describes the results of the benchmarks executed with MVAPICH2 and compares them to the results of DXNet. All \textit{osu} benchmarks (\S \ref{benchmarks}) were executed with MVAPICH2-2.3. Since MVAPICH2 supports MPI calls with multiple threads of the same process, some benchmarks were executed single and multi-threaded. We set the following environmental variables for optimal performance and comparability: 
\begin{itemize}
 \item MV2\_DEFAULT\_MAX\_SEND\_WQE=128
 \item MV2\_DEFAULT\_MAX\_RECV\_WQE=128
 \item MV2\_SRQ\_SIZE=1024
 \item MV2\_USE\_SRQ=1
 \item MV2\_ENABLE\_AFFINITY=1
\end{itemize}

Additionally for the multi-threaded benchmarks, the following environmental variables were set: 
\begin{itemize}
 \item MV2\_CPU\_BINDING\_POLICY=hybrid
 \item MV2\_THREADS\_PER\_PROCESS=X (where X equals the number of threads we used when executing the benchmark)
 \item MV2\_HYBRID\_BINDING\_POLICY=linear
\end{itemize}

%%%%%%%%%%%%%%%%%%%%%%%%%%%%%%%%%
\subsubsection{Uni-directional Throughput}

\begin{figure}[!t]
	\centering
	\includegraphics[width=3.4in]{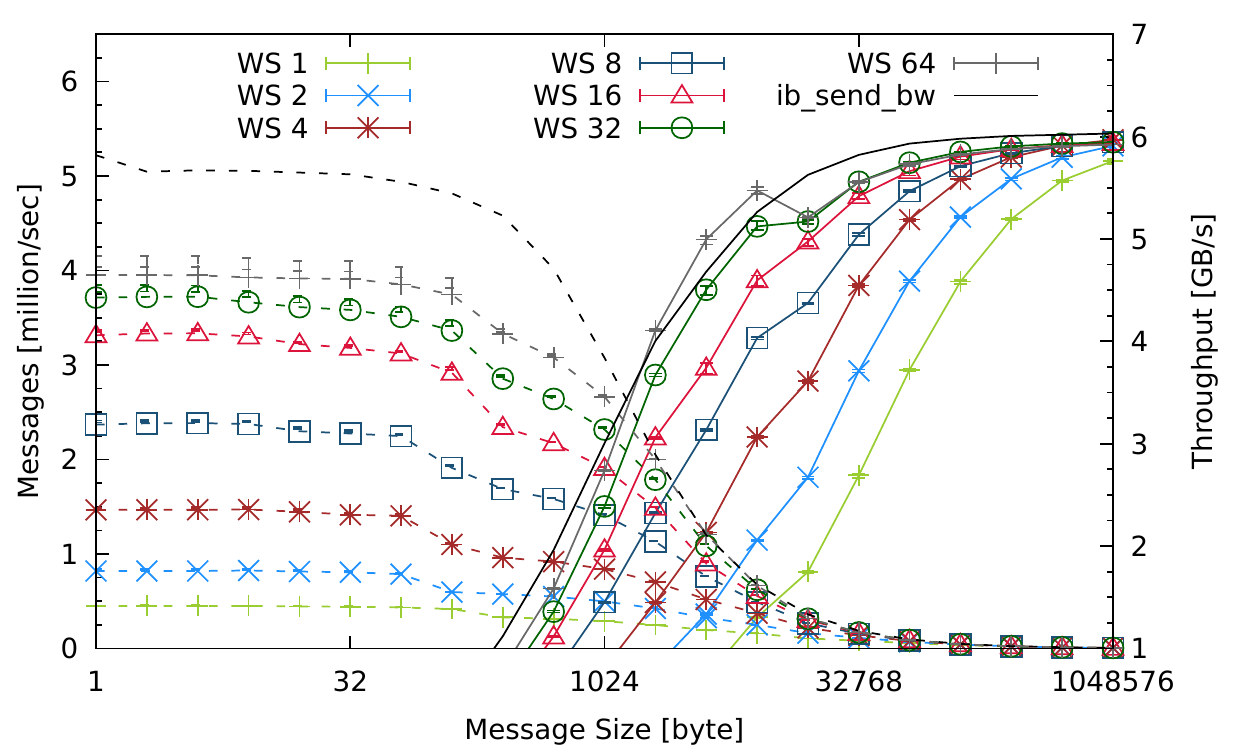}
	\caption{\textbf{MVAPICH2}: 2 nodes, uni-directional throughput and message rate, single threaded with increasing message and window size}
	\label{eval_mva_uni_st}
\end{figure}

The results of the uni-directional single threaded benchmark are depicted in figure \ref{eval_mva_uni_st}. The overall throughput increases with increasing message size, peaking at 5.9 GB/s with multiple WS on large messages: 512 kb with 64 WS, 256 kb with 32 WS, 512 KB with 8 WS, 512 kb with 4 WS and 1 MB with 2 WS. Bandwidth saturation starts at approx. 64 kb to 128 kb for WSs of 16 or greater. This also applies to smaller messages with up to 64 bytes. Reaching a peak of 4.0 mmps is only possible with WS 64. If send calls are not batched explicitly, message rates are rather low (0.45 mmps for WS 1).

Compared to the baseline performance of \textit{ib\_send\_bw}, MVAPICH2's peak performance is approx. 1.0 mmps less for small messages. With increasing message size, on a WS of 64, the performance comes close to the baseline and even exceeds it for 2 kb to 8 kb messages. MVAPICH2 peaks very close to the baseline's peak performance of 6.0 GB/s.

DXNet achieves very similar results (\S \ref{eval_dxnet_uni_tp}) compared to MVAPICH2 but without relying on explicit aggregation. DXNet's throughput saturates and peaks earlier at a message size of 16 kb with 5.9 GB/s. However, if using one MH, throughput drops for larger messages down to 5.4 GB/s due to increased message processing time (de-serialization). As already explained in Section \ref{eval_fmpj_uni_tp}, this can be resolved by using two MHs. For small messages of up to 64 bytes, DXNet achieves an equal to slightly higher message rate of 4.0 to 4.5 mmps.

%%%%%%%%%%%%%%%%%%%%%%%%%%%%%%%%%
\subsubsection{Bi-directional Throughput}
\label{eval_mva_bench_bi_st}

\begin{figure}[!t]
	\centering
	\includegraphics[width=3.4in]{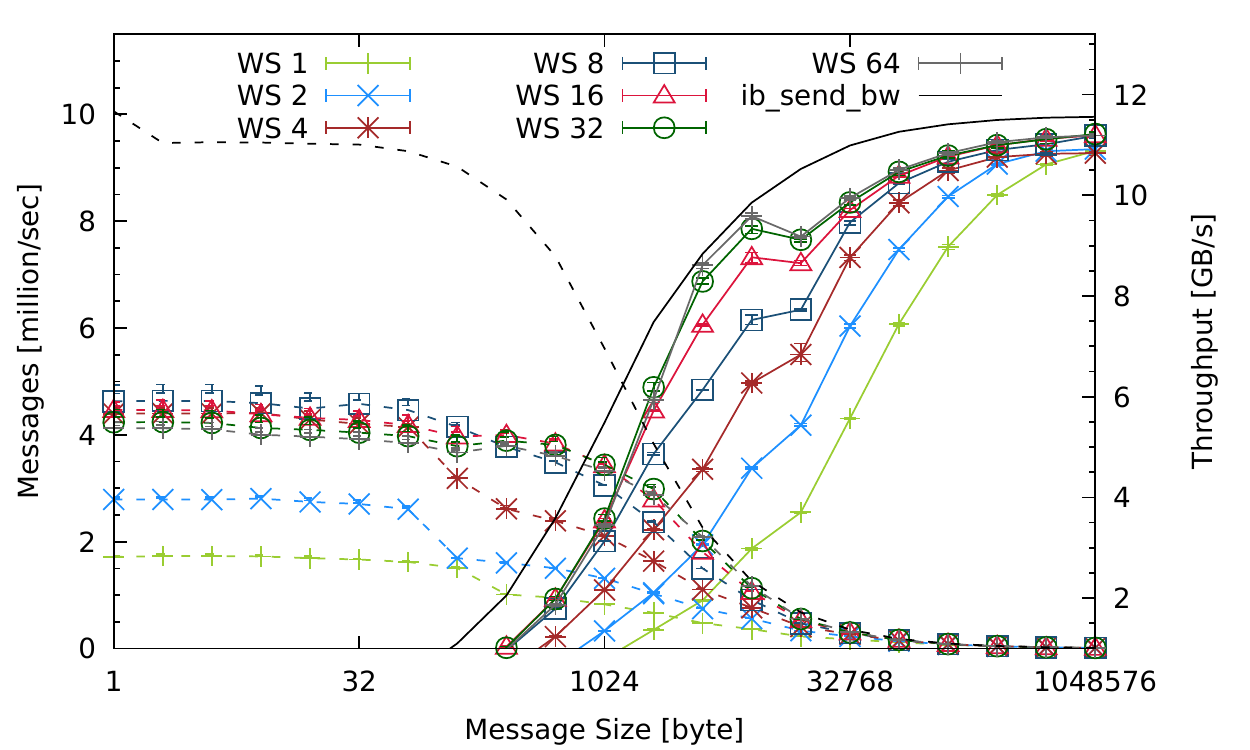}
	\caption{\textbf{MVAPICH2}: 2 nodes, bi-directional throughput and message rate, single threaded with increasing message and window size}
	\label{eval_mva_bi_st}
\end{figure}

The results of the bi-directional single threaded benchmark are depicted in figure \ref{eval_mva_bi_st}. Overall throughput increases with message size and, like on the uni-directional benchmark, benefits a lot from greater WSs. The aggregated throughput peaks at 11.1 GB/s with 512 kb messages on multiple WSs. Throughputs for 128 byte to 512 kb message sizes benefit from explicit aggregation. The message rate for small messages up to 64 bytes do not always profit from explicit aggregation. Message rate increases with WS 1 to 8 and peaks at 4.7 mmps with WS 8. However, greater WS degrade the message rate slightly compared to the optimal case.

Compared to the baseline performance of \textit{ib\_send\_bw}, MVAPICH2's peak performance for small messages is approx. half of \textit{ib\_send\_bw}'s 9.5 mmps. With increasing message size, the throughput of MVAPICH2 comes close \textit{ib\_send\_bw}'s with WS 64 and 32 for 4 and 8 kb messages, only. Peak throughput for large messages comes close to \textit{ib\_send\_bw}'s 11.5 GB/s.

Compared to DXNet's results (\S \ref{eval_dxnet_bi_tp}), the aggregated throughput is slightly higher than DXNet's (10.7 GB/s). However, DXNet outperforms MVAPICH2 for medium sized messages by reaching a peak throughput of 10.4 GB/s compared to 9.5 GB/s (on WS 64) for just 8 kb messages. Furthermore, DXNet offers a higher message rate of 6 to 7.2 mmps on small messages up to 64 bytes. DXNet achieves overall higher performance without relying on explicit message aggregation.

%%%%%%%%%%%%%%%%%%%%%%%%%%%%%%%%%
\subsubsection{Uni-directional Latency}

\begin{figure}[!t]
	\centering
	\includegraphics[width=3.4in]{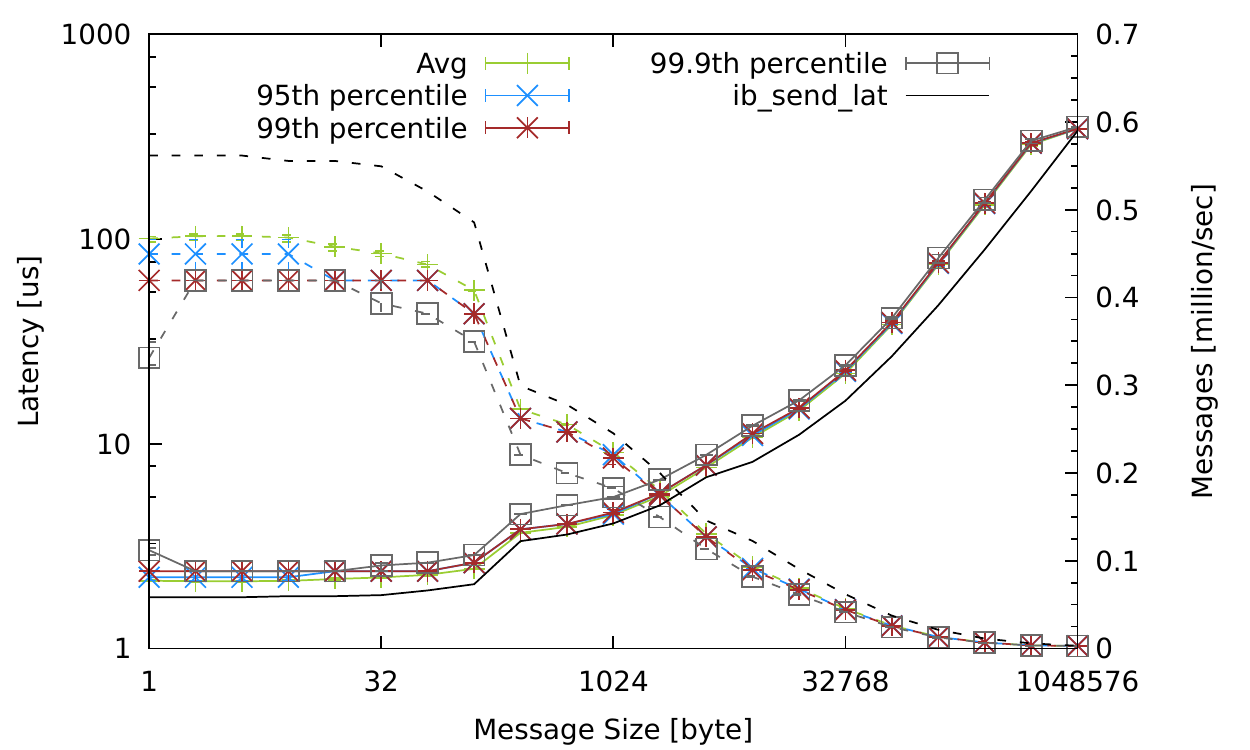}
	\caption{\textbf{MVAPICH2}: 2 nodes, uni-directional latency and message rate, single threaded with increasing message size}
	\label{eval_mva_uni_lat_st}
\end{figure}

Figure \ref{eval_mva_uni_lat_st} shows the results of the uni-directional single threaded latency benchmark. MVAPICH2 achieves a very low average RTT of 2.1 to 2.4 \textmu s for up to 64 byte messages and up to 3.9 \textmu s for up to 512 byte messages. The 95th, 99th and 99.9th percentile are just slightly higher than the average RTT with 2.2 to 4.0 \textmu s for the 95th, 2.4 to 4.0 \textmu s for the 99th and 2.4 to 5.0 \textmu s for the 99.9th (for up to 512 byte message size). This results in an average message rate of 0.43 to 0.47 mmps for up to 64 byte messages and 0.25 to 0.40 for 128 to 512 byte messages.

Compared to the baseline performance of \textit{ib\_send\_lat}, MVAPICH2's average, 95h, 99th, and 99.9th percentile RTT are very close to the baseline. With a minimum of 2.1 \textmu s for the average latency and maximum of 5.0 \textmu s for the 99.9th percentile on small messages, MVAPICH2 shows that its overall overhead is very low.

Compared to DXNet's results (\S \ref{fig_eval_dxnet_uni_lat}), MVAPICH2 achieves an overall lower latency. DXNet's average with 7.8 to 8.3 \textmu s is nearly four times higher. The 95h (8.5 to 8.9 \textmu s), 99th (8.9 to 9.2 \textmu s) and 99.9th percentile (11.8 to 12.7 \textmu s) are also at least two to three times higher. MVAPICH2 implements a very thin layer of abstraction, only. Application threads issuing MPI calls, are pinned to cores and are directly calling ibverbs functions after passing through these few layers of abstraction. DXNet however implements multiple pipeline stages with de/-serialization and multiple (JNI) context/thread switches. Naturally, data passing through such a long pipeline takes longer to process which impacts overall latency. However, DXNet traded latency for multithreading support and performance as well as efficient handling of small messages.

%%%%%%%%%%%%%%%%%%%%%%%%%%%
\subsubsection{All-to-all Throughput with up to 8 Nodes}
\label{eval_mva_nodes}

\begin{figure}[!t]
	\centering
	\includegraphics[width=3.4in]{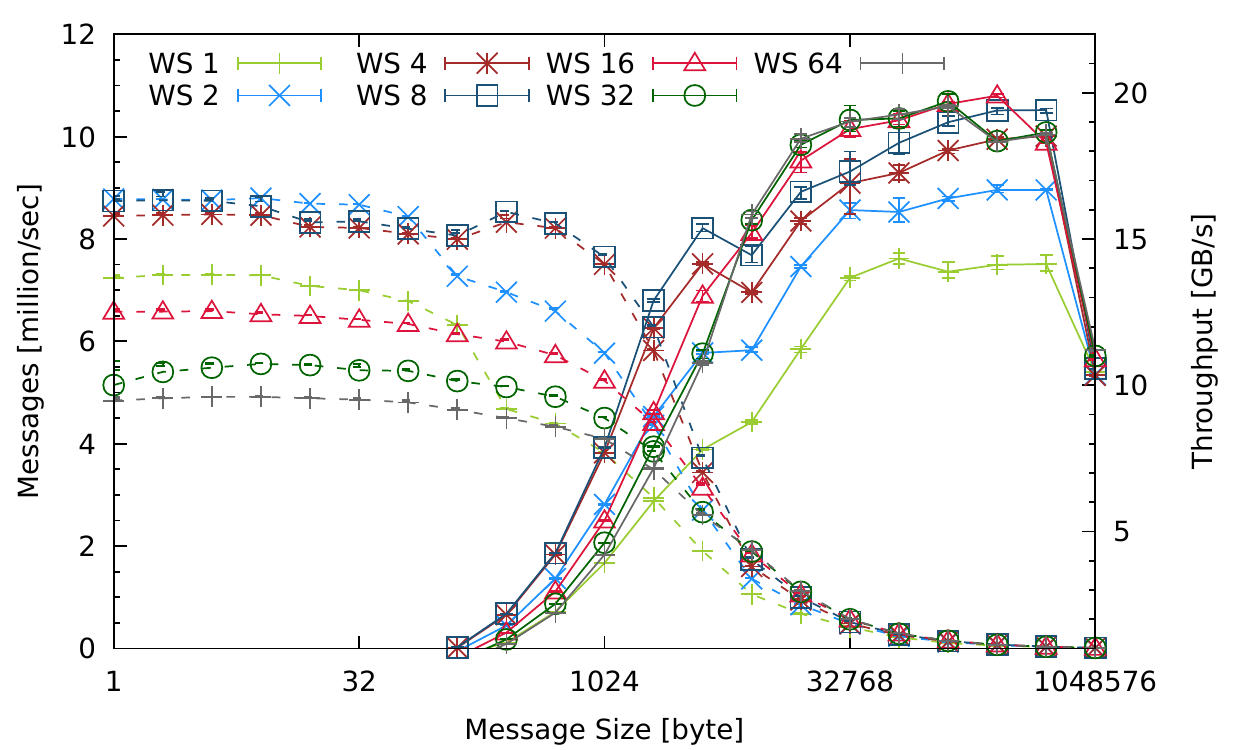}
	\caption{\textbf{MVAPICH2}: 4 nodes, aggregated send throughputs and message rates, single threaded with increasing message size}
	\label{eval_mva_nodes4}
\end{figure}

\begin{figure}[!t]
	\centering
	\includegraphics[width=3.4in]{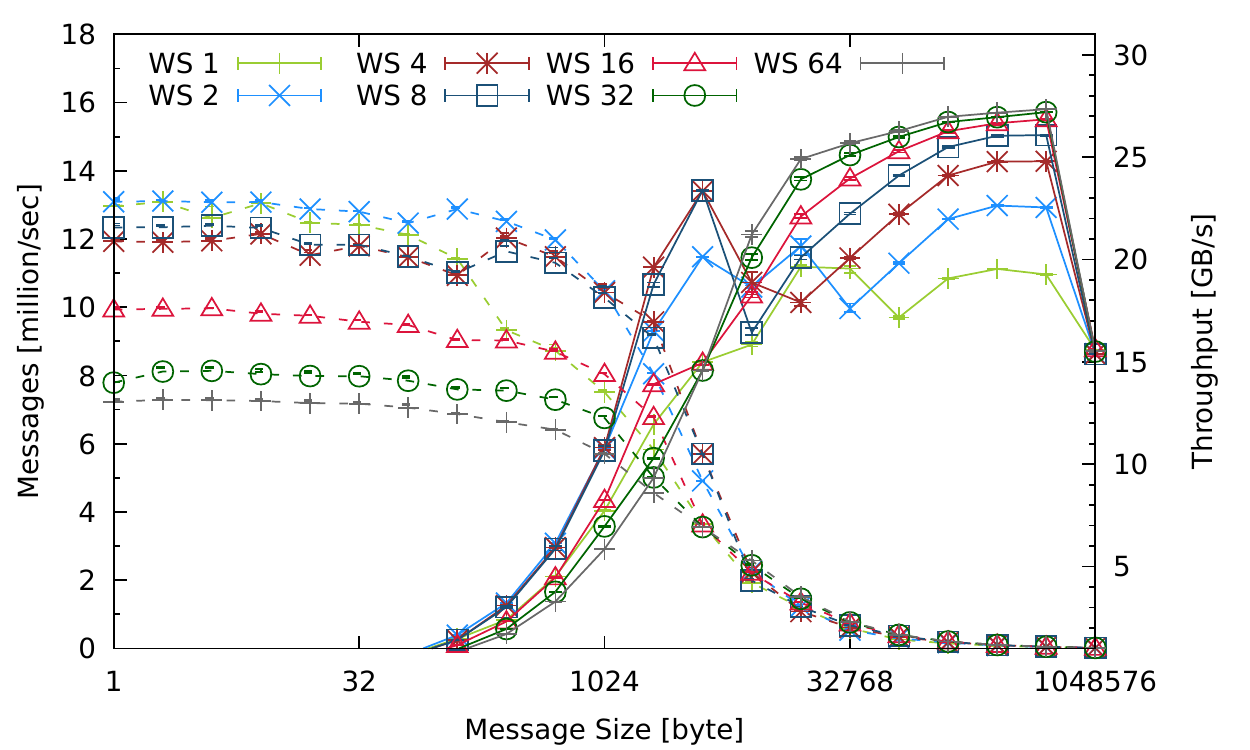}
	\caption{\textbf{MVAPICH2}: 6 nodes, aggregated send throughputs and message rates, single threaded with increasing message size}
	\label{eval_mva_nodes6}
\end{figure}

\begin{figure}[!t]
	\centering
	\includegraphics[width=3.4in]{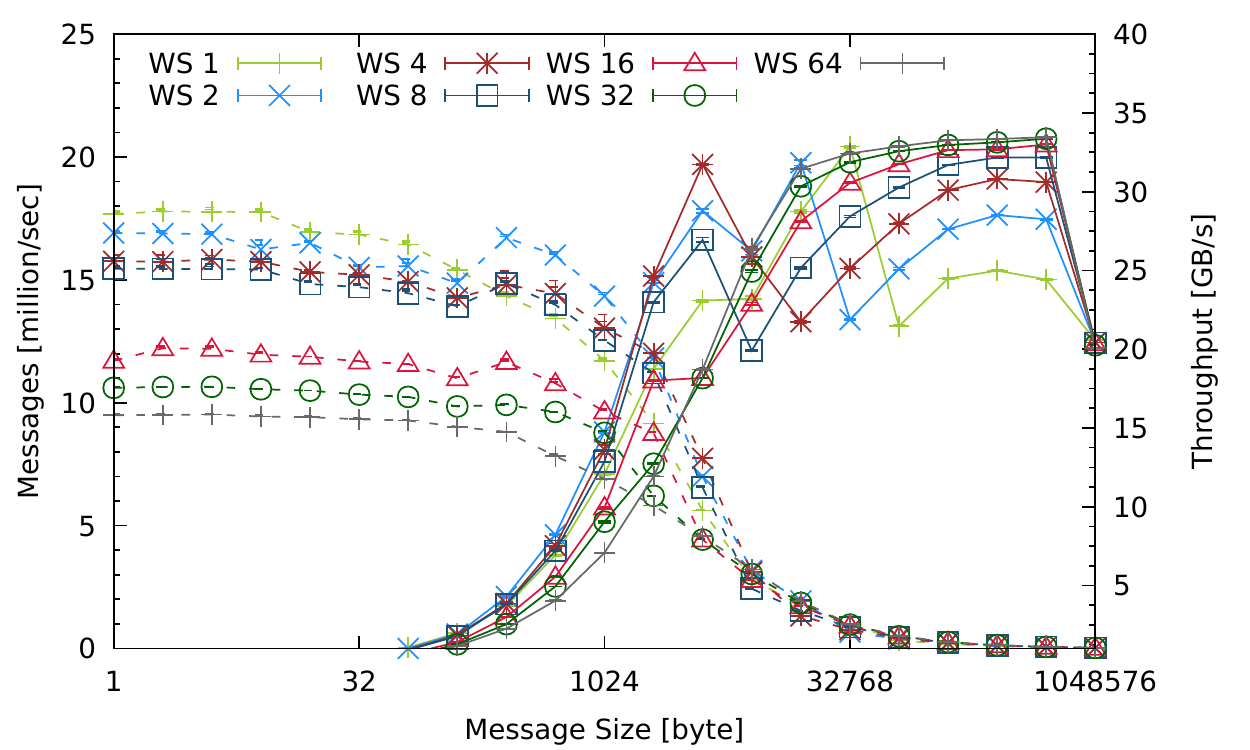}
	\caption{\textbf{MVAPICH2}: 8 nodes, aggregated send throughputs and message rates, single threaded with increasing message size}
	\label{eval_mva_nodes8}
\end{figure}

Figures \ref{eval_mva_nodes4}, \ref{eval_mva_nodes6} and \ref{eval_mva_nodes8} show the results of executing the all-to-all benchmark with 4, 6 and 8 nodes. The results for 2 nodes are depicted in figure \ref{eval_mva_bi_st}. 

MVAPICH2 achieves a peak throughput of 19.5 GB/s with 128 kb messages on WSs 16, 32 and 64 and starts at approx 32 kb message size. WS 8 gets close to the peak throughput as well but the remaining WSs peak lower for messages greater than 32 kb. Minor fluctuations appear for WS 1 to 16 for 1 kb to 16 kb messages. For small messages of up to 512 byte, the smaller the WS the better the performance. With WS 2, a message rate 8.4 to 8.8 mmps for up to 64 byte messages is achieved and 6.6 to 8.8 mmps for up to 512 byte. 

Running the benchmark with 6 nodes, MVAPICH2 hits a peak throughput of 27.3 GB/s with 512 kb messages on WSs 16, 32 and 64. Saturation starts with a message size of approx. 64 to 128 kb depending on the WS. For 1 kb to 32 kb messages, the fluctuations increased compared to executing the benchmark with 4 nodes. Again, message rate is degraded when using large WS for small messages. An optimal message rate of 11.9 to 13.1 is achieved with WS 2 for up to 64 byte messages.

With 8 nodes, the benchmark peaks at 33.3 GB/s with 64 kb messages on a WS of 64. Again, WS does matter for large messages as well with WS 16, 32 and 64 reaching the peak throughput and starting saturation at approx. 128 kb message size. The remaining WSs peak significantly lower. The fluctuations for mid range messages sizes of 1 kb to 64 kb increased further compared to 6 nodes. Most notable, the performance with 4 kb messages and WS 4 is nearly 10 GB/s better than 4 kb with WS 64. With up to 64 byte messages, a message rate of 16.5 to 17.8 mmps is achieved. For up to 512 byte messages, the message rate varies with 13.5 to 17.8 mmps. As with the previous node counts, a smaller WS increases the message rate significantly while larger WSs degrade performance by a factor of two.

MVAPICH2 has the same ``scalability issues'' as DXNet (\S \ref{eval_dxnet_nodes_tp}) and FastMPJ (\S \ref{eval_fmpj_nodes}). The maximum achievable bandwidth matches what was determined with the other systems. With the same results on three different systems, it's very unlikely that this is some kind of software issue like a bug or bad implementation but most likely a hardware limitation. So far, we haven't seen this issue discussed in any other publication and think it is noteworthy to know what the hardware is currently capable of.

Compared to DXNet (\S \ref{eval_dxnet_nodes_tp}), MVAPICH2 reaches slightly higher peak throughputs for large messages. However, this peak as well as saturation is reached later at 32 to 512 kb messages compared to DXNet with approx. 16 kb. The fluctuations for mid range size messages cannot be compared as DXNet does not rely on explicit aggregation. For small messages up to 64 byte, DXNet achieves significantly higher message rates, with peaks at 7.0 mmps, 15.0 mmps, 21.1 mmps and 27.3 mmps for 2 to 8 nodes, compared to MVAPICH2.

%%%%%%%%%%%%%%%%%%%%%%%%%%%%%%%%%
\subsubsection{Bi-directional Throughput Multi-threaded}

\begin{figure}[!t]
	\centering
	\includegraphics[width=3.4in]{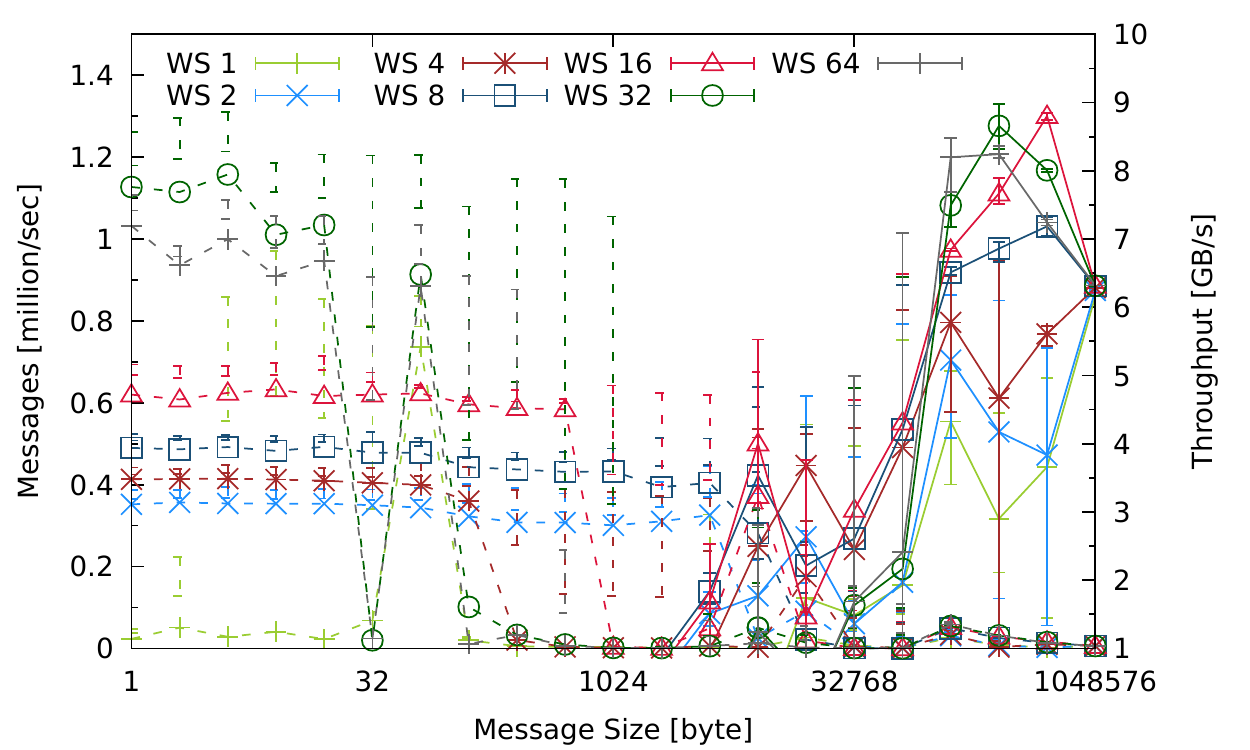}
	\caption{\textbf{MVAPICH2}: 2 nodes, bi-directional throughput and message rate, multi-threaded with one send and one recv thread with increasing message and window size}
	\label{eval_mva_bi_mt}
\end{figure}

Figure \ref{eval_mva_bi_mt} shows the results of the bi-directional multi-threaded benchmark with two threads (on each node): a separate thread for sending and receiving each. In our case, this is the simplest multi-threading configuration to utilize more than one thread for MPI calls. The plot shows highly fluctuating results of the three runs executed as well as overall low throughput compared to the single threaded results (\S \ref{eval_mva_bench_bi_st}). Throughput peaks at 8.8 GB/s with a message size of 512 kb for WS 16. A message rate of 0.78 to 1.19 mmps is reached for for up to 64 byte messages for WS 32.

We tried varying the configuration values (e.g. queue sizes, buffer sizes, buffer counts) but could not find configuration parameters that yielded significantly better, especially less fluctuating, results. Furthermore, the benchmarks could not be finished with sending 100,000,000 messages. When using \textit{MPI\_THREAD\_MULTIPLE}, the memory consumption increases continuously and exhausts the total memory available on our machine (64 GB). We reduced the number of messages to 1,000,000 which still consumes approx. 20\% of the total main memory but at least executes and finishes within a reasonable time. This does not happen with the widely used \textit{MPI\_THREAD\_SINGLE} mode.

MVAPICH2 implements multi-threading support using a single global lock for various MPI calls which includes \textit{MPI\_Isend} and \textit{MPI\_Irecv} used in the benchmark. This fulfils the requirements described in the MPI standard and avoids a complex architecture with lock-free data structures. However, a single global lock reduces concurrency significantly and does not scale well with increasing thread count \cite{MPIMultithreading17}. This effect impacts performance less on applications with short bursts and low thread count. However, for multi-threaded applications under high load, a single-threaded approach with one dedicated thread driving the network decoupled from the application threads, might be a better solution. Data between application threads and the network thread can be exchanged using data structures such as buffers, queues or pools like provided by DXNet.

MVAPICH2's implementation of multi-threading does not allow to improve performance by increasing the send or receive thread counts. Thus, further multi-threaded experiments using MVAPICH2 are not reasonable.

%%%%%%%%%%%%%%%%%%%%%%%%%%%
\subsubsection{Summary Results}

This section briefly summerizes the most important results and numbers of the previous benchmarks. All values are considered ``up to'' and show the possible peak performance in the given benchmark.
\textbf{Single-threaded}:
\begin{itemize}
 \item \textbf{Uni-directional throughput} Saturation with 64 kb to 128 kb message size, peak at 5.9 GB/s; Message rate of 4.0 mmps for up 64 byte messages
 \item \textbf{Bi-directional throughput} Saturation at 512 kb message size, peak at 11.1 GB/s; Message rate of 4.7 mmps for up to 64 byte messages
 \item \textbf{Uni-directional latency} For up to 64 byte message size: 2.1 to 2.4 \textmu s average latency and 2.4 to 5.0 \textmu s for 99.9th percentile; 0.43 to 0.47 mmps message rate 
 \item \textbf{All-to-all nodes} For 8 nodes: peak at 33.3 GB/s with 64 kb message size on WS 64, WS matters for large messages; Message rate of 16.5 to 17.8 mmps for up to 64 byte messages
 \item \textbf{Bi-directional throughput multi-threaded}: High fluctuations with low throughputs caused by global locking, 8.8 GB/s peak throughput at 512 kb message size; Message rate of 0.78 to 1.19 mmps for up to 64 byte messages
\end{itemize}

Compared to DXNet, the uni-directional results are similar but DXNet does not require explicit message aggregation to deliver high throughput. On bi-directional communication, MVAPICH2 achieves a slightly higher aggregated peak throughput than DXNet but DXNet performs better by approx 0.9 GB/s on medium sized messages. DXNet outperforms MVAPICH2 on small messages with a up to 1.8 times higher message rate. But, MVAPICH2 clearly outperforms DXNet on the uni-directional latency benchmark with an overall lower average, 95th, 99th and 99.9th percentile latency. On all-to-all communication with up to 8 nodes, MVAPICH2 reaches slightly higher peak throughputs for large messages but DXNet reaches its saturation earlier and performs significantly better on small message sizes up to 64 bytes.

The low multi-threading performance of MVAPICH2 cannot be compared to DXNet's due to the following reasons: First, MVAPICH2 implements synchronization using a global lock which is the most simplest but very often least performant method to ensure thread safety. Second, MVAPICH2, like many other MPI implementations, typically create multiple processes (one process per core) to enable concurrency on a single processor socket. However, as already discussed in related work (\S \ref{related_work}), this programming model is not suitable for all application domains, especially in big data applications.

\textbf{DXNet is better for small messages and multi-threaded access like required in big-data applications.}

%%%%%%%%%%%%%%%%%%%%%%%%%%%%%%%%%%%%%%%%%%%%%%%%%%%%%%%%%%%%%%%%%%%%%%%%
%%%%%%%%%%%%%%%%%%%%%%%%%%%%%%%%%%%%%%%%%%%%%%%%%%%%%%%%%%%%%%%%%%%%%%%%
%%%%%%%%%%%%%%%%%%%%%%%%%%%%%%%%%%%%%%%%%%%%%%%%%%%%%%%%%%%%%%%%%%%%%%%%
%%%%%%%%%%%%%%%%%%%%%%%%%%%%%%%%%%%%%%%%%%%%%%%%%%%%%%%%%%%%%%%%%%%%%%%%
%%%%%%%%%%%%%%%%%%%%%%%%%%%%%%%%%%%%%%%%%%%%%%%%%%%%%%%%%%%%%%%%%%%%%%%%
%%%%%%%%%%%%%%%%%%%%%%%%%%%%%%%%%%%%%%%%%%%%%%%%%%%%%%%%%%%%%%%%%%%%%%%%
\section{Conclusions}
\label{conclusions}

We presented Ibdxnet, a transport for the Java messaging library DXNet which allows multi-threaded Java applications to benefit from low latency and high-throughput using InfiniBand hardware. DXnet provides transparent connection management, concurrency handling, message serialization and hides the transport which allows the application to switch from Ethernet to InfiniBand hardware transparently, if the hardware is available. Ibdxnet's native subsystem provides dynamic, scalable, concurrent and automatic connection management and the msgrc messaging engine implementation. The msgrc engine uses a dedicated send and receive thread and to drive RC QPs asynchronously which ensures scalability with many nodes. Load adaptive parking avoids high loads on idle but ensures low latency when busy. SGEs are used to simplify buffer handling and increase buffer utilization when sending data provided by the higher level DXNet core. A carefully crafted architecture minimizes context switching between Java and the native space as well us exchanging data using shared memory buffers. The evaluation shows that DXNet with the Ibdxnet transport can keep up with FastMPJ and MVAPICH2 on single threaded applications and even exceed them in multi-threaded applications on high load applications. DXNet with Ibdxnet is capable of handling concurrent connections and data streams with up to 8 nodes. Furthermore, multi-threaded applications benefit significantly from the multi-threaded aware architecture.

The following topics are of interest for future research with DXnet and Ibdxnet:
\begin{itemize}
 \item Experiments with more than 100 nodes on our university's cluster
 \item Evaluate DXNet with the key-value store DXRAM using the YCSB and compare it to RAMCloud
 \item Implementation and evaluation of a UD QP based transport engine
 \item Hybrid mode for DXNet: Analyze if applications benefit from using Ethernet and InfiniBand if both are available
 \item RDMA path: Boost performance for applications like key-value storages
\end{itemize}

%%%%%%%%%%%%%%%%%%%%%%%%%%%%%%%%%%%%%%%%%%%%%%%%%%%%%%%%%%%%%%%%%%%%%%%%

\bibliography{paper}
\bibliographystyle{abbrv} 

\end{document}